\documentclass[twocolumn]{aastex63}

\usepackage{afterpage}

\newcommand{\muasyr}{$\mu$as yr$^{-1}$}
\newcommand{\muasyrsqr}{$\mu$as yr$^{-2}$}
\newcommand{\masyr}{mas yr$^{-1}$}
\newcommand{\masyrsqr}{mas yr$^{-2}$}
\newcommand{\kms}{km s$^{-1}$}

\newcommand{\kmsyr}{km s$^{-1}$ yr$^{-1}$}

\shorttitle{SiO Masers in the Galactic Center}
\shortauthors{Paine \& Darling}

\begin{document}

\title{3D Kinematics of Stellar SiO Masers in the Galactic Center}

\correspondingauthor{Jennie Paine}
\email{Jennie.Paine@colorado.edu}

\author[0000-0002-7517-9223]{Jennie Paine}
\affiliation{Center for Astrophysics and Space Astronomy, \\
Department of Astrophysical and Planetary Sciences, \\
University of Colorado, 389 UCB, Boulder, CO 80309-0389, USA}

\author{Jeremy Darling}
\affiliation{Center for Astrophysics and Space Astronomy, \\
Department of Astrophysical and Planetary Sciences, \\
University of Colorado, 389 UCB, Boulder, CO 80309-0389, USA}

\begin{abstract}

We present 3D velocity measurements and acceleration limits for stars within a few parsec of the Galactic Center (GC) black hole, Sgr A*, based on observations of 43 and 86 GHz circumstellar maser emission. Observations were taken with the Very Large Array (VLA) in 2013, 2014, and 2020 and with the Atacama Large Millimeter/submillimeter Array (ALMA) in 2015 and 2017. 
We detect 28 masers in total, of which four are new detections. 
Combining these data with extant maser astrometry, we calculate stellar proper motions and accelerations with uncertainties as low as $\sim$ 10 \muasyr~ and 0.5 \muasyrsqr, respectively, corresponding to approximately 0.5 \kms~ and 0.04 \kmsyr~ at a distance of 8 kpc. 
We measure radial velocities from maser spectra with $\sim 0.5$ \kms~ uncertainties, though the precision and accuracy of such measurements for deducing the underlying stellar velocities are limited by the complex spectral profiles of some masers. We therefore measure radial acceleration limits with typical uncertainties of $\sim 0.1$ \kmsyr.
We analyze the resulting 3D velocities and accelerations with respect to expected motions resulting from models of the mass distribution in the GC.

\end{abstract}

\keywords{Silicon monoxide masers --- Galactic center --- Stellar kinematics}

\section{Introduction} \label{sec:intro}

Observations of Galactic Center (GC) over the past several decades have identified stellar SiO masers within a few parsecs of the central galactic black hole Sgr A*. The SiO maser emission originates in the extended envelopes of late type stars and provides high brightness temperature sources for astrometry and narrow spectral profiles for Doppler velocity tracking. Thus, stellar masers are useful as high spatial and spectral resolution tracers of 3D stellar kinematics in the GC. 

Circumstellar SiO masers are associated with asymptotic giant branch (AGB) stars, which are late-type red giant stars (see \citealt{Kemball2007} for review). Several lines are observed at frequencies around 43 and 86 GHz, which are rotational transitions in  predominantly excited vibrational states, though a smaller number in ground vibrational states have been observed. Stellar SiO masers have been resolved with Very Long Baseline Interferometry (VLBI) observations, showing that the maser emission originates from discrete regions typically within about a few AU around the star (e.g. \citealt{Gonidakis2010}). When the emission is not resolved, SiO masers can be treated as point sources which track the stellar position. The cumulative spectrum typically shows maser emission in a range around $10$ \kms~ relative to the systemic stellar velocity and may have multiple distinct peaks. 

Astrometry and velocity tracking of SiO masers in the GC has been ongoing for several decades (e.g. \citealt{Menten1997,Reid2003,Reid2007,Li2010,Borkar2019}). Several  masers associated with infrared-bright stars have also been utilized to construct an astrometric reference frame to improve measurements of infrared (IR) stars on short-period orbits around Sgr A* (\citealt{Yelda2010}; \citealt{Plewa2015}; \citealt{Boehle2016}; \citealt{Sakai2019}). The orbits of these IR stars provided some of the first evidence for the existence of black holes and revealed the presence of a $4.15\times10^6$ M$_\odot$ mass black hole \citep{Ghez2008,Genzel2010}. The recent pericentre passage of the star S2 showed a detectable gravitational redshift \citep{GRAVITYCollaboration2018} and Schwarzschild precession \citep{GRAVITYCollaboration2020}. Such precision measurements depend on improvements to the astrometric reference frame derived from SiO masers. 

SiO maser stars may also directly probe Sgr A* and the surrounding environment. Stellar kinematics trace the underlying mass distribution, and therefore enable mapping of the total stellar and dark matter mass profiles at various radii in the GC. Additionally, if high velocity stellar masers with short orbital periods are identified, masers may enable measurements of the metric around Sgr A*. 
The required accuracy assumed for most general relativistic spectroscopic tests of stars near the central black hole is 10 \kms~ \citep{Angelil2011}, but the spectral resolution for SiO masers is at least two orders of magnitude better, with statistical uncertainties $\sim 10$ - 100 m~s$^{-1}$.  
Accurately tracking the systemic stellar velocity using the maser spectra is often a challenge as we demonstrate in this work. However, if the systematics can be quantified, general relativistic tests may be possible with SiO maser orbit measurements, such as tests of the equivalence principle or “no hair” theorem.

We present new observations from the Karl G. Jansky Very Large Array \footnote{The National Radio Astronomy Observatory is a facility of the National Science Foundation operated under cooperative agreement by Associated Universities, Inc.} (VLA) and Atacama Large Millimeter/submillimeter Array (ALMA) of stellar SiO masers within $\sim 2$ pc of Sgr A*. 
Observations of the 43 and 86 GHz masers and data reduction methods are detailed in Section \ref{sec:observations}. Results are presented in Section \ref{sec:results}, including stellar proper motions, proper accelerations, radial velocities, and radial accelerations. Analysis of the 3D stellar kinematics is given in Section \ref{sec:kinematics}. Finally, we discuss the implications of the stellar motions in Section \ref{sec:discussion} and give conclusions in Section \ref{sec:conclusions}.

\begin{deluxetable*}{lllccc}
\tablecaption{VLA and ALMA observations \label{tab:obs}}
\tablehead{
\colhead{Telescope} & \colhead{Date} & \colhead{MJD} & \colhead{Lines observed} & \colhead{Beam} & \colhead{Channel width}   \\
\colhead{} & \colhead{} & \colhead{} & \colhead{} & \colhead{(" $\times$ ")}  & \colhead{(\kms)} 
}
\startdata
VLA & 2013 Feb 14 & 2013.13 & $J=1-0$, $v=1$ and $v=2$ &  $3.398 \times 1.444$ & 1.0\\
VLA & 2014 Mar 7 & 2014.18  &  $J=1-0$, $v=1$ and $v=2$ & $0.1894 \times 0.1032$ & 1.0  \\
ALMA & 2015 Apr 10  & 2015.27 &  $J=2-1$, $v=1$ & $4.112 \times 2.512$ & 1.75 \\
ALMA & 2017 Sep 19 & 2017.72 &  $J=2-1$, $v=1$ & $0.625 \times 0.3125$ & 0.25 \\
VLA & 2020 Dec 27 & 2020.99 &  $J=1-0$, $v=1$ and $v=2$ &  $0.1230 \times 0.0483$ & 0.5  
\enddata

\end{deluxetable*}

\section{Observations and Data Reduction}\label{sec:observations}

Observations of the inner few parsec of the GC around Sgr A* were carried out over the course of seven years with VLA and ALMA. Three VLA sessions were used to observe the $v=1$ and $v=2$, $J=1-0$ lines at 43.12203 and 42.82048 GHz, and two  ALMA sessions were used to observe the $v=1$, $J=2-1$ line at 86.24337 GHz.
The ground state transitions $v=0$, $J=1-0$ and $v=0$, $J=2-1$ were also covered in our observations, but no ground state masers were observed in any VLA or ALMA sessions.
Initial observations with both VLA and ALMA were intended for high velocity maser searches and therefore have lower angular and spectral resolution than followup observations.  We detail the observations, calibration, and data reduction in the following subsections.
A summary of the observing sessions is given in Table \ref{tab:obs}.

\subsection{VLA Observations}\label{sec:vla}

We observed the GC with the VLA in the D configuration on 2013 February 14 (13A-071), and in the A configuration on 2014 March 7 (14A-440) and 2020 December 27 (19A-310). The phase center for all observations was Sgr A*. During the 2013 observation, the VLA had 26 antennas in operation with baselines ranging from $\sim$40 to 1486 m, resulting in a  $3.398" \times 1.444"$ synthesized beam.
The spectral setup consisted of 32 contiguous spectral windows, each with 1024 channels with 125 kHz channel width in dual polarization. The spectral windows covered frequencies from 42.1121 to 44.1276 GHz. 
During calibration, channels between frequencies 43.1171 and 43.1235 GHz  had to be discarded due to bandpass edge roll-off, which corresponds to velocities of about $-13$ - $+34$ \kms\  for  $v=1$, $J=1-0$ (43.12203 GHz) masers. 

The 2014 observation was conducted with 27 antennas in operation and baselines ranging from 793 to 36623 m. The synthesized beam was $0.1894" \times 0.1032"$. The spectral setup was similar to that of 2013, but covered a smaller range of frequencies. The observation had 16 contiguous spectral windows in dual polarization mode, each with 1024 channels with 125 kHz channel width, and covered frequencies from 42.6307  to 43.685 GHz.

The 2020 observation was conducted with 26 antennas in operation and baselines ranging from 793 to 36623 m. The synthesized beam was $0.1230" \times 0.0483"$. The spectral configuration included four spectral windows, each with 62.5 kHz wide channels in dual polarization mode. Three spectral windows contained 2048 channels and were centered on 42.612, 43.124, and 43.380 GHz with 128 MHz bandwidth each. A fourth smaller spectral window contained 1024 channels and was centered on 42.836 GHz with 64 MHz bandwidth.

We performed standard calibration using the Common Astronomy Software Application (CASA; \citealt{McMullin2007}). For all observations, 1733-1304 was used for bandpass and initial phase calibration and 1331+305 (3C286) for the flux density scale. Phases and amplitudes were self-calibrated with the Sgr A* continuum emission. The continuum emission was then subtracted in the $uv$-plane by fitting to the line-free channels. Initial spectral maps were made using the CLEAN algorithm, and these  images were used to identify masers by eye.
The measurement sets were regridded to the LSRK frame and channel widths of 1.0 \kms~ for the 2013 and 2014 data, and 0.5 \kms~ for the 2020 data.  
We then fit maser positions and spectra to the calibrated, continuum subtracted visibility data directly using the package \emph{uvmultifit} \citep{Marti-Vidal2014}. Positions are measured as offsets from the phase center, which is the Sgr A* continuum. Visibility fitting, rather than image fitting, was done in order to mitigate any systematics introduced between different telescopes, frequencies, and epochs.

\subsection{ALMA Observations}\label{sec:alma}

We observed the 86 GHz masers in the GC with ALMA on 2015 April 10 (2013.1.00834) and 2017 September 19 (2016.1.00940). The 2015 observation was conducted in the C34-1/(2) configuration, with 35 antennas in operation and baselines ranging from 15 to 348 m. The resulting synthesized beam was $4.112" \times 2.512"$. The spectral configuration consisted of four contiguous spectral windows in dual linear polarization, each with 1920 channels with 488.281 kHz channel width. The spectral windows covered frequencies from 84.842 to 88.5688 GHz. 

The 2017 observation was conducted in the more extended C40-8/9 configuration, with 44 antennas in operation and baselines ranging from 41 to 12145 m. The resulting synthesized beam was $0.625" \times 0.3125"$. The spectral setup contained four spectral windows in dual linear polarization which were intended for higher spectral resolution than the previous ALMA setup.  One window contained 3840 channels with 61.035 kHz width and was centered on 86.355026  GHz. The other three windows each contained 1920 channels with 122.07 kHz width and were centered on 85.752122, 86.958551, and 85.980837 GHz.

Calibrations for both observations were performed initially using the standard ALMA pipeline in CASA. Further self-calibration was done using the Sgr A* continuum emission. The remaining data reduction followed in a similar manner to that described above for the VLA observations. The Sgr A* continuum emission was subtracted from the visibility data by fitting to the line-free channels. Masers were initially identified by eye from spectral line maps created using the CLEAN algorithm to determine initial positions for visibility fitting. The measurement sets were regridded to the LSRK frame and channel widths of 1.75 \kms~ for the 2015 data and 0.25 \kms~ for the 2017 data. We then fit maser offset positions and spectra to the visibility data using the \emph{uvmultifit} package. 
 
\startlongtable
\begin{deluxetable*}{llcccc}
\tabletypesize{\scriptsize}
\tablecaption{Maser positions and velocities \label{tab:masers}}
\tablehead{
\colhead{Star} & \colhead{Epoch} & \colhead{Line} & \colhead{R.A. offset} & \colhead{Dec. offset} & \colhead{Radial velocity} \\
\colhead{} & \colhead{(year)} & \colhead{} & \colhead{(arcsec)} & \colhead{(arcsec)} & \colhead{(\kms)}
}
\startdata
IRS9 & 2013.13 & $J=1-0$, $v=1$ & $5.69 \pm 0.04$ & $-6.2 \pm 0.1$ & $-342.7 \pm 0.2$ \vspace{3pt} \\ 
 & 2014.18 & $J=1-0$, $v=1$ & $5.6986 \pm 0.0002$ & $-6.3157 \pm 0.0006$ & $-339.26 \pm 0.07$ \\ 
 & 2014.18 & $J=1-0$, $v=2$ & $5.6983 \pm 0.0003$ & $-6.3155 \pm 0.0006$ & $-337.82 \pm 0.07$ \vspace{3pt} \\ 
 & 2015.27 & $J=2-1$, $v=1$ & $5.98 \pm 0.05$ & $-6.30 \pm 0.03$ & $-342.0 \pm 0.1$ \vspace{3pt} \\ 
 & 2017.72 & $J=2-1$, $v=1$ & $5.7111 \pm 0.0002$ & $-6.3056 \pm 0.0002$ & $-341.36 \pm 0.01$ \vspace{3pt} \\ 
 & 2020.99 & $J=1-0$, $v=1$ & $5.7188 \pm 0.0003$ & $-6.3009 \pm 0.0006$ & $-341.57 \pm 0.04$ \\ 
\hline
IRS7 & 2014.18 & $J=1-0$, $v=1$ & $0.0323 \pm 0.0007$ & $5.487 \pm 0.002$ & $-106.8 \pm 0.2$ \\ 
 & 2014.18 & $J=1-0$, $v=2$ & $0.0334 \pm 0.0007$ & $5.487 \pm 0.002$ & $-104.96 \pm 0.08$ \vspace{3pt} \\ 
 & 2015.27 & $J=2-1$, $v=1$ & $0.034 \pm 0.003$ & $5.526 \pm 0.002$ & $-119.50 \pm 0.03$ \vspace{3pt} \\ 
 & 2017.72 & $J=2-1$, $v=1$ & $0.03304 \pm 0.00009$ & $5.47053 \pm 0.00009$ & $-120.28 \pm 0.03$ \vspace{3pt} \\ 
 & 2020.99 & $J=1-0$, $v=1$ & $0.038 \pm 0.001$ & $5.459 \pm 0.003$ & $-108.4 \pm 0.1$ \\ 
\hline
SiO-14 & 2013.13 & $J=1-0$, $v=1$ & $-7.57 \pm 0.04$ & $-28.2 \pm 0.1$ & $-112.2 \pm 0.2$ \\ 
 & 2013.13 & $J=1-0$, $v=2$ & $-7.56 \pm 0.03$ & $-28.39 \pm 0.08$ & $-109.5 \pm 0.2$ \vspace{3pt} \\ 
 & 2014.18 & $J=1-0$, $v=2$ & $-7.6321 \pm 0.0004$ & $-28.4614 \pm 0.0008$ & $-112.42 \pm 0.07$ \\ 
 & 2014.18 & $J=1-0$, $v=1$ & $-7.6312 \pm 0.0003$ & $-28.4612 \pm 0.0008$ & $-111.96 \pm 0.06$ \vspace{3pt} \\ 
 & 2015.27 & $J=2-1$, $v=1$ & $-7.552 \pm 0.007$ & $-28.413 \pm 0.004$ & $-112.76 \pm 0.02$ \vspace{3pt} \\ 
 & 2017.72 & $J=2-1$, $v=1$ & $-7.6247 \pm 0.0002$ & $-28.4698 \pm 0.0002$ & $-113.06 \pm 0.02$ \vspace{3pt} \\ 
 & 2020.99 & $J=1-0$, $v=2$ & $-7.6185 \pm 0.0005$ & $-28.467 \pm 0.001$ & $-115.92 \pm 0.08$ \\ 
 & 2020.99 & $J=1-0$, $v=1$ & $-7.6185 \pm 0.0005$ & $-28.468 \pm 0.001$ & $-115.55 \pm 0.07$ \\ 
\hline
SiO-28\tablenotemark{a} & 2014.18 & $J=1-0$, $v=2$ & $-1.113 \pm 0.002$ & $-42.555 \pm 0.005$ & $-103.8 \pm 0.2$ \\ 
 & 2014.18 & $J=1-0$, $v=1$ & $-1.112 \pm 0.002$ & $-42.559 \pm 0.004$ & $-103.11 \pm 0.08$ \vspace{3pt} \\ 
 & 2015.27 & $J=2-1$, $v=1$ & $-0.98 \pm 0.08$ & $-42.54 \pm 0.05$ & $-104.5 \pm 0.1$ \vspace{3pt} \\ 
 & 2017.72 & $J=2-1$, $v=1$ & $-1.114 \pm 0.002$ & $-42.568 \pm 0.002$ & $-105.48 \pm 0.03$ \vspace{3pt} \\ 
 & 2020.99 & $J=1-0$, $v=2$ & $-1.117 \pm 0.002$ & $-42.544 \pm 0.005$ & $-103.67 \pm 0.06$ \\ 
 & 2020.99 & $J=1-0$, $v=1$ & $-1.117 \pm 0.002$ & $-42.545 \pm 0.004$ & $-102.83 \pm 0.06$ \\ 
\hline
SiO-29\tablenotemark{a} & 2015.27 & $J=2-1$, $v=1$ & $-26.7 \pm 0.1$ & $23.57 \pm 0.07$ & $-93.0 \pm 0.2$ \vspace{3pt} \\ 
 & 2017.72 & $J=2-1$, $v=1$ & $-26.836 \pm 0.002$ & $23.604 \pm 0.002$ & $-90.20 \pm 0.07$ \\ 
\hline
SiO-31\tablenotemark{a} & 2020.99 & $J=1-0$, $v=1$ & $9.359 \pm 0.002$ & $-18.338 \pm 0.004$ & $-81.1 \pm 0.1$ \\ 
\hline
SiO-18 & 2014.18 & $J=1-0$, $v=1$ & $-18.684 \pm 0.001$ & $-26.095 \pm 0.003$ & $-78.5 \pm 0.1$ \vspace{3pt} \\ 
 & 2015.27 & $J=2-1$, $v=1$ & $-18.48 \pm 0.03$ & $-25.98 \pm 0.02$ & $-79.02 \pm 0.08$ \vspace{3pt} \\ 
 & 2017.72 & $J=2-1$, $v=1$ & $-18.625 \pm 0.002$ & $-26.077 \pm 0.002$ & $-74.57 \pm 0.07$ \vspace{3pt} \\ 
 & 2020.99 & $J=1-0$, $v=1$ & $-18.715 \pm 0.003$ & $-26.094 \pm 0.007$ & $-76.9 \pm 0.2$ \\ 
 & 2020.99 & $J=1-0$, $v=2$ & $-18.710 \pm 0.002$ & $-26.096 \pm 0.004$ & $-76.49 \pm 0.07$ \\ 
\hline
SiO-26 & 2015.27 & $J=2-1$, $v=1$ & $22.31 \pm 0.05$ & $23.49 \pm 0.03$ & $-72.5 \pm 0.1$ \vspace{3pt} \\ 
 & 2017.72 & $J=2-1$, $v=1$ & $22.499 \pm 0.001$ & $23.477 \pm 0.001$ & $-72.23 \pm 0.05$ \\ 
\hline
IRS12N & 2013.13 & $J=1-0$, $v=2$ & $-3.23 \pm 0.03$ & $-6.86 \pm 0.07$ & $-64.6 \pm 0.1$ \\ 
 & 2013.13 & $J=1-0$, $v=1$ & $-3.27 \pm 0.03$ & $-6.91 \pm 0.09$ & $-64.6 \pm 0.1$ \vspace{3pt} \\ 
 & 2014.18 & $J=1-0$, $v=2$ & $-3.2714 \pm 0.0004$ & $-6.931 \pm 0.001$ & $-66.02 \pm 0.08$ \\ 
 & 2014.18 & $J=1-0$, $v=1$ & $-3.2702 \pm 0.0004$ & $-6.9318 \pm 0.0008$ & $-65.41 \pm 0.07$ \vspace{3pt} \\ 
 & 2015.27 & $J=2-1$, $v=1$ & $-3.254 \pm 0.007$ & $-6.898 \pm 0.005$ & $-64.78 \pm 0.03$ \vspace{3pt} \\ 
 & 2017.72 & $J=2-1$, $v=1$ & $-3.2755 \pm 0.0001$ & $-6.9412 \pm 0.0001$ & $-62.64 \pm 0.01$ \vspace{3pt} \\ 
 & 2020.99 & $J=1-0$, $v=1$ & $-3.2796 \pm 0.0002$ & $-6.9494 \pm 0.0005$ & $-63.87 \pm 0.04$ \\ 
 & 2020.99 & $J=1-0$, $v=2$ & $-3.2793 \pm 0.0002$ & $-6.9491 \pm 0.0004$ & $-63.86 \pm 0.04$ \\ 
\hline
IRS28 & 2013.13 & $J=1-0$, $v=2$ & $10.50 \pm 0.05$ & $-5.9 \pm 0.1$ & $-53.8 \pm 0.1$ \vspace{3pt} \\ 
 & 2014.18 & $J=1-0$, $v=2$ & $10.4919 \pm 0.0002$ & $-5.8675 \pm 0.0004$ & $-53.84 \pm 0.03$ \\ 
 & 2014.18 & $J=1-0$, $v=1$ & $10.4916 \pm 0.0002$ & $-5.8673 \pm 0.0004$ & $-53.41 \pm 0.03$ \vspace{3pt} \\ 
 & 2015.27 & $J=2-1$, $v=1$ & $10.48 \pm 0.01$ & $-5.829 \pm 0.008$ & $-53.89 \pm 0.04$ \vspace{3pt} \\ 
 & 2017.72 & $J=2-1$, $v=1$ & $10.4965 \pm 0.0003$ & $-5.8875 \pm 0.0003$ & $-52.53 \pm 0.02$ \vspace{3pt} \\ 
 & 2020.99 & $J=1-0$, $v=2$ & $10.502 \pm 0.002$ & $-5.911 \pm 0.004$ & $-55.37 \pm 0.08$ \\ 
 & 2020.99 & $J=1-0$, $v=1$ & $10.5029 \pm 0.0009$ & $-5.914 \pm 0.002$ & $-52.91 \pm 0.08$ \\ 
\hline
SiO-27 & 2014.18 & $J=1-0$, $v=1$ & $-19.930 \pm 0.002$ & $33.672 \pm 0.005$ & $-44.0 \pm 0.2$ \vspace{3pt} \\ 
 & 2015.27 & $J=2-1$, $v=1$ & $-19.82 \pm 0.04$ & $33.68 \pm 0.03$ & $-43.67 \pm 0.07$ \vspace{3pt} \\ 
 & 2017.72 & $J=2-1$, $v=1$ & $-19.9361 \pm 0.0005$ & $33.6805 \pm 0.0005$ & $-43.50 \pm 0.03$ \vspace{3pt} \\ 
 & 2020.99 & $J=1-0$, $v=2$ & $-19.940 \pm 0.002$ & $33.697 \pm 0.005$ & $-44.3 \pm 0.4$ \\ 
 & 2020.99 & $J=1-0$, $v=1$ & $-19.940 \pm 0.002$ & $33.695 \pm 0.004$ & $-42.9 \pm 0.2$ \\ 
\hline
SiO-30\tablenotemark{a} & 2015.27 & $J=2-1$, $v=1$ & $-23.13 \pm 0.08$ & $21.38 \pm 0.05$ & $-38.6 \pm 0.1$ \vspace{3pt} \\ 
 & 2017.72 & $J=2-1$, $v=1$ & $-23.024 \pm 0.002$ & $21.314 \pm 0.002$ & $-37.34 \pm 0.04$ \\ 
\hline
SiO-15 & 2014.18 & $J=1-0$, $v=2$ & $-12.4597 \pm 0.0003$ & $-11.0673 \pm 0.0008$ & $-35.35 \pm 0.05$ \\ 
 & 2014.18 & $J=1-0$, $v=1$ & $-12.4593 \pm 0.0007$ & $-11.067 \pm 0.002$ & $-35.2 \pm 0.1$ \vspace{3pt} \\ 
 & 2015.27 & $J=2-1$, $v=1$ & $-12.42 \pm 0.03$ & $-11.01 \pm 0.02$ & $-36.45 \pm 0.06$ \vspace{3pt} \\ 
 & 2017.72 & $J=2-1$, $v=1$ & $-12.4667 \pm 0.0008$ & $-11.0682 \pm 0.0008$ & $-33.91 \pm 0.04$ \vspace{3pt} \\ 
 & 2020.99 & $J=1-0$, $v=2$ & $-12.474 \pm 0.001$ & $-11.070 \pm 0.002$ & $-36.1 \pm 0.1$ \\ 
 & 2020.99 & $J=1-0$, $v=1$ & $-12.475 \pm 0.001$ & $-11.070 \pm 0.003$ & $-35.8 \pm 0.2$ \\ 
\hline
SiO-19 & 2013.13 & $J=1-0$, $v=2$ & $16.22 \pm 0.06$ & $-21.6 \pm 0.2$ & $-28.8 \pm 0.1$ \vspace{3pt} \\ 
 & 2014.18 & $J=1-0$, $v=2$ & $16.2447 \pm 0.0003$ & $-21.6647 \pm 0.0007$ & $-28.71 \pm 0.07$ \\ 
 & 2014.18 & $J=1-0$, $v=1$ & $16.2442 \pm 0.0004$ & $-21.664 \pm 0.001$ & $-27.80 \pm 0.08$ \vspace{3pt} \\ 
 & 2015.27 & $J=2-1$, $v=1$ & $16.25 \pm 0.01$ & $-21.605 \pm 0.009$ & $-28.58 \pm 0.04$ \vspace{3pt} \\ 
 & 2017.72 & $J=2-1$, $v=1$ & $16.2549 \pm 0.0008$ & $-21.6745 \pm 0.0008$ & $-31.29 \pm 0.04$ \vspace{3pt} \\ 
 & 2020.99 & $J=1-0$, $v=1$ & $16.2626 \pm 0.0008$ & $-21.659 \pm 0.002$ & $-26.18 \pm 0.08$ \\ 
 & 2020.99 & $J=1-0$, $v=2$ & $16.261 \pm 0.001$ & $-21.651 \pm 0.003$ & $-25.80 \pm 0.07$ \\ 
\hline
IRS10EE & 2013.13 & $J=1-0$, $v=2$ & $7.68 \pm 0.01$ & $4.19 \pm 0.04$ & $-27.26 \pm 0.07$ \vspace{3pt} \\ 
 & 2014.18 & $J=1-0$, $v=2$ & $7.6838 \pm 0.0002$ & $4.1846 \pm 0.0004$ & $-27.48 \pm 0.02$ \\ 
 & 2014.18 & $J=1-0$, $v=1$ & $7.6839 \pm 0.0003$ & $4.1843 \pm 0.0006$ & $-26.98 \pm 0.03$ \vspace{3pt} \\ 
 & 2015.27 & $J=2-1$, $v=1$ & $7.68 \pm 0.02$ & $4.22 \pm 0.01$ & $-28.00 \pm 0.05$ \vspace{3pt} \\ 
 & 2017.72 & $J=2-1$, $v=1$ & $7.6844 \pm 0.0002$ & $4.1767 \pm 0.0002$ & $-28.56 \pm 0.01$ \vspace{3pt} \\ 
 & 2020.99 & $J=1-0$, $v=1$ & $7.68504 \pm 0.00009$ & $4.1701 \pm 0.0002$ & $-27.68 \pm 0.02$ \\ 
 & 2020.99 & $J=1-0$, $v=2$ & $7.68490 \pm 0.00008$ & $4.1702 \pm 0.0002$ & $-27.59 \pm 0.01$ \\ 
\hline
SiO-20 & 2014.18 & $J=1-0$, $v=1$ & $-13.8629 \pm 0.0004$ & $20.3643 \pm 0.0009$ & $-17.11 \pm 0.07$ \\ 
 & 2014.18 & $J=1-0$, $v=2$ & $-13.8625 \pm 0.0006$ & $20.362 \pm 0.001$ & $-16.9 \pm 0.1$ \vspace{3pt} \\ 
 & 2015.27 & $J=2-1$, $v=1$ & $-13.84 \pm 0.03$ & $20.41 \pm 0.02$ & $-17.13 \pm 0.08$ \vspace{3pt} \\ 
 & 2017.72 & $J=2-1$, $v=1$ & $-13.8604 \pm 0.0006$ & $20.3502 \pm 0.0006$ & $-16.87 \pm 0.02$ \vspace{3pt} \\ 
 & 2020.99 & $J=1-0$, $v=1$ & $-13.858 \pm 0.002$ & $20.343 \pm 0.005$ & $-20.8 \pm 0.1$ \\ 
 & 2020.99 & $J=1-0$, $v=2$ & $-13.857 \pm 0.003$ & $20.344 \pm 0.006$ & $-19.55 \pm 0.05$ \\ 
\hline
IRS15NE & 2014.18 & $J=1-0$, $v=2$ & $1.1968 \pm 0.0003$ & $11.2295 \pm 0.0007$ & $-11.81 \pm 0.05$ \\ 
 & 2014.18 & $J=1-0$, $v=1$ & $1.1970 \pm 0.0003$ & $11.2293 \pm 0.0006$ & $-11.04 \pm 0.07$ \vspace{3pt} \\ 
 & 2015.27 & $J=2-1$, $v=1$ & $1.24 \pm 0.03$ & $11.25 \pm 0.02$ & $-11.02 \pm 0.05$ \vspace{3pt} \\ 
 & 2017.72 & $J=2-1$, $v=1$ & $1.1903 \pm 0.0004$ & $11.2097 \pm 0.0004$ & $-11.78 \pm 0.03$ \vspace{3pt} \\ 
 & 2020.99 & $J=1-0$, $v=2$ & $1.1847 \pm 0.0005$ & $11.193 \pm 0.001$ & $-14.56 \pm 0.09$ \\ 
 & 2020.99 & $J=1-0$, $v=1$ & $1.1855 \pm 0.0008$ & $11.196 \pm 0.002$ & $-12.4 \pm 0.1$ \\ 
\hline
IRS14NE & 2014.18 & $J=1-0$, $v=1$ & $0.935 \pm 0.002$ & $-8.159 \pm 0.004$ & $-9.4 \pm 0.1$ \vspace{3pt} \\ 
 & 2015.27 & $J=2-1$, $v=1$ & $0.91 \pm 0.05$ & $-8.31 \pm 0.03$ & $-12.2 \pm 0.1$ \vspace{3pt} \\ 
 & 2017.72 & $J=2-1$, $v=1$ & $0.9468 \pm 0.0009$ & $-8.1701 \pm 0.0009$ & $-11.63 \pm 0.05$ \vspace{3pt} \\ 
 & 2020.99 & $J=1-0$, $v=2$ & $0.965 \pm 0.002$ & $-8.182 \pm 0.005$ & $-12.4 \pm 0.2$ \\ 
\hline
SiO-16 & 2015.27 & $J=2-1$, $v=1$ & $-26.33 \pm 0.03$ & $-34.39 \pm 0.02$ & $7.90 \pm 0.08$ \vspace{3pt} \\ 
 & 2017.72 & $J=2-1$, $v=1$ & $-26.4178 \pm 0.0009$ & $-34.4788 \pm 0.0009$ & $6.36 \pm 0.03$ \\ 
\hline
SiO-21 & 2015.27 & $J=2-1$, $v=1$ & $40.88 \pm 0.03$ & $-21.98 \pm 0.02$ & $13.46 \pm 0.06$ \vspace{3pt} \\ 
 & 2017.72 & $J=2-1$, $v=1$ & $40.911 \pm 0.001$ & $-22.043 \pm 0.001$ & $13.44 \pm 0.03$ \\ 
\hline
SiO-24 & 2015.27 & $J=2-1$, $v=1$ & $17.16 \pm 0.04$ & $-4.71 \pm 0.03$ & $18.10 \pm 0.08$ \vspace{3pt} \\ 
 & 2017.72 & $J=2-1$, $v=1$ & $17.1976 \pm 0.0008$ & $-4.8225 \pm 0.0008$ & $17.95 \pm 0.03$ \vspace{3pt} \\ 
 & 2020.99 & $J=1-0$, $v=1$ & $17.214 \pm 0.001$ & $-4.831 \pm 0.003$ & $19.0 \pm 0.1$ \\ 
 & 2020.99 & $J=1-0$, $v=2$ & $17.213 \pm 0.001$ & $-4.828 \pm 0.003$ & $19.5 \pm 0.1$ \\ 
\hline
SiO-22 & 2015.27 & $J=2-1$, $v=1$ & $41.44 \pm 0.03$ & $15.21 \pm 0.02$ & $32.79 \pm 0.06$ \vspace{3pt} \\ 
 & 2017.72 & $J=2-1$, $v=1$ & $41.4088 \pm 0.0007$ & $15.1811 \pm 0.0007$ & $35.41 \pm 0.03$ \\ 
\hline
SiO-6 & 2015.27 & $J=2-1$, $v=1$ & $35.15 \pm 0.02$ & $30.72 \pm 0.01$ & $51.58 \pm 0.09$ \vspace{3pt} \\ 
 & 2017.72 & $J=2-1$, $v=1$ & $35.2719 \pm 0.0005$ & $30.7032 \pm 0.0005$ & $52.17 \pm 0.05$ \\ 
\hline
SiO-17 & 2013.13 & $J=1-0$, $v=2$ & $8.111 \pm 0.009$ & $-27.47 \pm 0.02$ & $53.17 \pm 0.03$ \\ 
 & 2013.13 & $J=1-0$, $v=1$ & $8.10 \pm 0.02$ & $-27.51 \pm 0.04$ & $53.59 \pm 0.05$ \vspace{3pt} \\ 
 & 2014.18 & $J=1-0$, $v=2$ & $8.0811 \pm 0.0001$ & $-27.6578 \pm 0.0002$ & $53.37 \pm 0.02$ \\ 
 & 2014.18 & $J=1-0$, $v=1$ & $8.0811 \pm 0.0001$ & $-27.6585 \pm 0.0003$ & $53.73 \pm 0.02$ \vspace{3pt} \\ 
 & 2015.27 & $J=2-1$, $v=1$ & $8.07 \pm 0.02$ & $-27.64 \pm 0.02$ & $53.45 \pm 0.03$ \vspace{3pt} \\ 
 & 2017.72 & $J=2-1$, $v=1$ & $8.0900 \pm 0.0005$ & $-27.6538 \pm 0.0004$ & $53.40 \pm 0.02$ \vspace{3pt} \\ 
 & 2020.99 & $J=1-0$, $v=2$ & $8.0978 \pm 0.0005$ & $-27.641 \pm 0.001$ & $54.20 \pm 0.04$ \\ 
 & 2020.99 & $J=1-0$, $v=1$ & $8.0991 \pm 0.0007$ & $-27.641 \pm 0.002$ & $54.48 \pm 0.06$ \\ 
\hline
SiO-11 & 2015.27 & $J=2-1$, $v=1$ & $1.87 \pm 0.08$ & $40.36 \pm 0.05$ & $70.5 \pm 0.1$ \vspace{3pt} \\ 
 & 2017.72 & $J=2-1$, $v=1$ & $1.764 \pm 0.001$ & $40.316 \pm 0.001$ & $70.94 \pm 0.05$ \vspace{3pt} \\ 
 & 2020.99 & $J=1-0$, $v=2$ & $1.7714 \pm 0.0001$ & $40.3245 \pm 0.0003$ & $70.84 \pm 0.02$ \\ 
 & 2020.99 & $J=1-0$, $v=1$ & $1.7717 \pm 0.0002$ & $40.3241 \pm 0.0004$ & $71.85 \pm 0.02$ \\ 
\hline
IRS17 & 2014.18 & $J=1-0$, $v=2$ & $13.133 \pm 0.001$ & $5.550 \pm 0.002$ & $74.5 \pm 0.2$ \\ 
 & 2014.18 & $J=1-0$, $v=1$ & $13.132 \pm 0.002$ & $5.546 \pm 0.004$ & $75.6 \pm 0.3$ \vspace{3pt} \\ 
 & 2015.27 & $J=2-1$, $v=1$ & $13.13 \pm 0.02$ & $5.61 \pm 0.01$ & $74.13 \pm 0.05$ \vspace{3pt} \\ 
 & 2017.72 & $J=2-1$, $v=1$ & $13.1310 \pm 0.0003$ & $5.5480 \pm 0.0003$ & $71.81 \pm 0.02$ \vspace{3pt} \\ 
 & 2020.99 & $J=1-0$, $v=2$ & $13.126 \pm 0.001$ & $5.543 \pm 0.003$ & $72.31 \pm 0.06$ \\ 
 & 2020.99 & $J=1-0$, $v=1$ & $13.126 \pm 0.001$ & $5.543 \pm 0.003$ & $72.90 \pm 0.08$ \\ 
\hline
IRS19NW & 2013.13 & $J=1-0$, $v=1$ & $14.57 \pm 0.05$ & $-18.3 \pm 0.1$ & $84.4 \pm 0.1$ \\ 
 & 2013.13 & $J=1-0$, $v=2$ & $14.46 \pm 0.05$ & $-18.4 \pm 0.1$ & $84.4 \pm 0.1$ \vspace{3pt} \\ 
 & 2014.18 & $J=1-0$, $v=2$ & $14.5695 \pm 0.0003$ & $-18.4681 \pm 0.0007$ & $84.06 \pm 0.05$ \\ 
 & 2014.18 & $J=1-0$, $v=1$ & $14.5694 \pm 0.0003$ & $-18.4683 \pm 0.0006$ & $84.23 \pm 0.05$ \vspace{3pt} \\ 
 & 2015.27 & $J=2-1$, $v=1$ & $14.53 \pm 0.02$ & $-18.44 \pm 0.01$ & $84.08 \pm 0.05$ \vspace{3pt} \\ 
 & 2017.72 & $J=2-1$, $v=1$ & $14.5735 \pm 0.0008$ & $-18.4727 \pm 0.0008$ & $82.85 \pm 0.04$ \vspace{3pt} \\ 
 & 2020.99 & $J=1-0$, $v=2$ & $14.5772 \pm 0.0002$ & $-18.4724 \pm 0.0005$ & $82.60 \pm 0.02$ \\ 
 & 2020.99 & $J=1-0$, $v=1$ & $14.5774 \pm 0.0003$ & $-18.4728 \pm 0.0006$ & $83.68 \pm 0.04$ \\ 
\hline
SiO-25 & 2013.13 & $J=1-0$, $v=2$ & $-33.07 \pm 0.03$ & $-17.81 \pm 0.08$ & $118.19 \pm 0.07$ \\ 
 & 2013.13 & $J=1-0$, $v=1$ & $-33.05 \pm 0.03$ & $-17.71 \pm 0.08$ & $118.80 \pm 0.05$ \vspace{3pt} \\ 
 & 2014.18 & $J=1-0$, $v=2$ & $-33.056 \pm 0.002$ & $-17.716 \pm 0.005$ & $117.60 \pm 0.06$ \\ 
 & 2014.18 & $J=1-0$, $v=1$ & $-33.061 \pm 0.002$ & $-17.728 \pm 0.005$ & $118.62 \pm 0.06$ \vspace{3pt} \\ 
 & 2015.27 & $J=2-1$, $v=1$ & $-32.96 \pm 0.02$ & $-17.88 \pm 0.02$ & $118.01 \pm 0.06$ \vspace{3pt} \\ 
 & 2017.72 & $J=2-1$, $v=1$ & $-33.105 \pm 0.002$ & $-17.917 \pm 0.002$ & $117.63 \pm 0.03$ 
\enddata
\tablenotetext{a}{New detection}
\end{deluxetable*}

\begin{figure}[ht!]
    \centering
    \includegraphics[width=\columnwidth]{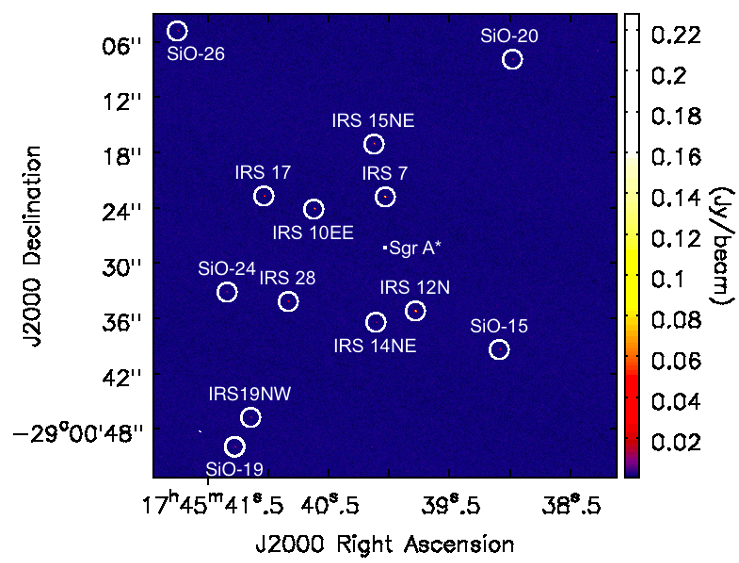}
	\caption{A subset of the 86 GHz SiO masers near Sgr A* detected by ALMA in 2017. The field of view spans $\pm1$ pc. The beam is $0.1617 \times 0.1176$ arcsec ($\sim 6.4 \times 4.7$ mpc) and is shown in the lower left.  \label{fig:ALMA masers}}
\end{figure}

\section{Results}\label{sec:results}

We required a $5\sigma$ significance flux measurement reported by \emph{uvmultifit} in at least one channel for detection. A subset of the masers within 1 pc projected distance from Sgr A* detected by ALMA in 2017 are shown in Figure \ref{fig:ALMA masers}. Maser positions where then determined by finding the error-weighted mean position for all channels with $>3\sigma$ measurements. The maser spectra, which are shown in Appendix~\ref{app:spectra}, were used to determine radial velocities by calculating the centroid of the $>3\sigma$ channels. 
Offset positions referenced to the Sgr A* continuum centroid and radial velocities of each maser are given in Table~\ref{tab:masers}. 

Two new masers were detected in the VLA data: we name these SiO-28 and SiO-31. Three new masers were detected in the ALMA data: SiO-28, SiO-29, and SiO-30. 
New masers were named using the convention of \cite{Reid2007},  \cite{Li2010}, and \cite{Borkar2019}. 
All new sources were detected in at least two epochs, except for SiO-31 which was only detected at 5$\sigma$-significance in two channels in the 2020 VLA data (see Figure~\ref{fig:spectra and fits}). Further observations will be required to confirm if SiO-31 is a genuine SiO maser source. 

Maser positions are measured with typical uncertainties $<1$ mas for the higher resolution epochs (2014.18, 2017.72, and 2020.99) and around 10 mas for the lower resolution epochs (2013.13 and 2015.27). These offset uncertainties are dominated by the maser position fit since the position of Sgr A* is well-measured in all epochs. Extended continuum emission from the GC mini-spiral was detected in both of the lower resolution epochs which was resolved out in the higher resolution data. We note that this emission may impact the resulting astrometry, but these epochs do not strongly affect the proper motion fits described in the next section since the statistical uncertainties of the offset positions are relatively large. 

The majority of our spectra show variability in time and between different lines (see Appendix \ref{app:spectra}). The most notable example is IRS 7, where the peak of maser emission from the 43 and 87 GHz spectra are significantly offset by more than 10 \kms, with strong SiO maser emission spread over $\sim 30$ \kms~ in the 87 GHz spectra. Since the VLA and ALMA spectra appear to be self-consistent over time, the variability of the IRS 7 spectra seems to be due to differences in the excitation conditions between the 43 and 87 GHz masers. Generally, however, SiO maser flux is known to be variable on timescales of $\sim 1$ year or longer following the underlying stellar pulsations (e.g. \citealt{Pardo2004}). Since the time interval between VLA and ALMA observations is $>1$ year, it is unclear if the smaller differences between 43 and 86 GHz spectra seen in the other stars in the sample are due to time-dependent variability in the star or differences in the excitation conditions. We also note that the peak and centroid velocities of the $J=1-0$, $v=1$ and $v-2$ spectra among stars with simultaneous detections of both lines often disagree, and we address this further in Section \ref{sec:doppler}.

The spectral variability of IRS 7 also indicates that its astrometry may not be as accurate as that of other stars in the sample. As noted by \cite{Reid2003}, the broad velocity range of the SiO emissions suggests that IRS 7 has a large maser shell that may span 10 mas. The maser position therefore may not match the stellar position to better than $\sim$5 mas. However, we report only the statistical uncertainties on the maser positions in Table \ref{tab:masers}.

\begin{deluxetable*}{lccccccc}
\tablecaption{Offsets, proper motions and proper accelerations of stellar masers \label{tab:pms}}
\tablewidth{700pt}
\tabletypesize{\scriptsize}

\tablehead{
\colhead{Star} & \colhead{$t_0$} & \colhead{$\Delta \alpha_0$} & \colhead{$\mu_\alpha$} &  \colhead{$a_\alpha$} & \colhead{$\Delta \delta_0$} &  \colhead{$\mu_\delta$} &  \colhead{$a_\delta$}  \\
\colhead{} & \colhead{(year)} & \colhead{(arcsec)} & \colhead{(\masyr)} &  \colhead{(\masyrsqr)} & \colhead{(arcsec)}  & \colhead{(\masyr)} &  \colhead{(\masyrsqr)} 
}
\startdata
IRS9 & $1998.39$ & $5.65051 \pm 0.00064$ & $3.057 \pm 0.042$ & $-0.0003 \pm 0.0036$ & $-6.3517 \pm 0.0011$ & $2.303 \pm 0.068$ & $-0.0001 \pm 0.0057$ \\ 
IRS7 & $1995.49$ & $0.0344 \pm 0.0050$ \tablenotemark{a} & $-0.04 \pm 0.11$ & $0.0037 \pm 0.0076$ & $5.5659 \pm 0.0050$\tablenotemark{a} & $-4.25 \pm 0.15$ & $-0.002 \pm 0.011$ \\ 
SiO-14 & $1998.41$ & $-7.66448 \pm 0.00034$ & $2.054 \pm 0.021$ & $-0.0007 \pm 0.0017$ & $-28.4517 \pm 0.0020$ & $-0.75 \pm 0.12$ & $-0.000 \pm 0.010$ \\ 
SiO-28 & $2014.18$ & $-1.11207 \pm 0.00037$ & $-0.712 \pm 0.080$ & $-0.006 \pm 0.023$ & $-42.5651 \pm 0.0087$ & $1.8 \pm 2.0$ & $0.31 \pm 0.54$ \\ 
SiO-29 & \nodata & \nodata & \nodata & \nodata & \nodata & \nodata & \nodata \\ 
SiO-31 & \nodata & \nodata & \nodata & \nodata & \nodata & \nodata & \nodata \\ 
SiO-18 & $2014.18$ & $-18.670 \pm 0.031$ & $-2.9 \pm 7.4$ & $-0.7 \pm 2.1$ & $-26.088 \pm 0.010$ & $1.0 \pm 2.7$ & $-0.43 \pm 0.77$ \\ 
SiO-26 & \nodata & \nodata & \nodata & \nodata & \nodata & \nodata & \nodata \\ 
IRS12N & $1996.41$ & $-3.25154 \pm 0.00038$ & $-1.120 \pm 0.023$ & $-0.0007 \pm 0.0018$ & $-6.88149 \pm 0.00039$ & $-2.778 \pm 0.023$ & $0.0004 \pm 0.0018$ \\ 
IRS28 & $1998.41$ & $10.4674 \pm 0.0010$ & $1.541 \pm 0.063$ & $-0.0005 \pm 0.0050$ & $-5.7813 \pm 0.0029$ & $-5.54 \pm 0.18$ & $-0.007 \pm 0.014$ \\ 
SiO-27 & $2014.18$ & $-19.93043 \pm 0.00061$ & $-1.52 \pm 0.13$ & $0.015 \pm 0.031$ & $33.6670 \pm 0.0028$ & $3.94 \pm 0.70$ & $0.08 \pm 0.15$ \\ 
SiO-30 & \nodata & \nodata & \nodata & \nodata & \nodata & \nodata & \nodata \\ 
SiO-15 & $2000.85$ & $-12.42632 \pm 0.00099$ & $-2.426 \pm 0.065$ & $0.0026 \pm 0.0053$ & $-11.0754 \pm 0.0031$ & $0.45 \pm 0.21$ & $-0.009 \pm 0.017$ \\ 
SiO-19 & $2014.18$ & $16.24461 \pm 0.00047$ & $2.63 \pm 0.12$ & $-0.011 \pm 0.035$ & $-21.6666 \pm 0.0045$ & $0.2 \pm 1.4$ & $0.22 \pm 0.42$ \\ 
IRS10EE & $1995.49$ & $7.68371 \pm 0.00024$ & $0.036 \pm 0.014$ & $0.0003 \pm 0.0010$ & $4.22082 \pm 0.00044$ & $-1.979 \pm 0.026$ & $-0.0001 \pm 0.0019$ \\ 
SiO-20 & $2014.18$ & $-13.86276 \pm 0.00018$ & $0.717 \pm 0.062$ & $0.005 \pm 0.020$ & $20.36321 \pm 0.00093$ & $-3.47 \pm 0.35$ & $0.07 \pm 0.13$ \\ 
IRS15NE & $1995.49$ & $1.23178 \pm 0.00039$ & $-1.852 \pm 0.023$ & $0.0006 \pm 0.0017$ & $11.33613 \pm 0.00079$ & $-5.661 \pm 0.048$ & $0.0013 \pm 0.0037$ \\ 
IRS14NE & $2014.18$ & $0.9334 \pm 0.0033$ & $4.29 \pm 0.85$ & $0.09 \pm 0.23$ & $-8.15823 \pm 0.00068$ & $-3.38 \pm 0.19$ & $-0.024 \pm 0.049$ \\ 
SiO-16 & $2008.42$ & $-26.44 \pm 0.13$ & $2 \pm 14$ & $-0.9 \pm 2.4$ & $-34.087 \pm 0.075$ & $-42.1 \pm 8.1$ & $0.2 \pm 1.0$ \\ 
SiO-21 & $2008.42$ & $40.754 \pm 0.022$ & $17.0 \pm 2.4$ & $-0.11 \pm 0.40$ & $-21.795 \pm 0.018$ & $-26.7 \pm 1.9$ & $-0.02 \pm 0.24$ \\ 
SiO-24 & $2015.27$ & $17.1853 \pm 0.0011$ & $4.99 \pm 0.26$ & $-0.002 \pm 0.063$ & $-4.8159 \pm 0.0084$ & $-2.5 \pm 2.4$ & $0.04 \pm 0.59$ \\ 
SiO-22 & $2008.42$ & $41.319 \pm 0.073$ & $9.7 \pm 7.8$ & $-0.5 \pm 1.4$ & $15.38 \pm 0.12$ & $-22 \pm 13$ & $0.4 \pm 1.6$ \\ 
SiO-6 & $2008.42$ & $34.98 \pm 0.12$ & $32 \pm 13$ & $0.7 \pm 1.9$ & $30.792 \pm 0.092$ & $-9.6 \pm 9.9$ & $0.1 \pm 1.2$ \\ 
SiO-17 & $1998.41$ & $8.04279 \pm 0.00044$ & $2.450 \pm 0.026$ & $0.0003 \pm 0.0022$ & $-27.7040 \pm 0.0027$ & $2.78 \pm 0.17$ & $-0.001 \pm 0.014$ \\ 
SiO-11 & $2008.42$ & $1.7424 \pm 0.0083$ & $2.32 \pm 0.69$ & $-0.002 \pm 0.064$ & $40.296 \pm 0.019$ & $2.3 \pm 1.5$ & $0.01 \pm 0.14$ \\ 
IRS17 & $2000.85$ & $13.1477 \pm 0.0012$ & $-1.052 \pm 0.078$ & $0.0018 \pm 0.0066$ & $5.5657 \pm 0.0015$ & $-1.083 \pm 0.097$ & $-0.0002 \pm 0.0093$ \\ 
IRS19NW & $1998.41$ & $14.55147 \pm 0.00028$ & $1.143 \pm 0.017$ & $-0.0000 \pm 0.0013$ & $-18.4621 \pm 0.0012$ & $-0.460 \pm 0.067$ & $-0.0011 \pm 0.0052$ \\ 
SiO-25 & $2013.13$ & $-33.0445 \pm 0.0054$ & $-13.2 \pm 1.8$ & $-0.02 \pm 0.64$ & $-17.669 \pm 0.015$ & $-54.1 \pm 3.9$ & $0.1 \pm 1.4$ \\ 
\enddata
\tablenotetext{a}{Position uncertainties for IRS 7 have been manually increased to 5 mas to reflect the fact that this star may have an extended maser shell as discussed in Section \ref{sec:results}.}
\end{deluxetable*}

\begin{deluxetable*}{lcccccccc}
\tablecaption{3D velocities and accelerations of stellar masers \label{tab:vels and accels}}
\tablewidth{700pt}
\tabletypesize{\scriptsize}

\tablehead{
\colhead{Star} & \colhead{$v_\alpha$} &  \colhead{$a_\alpha$} &  \colhead{$v_\delta$} &  \colhead{$a_\delta$} &  \colhead{$v_{LSR}$} &  \colhead{$a_{LSR}$} &  \colhead{$|v|$} &  \colhead{$|a|$}\\
\colhead{} &  \colhead{(\kms)} &  \colhead{(\kmsyr)}  & \colhead{(\kms)} &  \colhead{(\kmsyr)} & \colhead{(\kms)} &  \colhead{(\kmsyr) } & \colhead{(\kms)} &  \colhead{(\kmsyr) } 
}
\startdata
IRS9 & $115.9 \pm 1.6$ & $-0.01 \pm 0.14$ & $87.3 \pm 2.6$ & $-0.00 \pm 0.22$ & $-341.1 \pm 1.2$ & $-0.05 \pm 0.29$ & $370.7 \pm 1.4$ & $0.05 \pm 0.29$ \\ 
IRS7\tablenotemark{a} & $-1.4 \pm 4.3$ & $0.14 \pm 0.29$ & $-161.2 \pm 5.5$ & $-0.08 \pm 0.41$ & $-113.6 \pm 7.0$ & $0.1 \pm 1.8$ & $197.2 \pm 6.1$ & $0.19 \pm 0.90$ \\ 
SiO-14\tablenotemark{b} & $77.91 \pm 0.81$ & $-0.027 \pm 0.065$ & $-28.6 \pm 4.6$ & $-0.01 \pm 0.38$ & $-111.02 \pm 0.43$ & $-0.58 \pm 0.10$ & $138.6 \pm 1.1$ & $0.58 \pm 0.10$ \\ 
SiO-28 & $-27.0 \pm 3.0$ & $-0.23 \pm 0.87$ & $69 \pm 77$ & $12 \pm 20$ & $-104.16 \pm 0.98$ & $0.02 \pm 0.25$ & $128 \pm 41$ & $12 \pm 20$ \\ 
SiO-29 & \nodata & \nodata & \nodata & \nodata & $-91.58 \pm 0.98$ & \nodata & \nodata & \nodata \\ 
SiO-31 & \nodata & \nodata & \nodata & \nodata & $-81.10 \pm 0.13$ & \nodata & \nodata & \nodata \\ 
SiO-18 & $-110 \pm 280$ & $-28 \pm 80$ & $40 \pm 100$ & $-16 \pm 29$ & $-78.4 \pm 1.5$ & $0.42 \pm 0.38$ & $140 \pm 220$ & $33 \pm 71$ \\ 
SiO-26 & \nodata & \nodata & \nodata & \nodata & $-72.36 \pm 0.11$ & \nodata & \nodata & \nodata \\ 
IRS12N & $-42.47 \pm 0.85$ & $-0.026 \pm 0.066$ & $-105.35 \pm 0.88$ & $0.015 \pm 0.069$ & $-65.01 \pm 0.69$ & $0.22 \pm 0.16$ & $130.88 \pm 0.83$ & $0.22 \pm 0.16$ \\ 
IRS28 & $58.5 \pm 2.4$ & $-0.02 \pm 0.19$ & $-209.9 \pm 6.9$ & $-0.25 \pm 0.52$ & $-53.60 \pm 0.50$ & $-0.01 \pm 0.12$ & $224.4 \pm 6.5$ & $0.25 \pm 0.52$ \\ 
SiO-27 & $-57.5 \pm 5.1$ & $0.6 \pm 1.2$ & $150 \pm 27$ & $2.9 \pm 5.8$ & $-43.846 \pm 0.080$ & $0.099 \pm 0.025$ & $166 \pm 24$ & $3.0 \pm 5.7$ \\ 
SiO-30 & \nodata & \nodata & \nodata & \nodata & $-37.96 \pm 0.45$ & \nodata & \nodata & \nodata \\ 
SiO-15 & $-92.0 \pm 2.5$ & $0.10 \pm 0.20$ & $17.3 \pm 8.0$ & $-0.33 \pm 0.65$ & $-35.4 \pm 1.0$ & $0.01 \pm 0.26$ & $100.1 \pm 2.7$ & $0.34 \pm 0.62$ \\ 
SiO-19 & $99.7 \pm 4.5$ & $-0.4 \pm 1.3$ & $8 \pm 53$ & $9 \pm 16$ & $-29.3 \pm 1.4$ & $0.21 \pm 0.32$ & $104.2 \pm 6.0$ & $9 \pm 16$ \\ 
IRS10EE & $1.36 \pm 0.53$ & $0.010 \pm 0.039$ & $-75.07 \pm 0.98$ & $-0.005 \pm 0.072$ & $-27.53 \pm 0.38$ & $-0.070 \pm 0.091$ & $79.97 \pm 0.93$ & $0.071 \pm 0.090$ \\ 
SiO-20 & $27.2 \pm 2.3$ & $0.20 \pm 0.75$ & $-132 \pm 13$ & $2.7 \pm 5.0$ & $-16.62 \pm 0.67$ & $-0.37 \pm 0.17$ & $136 \pm 13$ & $2.8 \pm 5.0$ \\ 
IRS15NE\tablenotemark{b} & $-70.25 \pm 0.86$ & $0.021 \pm 0.063$ & $-214.7 \pm 1.8$ & $0.05 \pm 0.14$ & $-10.99 \pm 0.41$ & $-0.34 \pm 0.11$ & $226.2 \pm 1.7$ & $0.35 \pm 0.11$ \\ 
IRS14NE & $163 \pm 32$ & $3.4 \pm 8.7$ & $-128.1 \pm 7.0$ & $-0.9 \pm 1.8$ & $-10.50 \pm 0.95$ & $-0.32 \pm 0.25$ & $207 \pm 26$ & $3.5 \pm 8.4$ \\ 
SiO-16 & $90 \pm 520$ & $-33 \pm 91$ & $-1600 \pm 310$ & $8 \pm 39$ & $7.13 \pm 0.55$ & \nodata & $1600 \pm 310$ & \nodata \\ 
SiO-21 & $643 \pm 91$ & $-4 \pm 15$ & $-1012 \pm 72$ & $-0.8 \pm 9.0$ & $13.447 \pm 0.033$ & \nodata & $1199 \pm 78$ & \nodata \\ 
SiO-24 & $189.3 \pm 9.8$ & $-0.1 \pm 2.4$ & $-96 \pm 91$ & $2 \pm 22$ & $17.84 \pm 0.47$ & $0.22 \pm 0.13$ & $213 \pm 42$ & $2 \pm 22$ \\ 
SiO-22 & $370 \pm 300$ & $-18 \pm 53$ & $-820 \pm 480$ & $15 \pm 62$ & $34.10 \pm 0.93$ & \nodata & $900 \pm 450$ & \nodata \\ 
SiO-6 & $1200 \pm 490$ & $27 \pm 72$ & $-360 \pm 370$ & $5 \pm 47$ & $51.87 \pm 0.21$ & \nodata & $1250 \pm 480$ & \nodata \\ 
SiO-17 & $92.9 \pm 1.0$ & $0.011 \pm 0.083$ & $105.3 \pm 6.4$ & $-0.06 \pm 0.52$ & $53.27 \pm 0.17$ & $0.105 \pm 0.040$ & $150.2 \pm 4.5$ & $0.12 \pm 0.24$ \\ 
SiO-11\tablenotemark{b} & $88 \pm 26$ & $-0.1 \pm 2.4$ & $86 \pm 59$ & $0.5 \pm 5.4$ & $70.60 \pm 0.12$ & $0.105 \pm 0.029$ & $142 \pm 39$ & $0.5 \pm 5.2$ \\ 
IRS17 & $-39.9 \pm 3.0$ & $0.07 \pm 0.25$ & $-41.1 \pm 3.7$ & $-0.01 \pm 0.35$ & $74.36 \pm 0.80$ & $-0.36 \pm 0.21$ & $93.8 \pm 2.1$ & $0.37 \pm 0.21$ \\ 
IRS19NW & $43.34 \pm 0.63$ & $-0.001 \pm 0.049$ & $-17.4 \pm 2.5$ & $-0.04 \pm 0.20$ & $84.37 \pm 0.22$ & $-0.226 \pm 0.052$ & $96.44 \pm 0.57$ & $0.230 \pm 0.063$ \\ 
SiO-25\tablenotemark{b} & $-501 \pm 68$ & $-1 \pm 24$ & $-2050 \pm 150$ & $3 \pm 52$ & $118.48 \pm 0.10$ & $-0.196 \pm 0.038$ & $2120 \pm 140$ & $3 \pm 50$ \\ 
\enddata
\tablenotetext{a}{The velocity centroids of the VLA and ALMA spectra for IRS 7 are offset by $\sim$15 \kms. The radial velocity and acceleration measurements therefore may not accurately track the motion of the star.}
\tablenotetext{b}{Stars with $>3\sigma$ radial acceleration measurements, which may be impacted by maser spectral variability.}
\end{deluxetable*}

\subsection{Proper motions and proper accelerations}\label{sec:pm}

We combined our five epochs with extant maser astrometry of \citet{Reid2007} and \citet{Li2010} to calculate stellar proper motions and accelerations. 
For the majority of masers, we included only our three higher resolution epochs (VLA 2014, ALMA 2017, and VLA 2020) and the \cite{Reid2007} astrometry to perform fits. However, for masers fewer than three high resolution epochs to fit, we also included our lower resolution data (VLA 2013 and ALMA 2015) and the \cite{Li2010} astrometry.   \cite{Li2010} include observations of the 86 GHz masers from the Australia Telescope Compact Array (ATCA), which has lower resolution than VLA or ALMA astrometry and often showed large offsets from the trends in the masers' positions. 
17 masers were fit with only the high resolution data, 7 masers included the lower resolution data, and 4 masers did not have enough epochs of observations for a fit.

All maser positions, including those from literature, are referenced to the in-beam Sgr A* continuum position. The resulting proper motions are therefore referenced to Sgr A* and should be independent of systematic differences in absolute astrometry between epochs. We first find proper motions by fitting a straight line to the maser positions over time, following the equation
\begin{equation}
\Delta x = \Delta x_0 + \mu_x \left( t - t_0 \right), \label{eq:proper motion}
\end{equation}
where $x$ denotes either the R.A. or Dec. coordinate directions, $\alpha$ and $\delta$, respectively. $\Delta x$ is the offset position from Sgr A*,  $\mu_x$ is the proper motion, and $\Delta x_0$ and $t_0$ are the first position measurement and date used in the fit, which varies per maser depending on the earliest detection. 

Proper accelerations were then measured by fitting the residuals of the proper motion fit, following the equation
\begin{equation}
r_{\Delta x} = r_{\Delta x_0} + \frac{1}{2} a_x \left( t - t_0 \right)^2, \label{eq:proper acceleration}
\end{equation}
where $r_{\Delta x}$ and $r_{\Delta x_0}$ are the residuals of the offset positions and initial positions, and $a_x$ is the proper acceleration. We measure proper motions and accelerations for 24 masers with at least three epochs of observations. Figures showing the proper motion fits are given in Appendix~\ref{app:pm fits} and the results are given in Table~\ref{tab:pms}. 
We do not significantly measure proper accelerations for any star in our sample using the residual fitting method. As shown in Table~\ref{tab:pms}, all measurements are consistent with zero transverse acceleration in each coordinate and in magnitude. We also checked for acceleration by simultaneously fitting proper motion and acceleration to the position time series for masers with at least four epochs, however the results gave unrealistically large accelerations for some masers given their projected distance from Sgr A*.  

\subsection{Transverse velocities and accelerations}\label{sec:transverse}

We converted the proper motions and proper accelerations to velocities and accelerations using an assumed distance of 8 kpc since the precise distances of the stars in our sample are not known. More precise stellar distances may be constrained with future measurements if 3D acceleration is measured, and therefore the distance from the star to Sgr A* can be calculated. The resulting velocities and accelerations are listed in Table~\ref{tab:vels and accels}.

\begin{figure*}[t!]
\centering
\includegraphics[width=\textwidth]{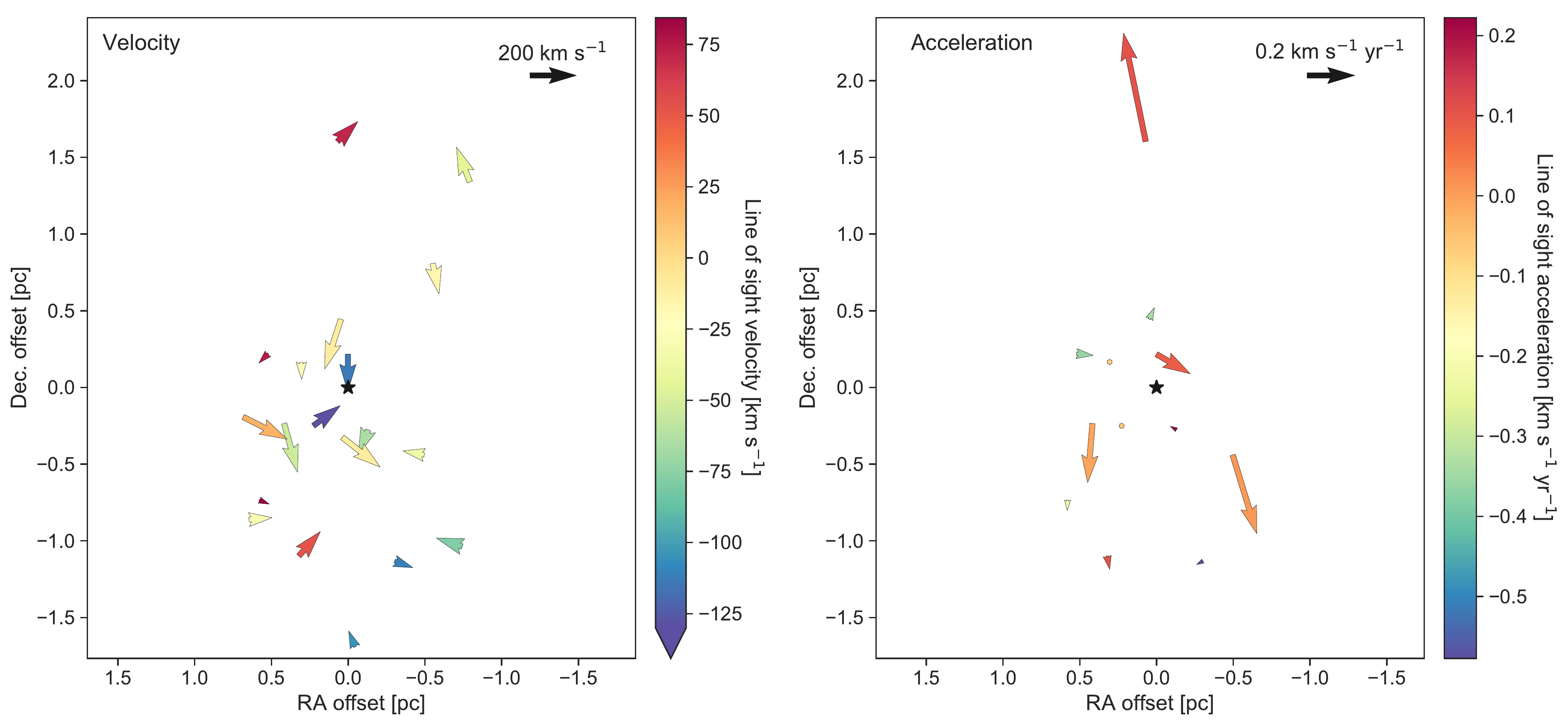}
\caption{Stellar velocities (left) and accelerations (right) assuming a distance of 8 kpc to convert proper motions to physical velocities and accelerations. Masers with proper velocities $>900$ \kms~ or proper accelerations $>0.5$ \kmsyr~ are not shown. The majority of 2D velocities shown are significant, with the exception of SiO-11, SiO-18, and SiO-28, but all of the radial velocities are significant. Note that none of the accelerations in the plane of the sky are significant, and only SiO-14 has a radial acceleration measured with $>5 \sigma$ significance. Positions are indicated by the arrow tails. The star IRS 9 at ($\Delta x$, $\Delta y$) = (0.22 pc, -0.25 pc) has a radial velocity of -340 \kms, which is below the range of the velocity color bar on the left. \label{fig:arrows}}
\end{figure*}

\subsection{Doppler velocities and accelerations}\label{sec:doppler}

As described previously, we find radial velocities by calculating the centroid of each maser spectrum. However, very long baseline interferometry (VLBI) observations of stellar SiO masers shows that the maser emission is distributed in patches around the star and the velocity inferred from the cumulative spectrum may not match the true stellar velocity (e.g. \citealt{Gonidakis2010}). As can be seen in the spectra shown in Appendix~\ref{app:spectra}, many of the masers in our sample have complicated spectral profiles and the inferred velocities can differ between two line profiles taken during the same observation. We therefore estimated an additional systematic uncertainty on the stellar velocities. First, in cases where two lines were detected in a single epoch, we calculated the weighted mean of the two independent velocity measurements so we have one velocity per epoch. 
We then performed a linear fit to the velocities, and calculated a systematic uncertainty from standard deviation of the residuals of the linear fit. The systematic was added in quadrature to the statistical uncertainties. 

Radial accelerations were calculated by fitting a slope to the velocity measurements for stars with at least three epochs of observations. Velocity and acceleration fit results are given in Table~\ref{tab:vels and accels}, and figures showing the fits are given in Appendix~\ref{app:spectra}.  
For stars with too few measurements to fit an acceleration, we report the error weighted mean velocity in Table~\ref{tab:vels and accels}. Radial velocities and accelerations are expressed in the local standard of rest (LSR) frame. Note that radial velocity and acceleration measurements are only based on the observations detailed in this work and did not include any measurements from literature, unlike the proper motion analysis. 

The majority of stars have insignificant radial acceleration limits, with uncertainties ranging from $\sim 0.02$--1 \kmsyr.  However radial acceleration was measured at the $>3\sigma$ level for SiO-14 ($-0.58\pm0.10$ \kmsyr),  SiO-27 ($0.099\pm0.025$ \kmsyr), IRS15NE ($-0.34\pm0.11$ \kmsyr), SiO-11 ($0.105\pm0.029$ \kmsyr), IRS19NW ($-0.226\pm0.052$ \kmsyr), and SiO-25 ($-0.196\pm0.038$ \kmsyr). We note that maser variability over time and between lines may introduce a systematic acceleration that does not match the true acceleration of the star. 

Velocities from the 2017 ALMA data are often but not always the largest outliers from the fit compared other epochs, with average residuals around $\pm 1.2$ \kms. We verified that there is no systematic velocity shift being introduced by our reduction or spectra fitting methods by comparing the Sgr A* absorption spectrum observed with two ALMA epochs. By cross-correlating the two spectra, we found that the cross-correlation is maximized at $-0.4$ \kms. It is therefore unlikely that the 2017 ALMA velocities are being offset by a significant systematic error.

We also attempted to measure radial accelerations by simultaneously fitting spectral profiles to all spectra for a given star following a method detailed in \cite{Pesce2020} and \cite{Reid2013}, who measured accelerations of extragalactic water megamasers. Per star, we performed 100 fits with varying number of Gaussian components and randomized initial values for the velocities of each component. However, we found that the spectra were poorly fit unless we allowed each Gaussian component to have an independent acceleration measurement, which is not useful for determining the underlying systemic acceleration. This is likely due to the variable nature of stellar SiO masers and the fact that we include spectra from multiple maser transitions, which complicates tracking maser peaks between epochs. This method may have better utility for future works with a higher number of spectra of the same transition.  

\section{Stellar kinematics}\label{sec:kinematics}

The 3D stellar velocities and accelerations are shown in Figure~\ref{fig:arrows}. 
Masers with proper velocities $>900$ \kms~ or proper accelerations $>0.5$ \kmsyr~ are not shown in the figure. Note that transverse motions are referenced to the Sgr A* continuum, whereas radial motions are in the LSR frame and do not account for any radial velocity of Sgr A*.

\subsection{Enclosed mass limits}\label{sec:velocities}
Stellar kinematics may be utilized to constrain the stellar and dark matter mass distribution in the GC. We estimate enclosed mass limits using the 3D velocities for each maser with a $>5\sigma$ significance 3D velocity measurement. \citet{Reid2003} derived a strict lower limit on the enclosed mass for a star given its total velocity magnitude, $v$, and projected distance from Sgr A*, $r_{proj}$: 
\begin{equation}
M_{encl} \geq \frac{v^2 r_{proj}}{2G}. \label{eq:Mencl}
\end{equation}
The lower limit approaches an equality as $r_{proj}$ approaches the true distance, when the star is near pericenter, and if the eccentricity of the orbit is close to 1. Following \cite{Reid2003}, we subtract $2 \sigma$ from the stellar velocity to obtain a conservatively small velocity allowed by the measurement uncertainties prior to calculating the enclosed mass limit. 

\begin{figure}[]
    \centering
    \includegraphics[width=\columnwidth]{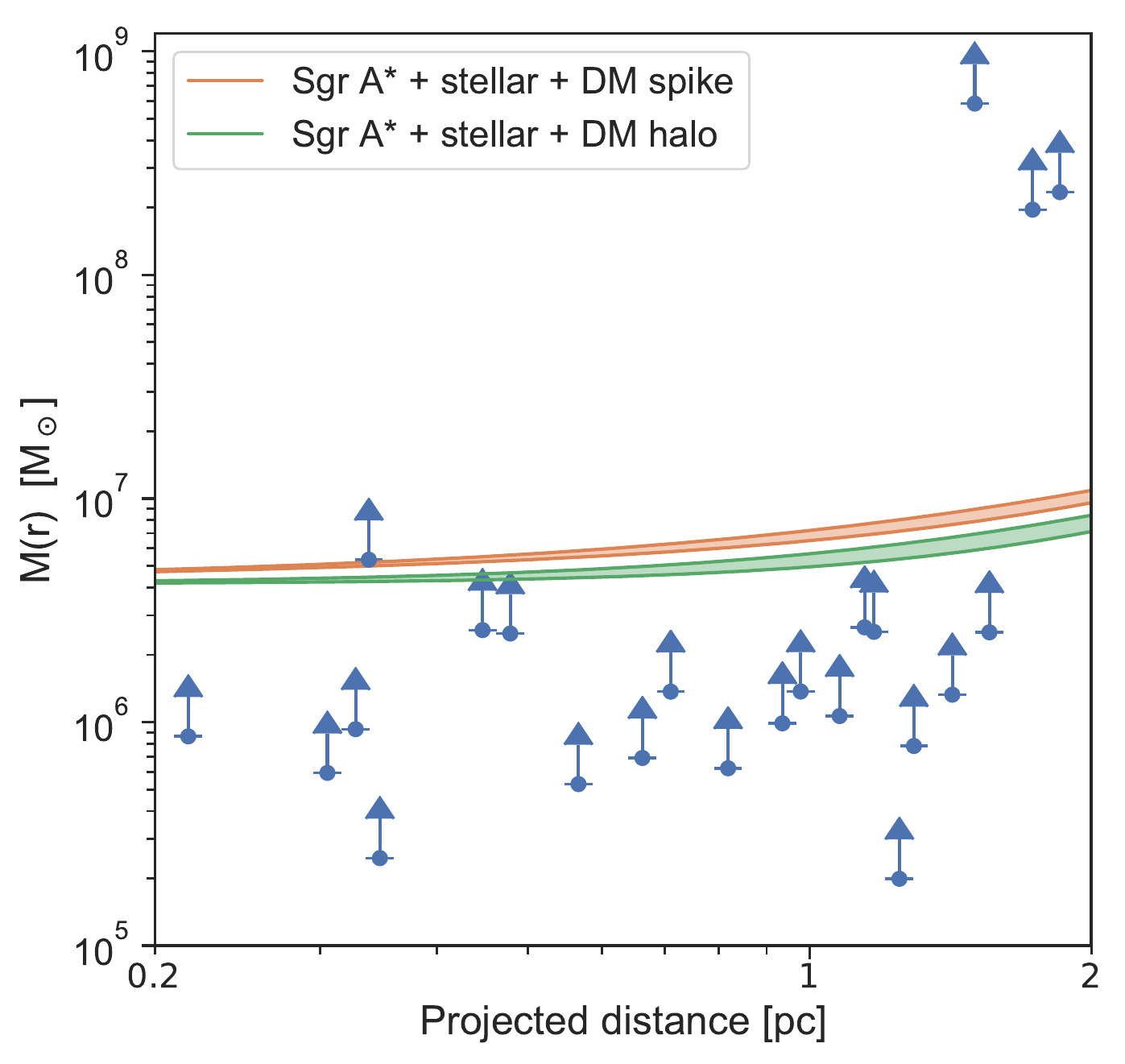}
	\caption{Enclosed mass lower limits based on the stellar 3D velocities as a function of projected distance. The dark matter halo model (green region) is a generalized NFW profile \citep{McMillan2016}, and the maximal dark matter spike model (orange region) includes an additional sharp spike in the dark matter density profile \citep{Lacroix2018}.\label{fig:mass lims}}
\end{figure}

Figure~\ref{fig:mass lims} shows the mass limits calculated for each star compared to two models of the dark matter distribution in the Galactic Center. The first is a dark matter halo based on a generalized Navarro-Frenk-White (NFW) profile, using the main model parameters of \cite{McMillan2016}:  a slope index of $\gamma=1$, scale radius of $r_s = 18.8$ kpc, local dark matter density of $\rho_\odot =0.0101$ M$_\odot$~pc$^{-3}$, and distance to the GC of $R_0=8.2$ kpc. 
The second model includes a sharp spike in the dark matter density profile scaling as $r^{-7/3}$ within the spike scale radius of $R_{sp}=100$ pc, which is the maximal dark matter spike obtained by \cite{Lacroix2018}. 
Both models include a central black hole mass of $4.15\times10^6$ M$_\odot$ \citep{GRAVITYCollaboration2019} and stellar density profiles following $r^{\gamma}$. We display a range of stellar profiles with $\gamma$ values between 1.1 to 1.5 and stellar mass density normalizations between 1.2 to $1.8 \times 10^5$ M$_\odot$~pc$^{-3}$ at 1 pc \citep{Schodel2018}. 
All but four of the stars have $M_{encl}$ lower limits consistent with the expected values: IRS 9, SiO-25, SiO-16, and SiO-21, in order of projected distance. We discuss possible reasons that the enclosed mass may be over estimated for these stars in Section~\ref{sec:high vel stars}.

\subsection{Accelerations}\label{sec:accelerations}

\begin{figure}[t!]
    \centering
    \includegraphics[width=\columnwidth]{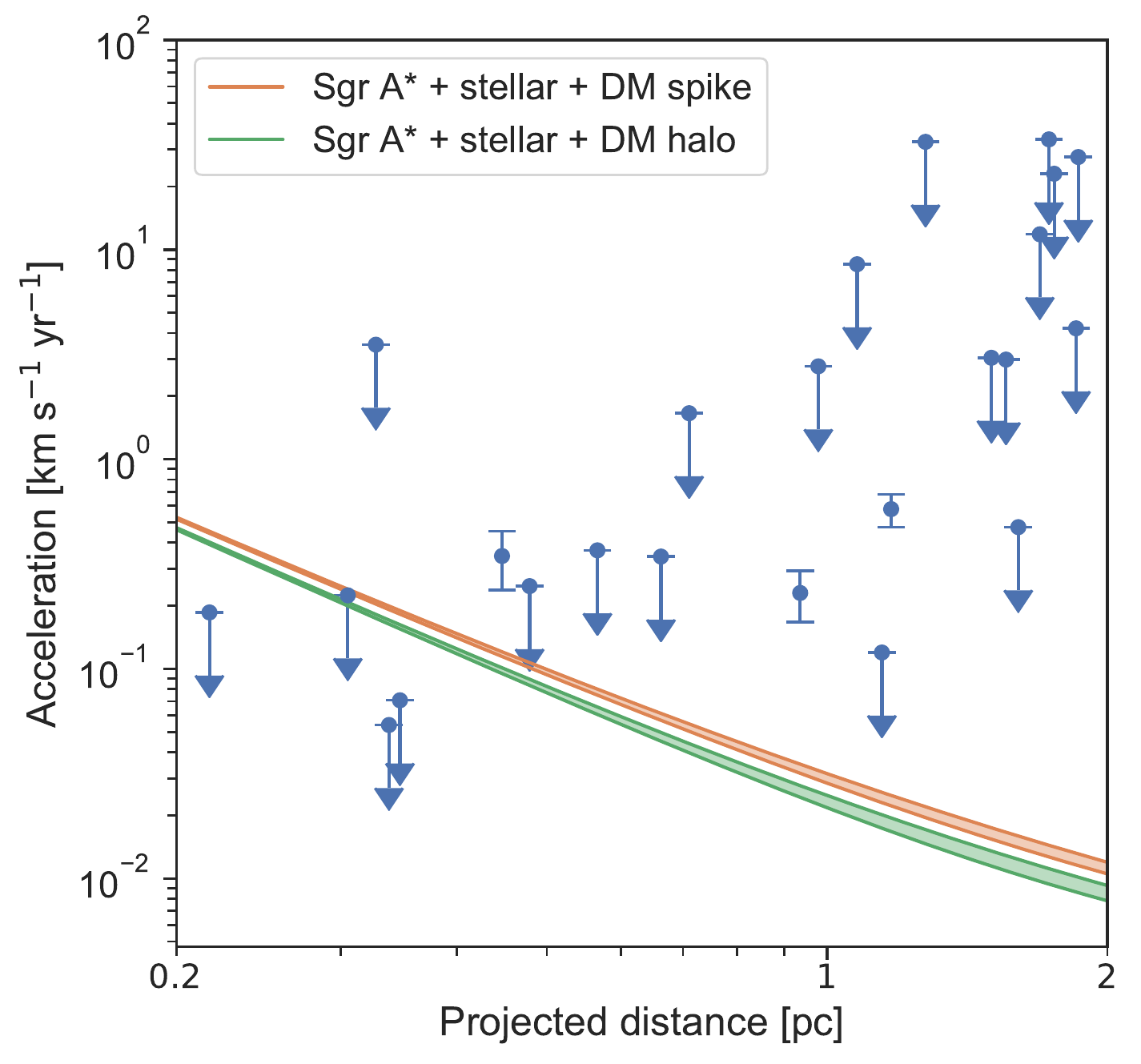}
	\caption{3D stellar acceleration magnitude limits as a function of projected distance. Shaded regions show expected accelerations due to models of the interior stellar and dark matter mass. The dark matter halo model (green) is a generalized NFW profile \citep{McMillan2016}, and the maximal dark matter spike model (orange) includes an additional sharp spike in the dark matter density profile \citep{Lacroix2018}.\label{fig:accel lims}}
\end{figure}

Figure~\ref{fig:accel lims} shows the acceleration estimates for our masers as a function of projected distance from Sgr A* compared to expected accelerations for the two mass models described in Section~\ref{sec:velocities}, assuming circular orbits.  Approximately 0.04 \kmsyr~ precision would be required to differentiate between the two dark matter models in the range of radii where our sample of masers are found. 

The upper limit accelerations for the majority of masers are at least an order of magnitude higher than the expected acceleration and the difference tends to be larger for masers at larger projected distance. This is due the fact that masers closer to Sgr A* are more likely to have more position measurements and therefore lower acceleration uncertainties than masers farther from Sgr A*. 
Three masers, SiO-14, IRS15 and IRS19NW, have at least $3\sigma$-significance acceleration magnitudes, indicated by error bars rather than upper limits in Figure~\ref{fig:accel lims}, which disagree with the model accelerations. These accelerations are dominated by the radial acceleration and may be affected by intrinsic spectral variation as discussed in Section~\ref{sec:doppler} (see also Figure~\ref{fig:spectra and fits}).

\section{Discussion}\label{sec:discussion}

\subsection{High velocity stars}\label{sec:high vel stars}

From Figure \ref{fig:mass lims}, we identified four stars with velocities that exceed the expectation for stars within their projected distance from Sgr A*. The first is IRS 9, which has a 3D velocity of 370 \kms. This result is consistent with that of \citet{Reid2007}, who discussed several possible explanations for the star's high velocity. However, \cite{Trippe2008} found that this velocity is not excessively high compared to the global distribution of 3D stellar velocities in the GC, the tail end of which extends up to about 500 \kms.  

SiO-25, SiO-16, and SiO-21 each have significantly higher velocity measurements than predicted for stars in orbit in the GC: approximately 2100, 1600, and 1200 \kms, respectively.  In the case of these three masers, however, the total velocity is dominated by the 2D velocity inferred from the proper motion. Observational limitations may be causing spurious large proper motion measurements since these masers have a relative paucity of observations compared to other masers in our sample. In fact, SiO-16 and SiO-21 are outside of the VLA field of view and therefore only detected in the 86 GHz line by ALMA. 

However, if the proper motions for these stars are accurate but the stars are closer than the assumed 8 kpc distance to the GC, we may have overestimated the 2D velocity. To estimate the likelihood that three foreground stars would contaminate our sample of masers, we used the model for the distribution of AGB stars in the Galaxy from \citet{Jackson2002}. We integrated their number density model over the volume observed in our ALMA observations towards the GC, resulting in $\sim 0.1$ foreground AGB stars expected in our observations. We therefore conclude it is unlikely that all three masers are actually foreground stars. 

\subsection{GC reference frame}\label{reference frame}

Proper motions of SiO masers associated with bright IR stars have been used to derive the GC astrometric reference frame (\citealt{Yelda2010}; \citealt{Plewa2015}; \citealt{Boehle2016}; \citealt{Sakai2019}), which is necessary to mitigate systematic uncertainties in precision measurements of the orbits of IR stars. \cite{Sakai2019} constructed the astrometric reference frame using maser astrometry from 2005 -- 2017 and reported uncertainties on the SiO-based proper motions ranging from about 0.02  -- 0.6  \masyr~ for seven stars: IRS 9, IRS 7, IRS 12N, IRS 28, IRS 10EE, IRS 15NE, and IRS 17.  

Our proper motion analysis incorporates several more epochs of observations and a longer time baseline of about 25 years from 1995 -- 2020. For the seven reference frame stars, we measure proper motions with similar or better precision than those reported in \cite{Sakai2019}, with uncertainties ranging from about 0.01 -- 0.2 \masyr. The astrometry reported in this work could be used to further improve the GC reference frame since the proper motion fit parameters in Table \ref{tab:pms} may be used with Equation \ref{eq:proper motion} to predict the stellar position at any future epoch. 

\section{Conclusions}\label{sec:conclusions}

We presented five epochs of SiO maser astrometry for stars in the GC from 2013 to 2020. Masers were observed at 43 GHz with the VLA in 2013, 2014, and 2020, and at 86 GHz with ALMA in 2015 and 2017. We observed 28 masers total, of which four are new detections. 
These masers act as high brightness temperature probes of the underlying stellar kinematics within a few parsec of Sgr A*. By incorporating archival data since 1995 in our maser position time series, we calculated proper motions and proper accelerations with $\sim 0.1$ \masyr~ and 0.01 \masyrsqr~ precision, respectively. The proper motions have similar or better precision than those used to derived the GC astrometric reference frame for IR stellar orbits, and may therefore be used to further improve the precision of the reference frame.  

The proper motions and proper accelerations correspond to 2D velocities and accelerations with uncertainties of about 4 \kms~ and 0.4 \kmsyr, though we note that accurate determination of the stellar velocities and accelerations depend on the radial distances to these stars which are currently unknown and assumed to be 8 kpc for all masers. From spectra, we measure radial velocities and accelerations with $\sim$0.5 \kms~ and 0.1 \kmsyr~ precision, respectively.

We also studied the stellar kinematics in the context of expected stellar velocities and accelerations in the GC based on models of the combined mass of Sgr A*, the stellar population, and the dark matter distribution. Lower limits on the enclosed mass were calculated for each maser based on their 3D velocities and projected distances. Several stars were found to yield enclosed mass limits 1--2 orders of magnitude higher than those predicted due to particularly high transverse velocity measurements. These may caused by inaccurate proper motion measurements due to a paucity of observations or by incorrectly translating proper motions to velocities if the stars are closer than the assumed 8 kpc. 

The primary limitation on the analysis presented here is the variability of SiO masers. The majority of our masers show spectral variability in time and between lines, which may degrade the accuracy of measuring the systemic radial velocities and accelerations of the stars themselves. Further high resolution observations will improve the statistical uncertainties in proper motion and radial measurements, as well as potentially help resolve any systematics introduced by maser variability. 
Higher precision measurements will enable improved mapping of the stellar and dark matter mass distributions, as well as probes of the metric around Sgr A* and general relativistic tests, which we intend to pursue in future work.  

\acknowledgments

We thank the anonymous referee for very helpful comments. The authors acknowledge support from the NSF Graduate Research Fellowship Program under grant DGE-1650115 and the NSF grant AST-1908122. This paper makes use of the following ALMA data: ADS/JAO.ALMA\#2013.1.00834.S, ADS/JAO.ALMA\#2016.1.00940.S. ALMA is a partnership of ESO (representing its member states), NSF (USA) and NINS (Japan), together with NRC (Canada), MOST and ASIAA (Taiwan), and KASI (Republic of Korea), in cooperation with the Republic of Chile. The Joint ALMA Observatory is operated by ESO, AUI/NRAO and NAOJ.

\facilities{ALMA, VLA}

\software{astropy \citep{Astropy2013, Astropy2018}, CASA \citep{McMullin2007}, numpy \citep{Harris2020}, matplotlib \citep{Hunter2007}, scipy \citep{Virtanen2020}, uvmultifit \citep{Marti-Vidal2014}}

\vspace{10cm}

\bibliography{references}{}

\begin{thebibliography}{}
\expandafter\ifx\csname natexlab\endcsname\relax\def\natexlab#1{#1}\fi
\providecommand{\url}[1]{\href{#1}{#1}}
\providecommand{\dodoi}[1]{doi:~\href{http://doi.org/#1}{\nolinkurl{#1}}}
\providecommand{\doeprint}[1]{\href{http://ascl.net/#1}{\nolinkurl{http://ascl.net/#1}}}
\providecommand{\doarXiv}[1]{\href{https://arxiv.org/abs/#1}{\nolinkurl{https://arxiv.org/abs/#1}}}

\bibitem[{{Ang{\'e}lil} \& {Saha}(2011)}]{Angelil2011}
{Ang{\'e}lil}, R., \& {Saha}, P. 2011, \apjl, 734, L19,
  \dodoi{10.1088/2041-8205/734/1/L19}

\bibitem[{{Astropy Collaboration} {et~al.}(2013){Astropy Collaboration},
  {Robitaille}, {Tollerud}, {Greenfield}, {Droettboom}, {Bray}, {Aldcroft},
  {Davis}, {Ginsburg}, {Price-Whelan}, {Kerzendorf}, {Conley}, {Crighton},
  {Barbary}, {Muna}, {Ferguson}, {Grollier}, {Parikh}, {Nair}, {Unther},
  {Deil}, {Woillez}, {Conseil}, {Kramer}, {Turner}, {Singer}, {Fox}, {Weaver},
  {Zabalza}, {Edwards}, {Azalee Bostroem}, {Burke}, {Casey}, {Crawford},
  {Dencheva}, {Ely}, {Jenness}, {Labrie}, {Lim}, {Pierfederici}, {Pontzen},
  {Ptak}, {Refsdal}, {Servillat}, \& {Streicher}}]{Astropy2013}
{Astropy Collaboration}, {Robitaille}, T.~P., {Tollerud}, E.~J., {et~al.} 2013,
  \aap, 558, A33, \dodoi{10.1051/0004-6361/201322068}

\bibitem[{{Astropy Collaboration} {et~al.}(2018){Astropy Collaboration},
  {Price-Whelan}, {Sip{\H{o}}cz}, {G{\"u}nther}, {Lim}, {Crawford}, {Conseil},
  {Shupe}, {Craig}, {Dencheva}, {Ginsburg}, {VanderPlas}, {Bradley},
  {P{\'e}rez-Su{\'a}rez}, {de Val-Borro}, {Aldcroft}, {Cruz}, {Robitaille},
  {Tollerud}, {Ardelean}, {Babej}, {Bach}, {Bachetti}, {Bakanov}, {Bamford},
  {Barentsen}, {Barmby}, {Baumbach}, {Berry}, {Biscani}, {Boquien}, {Bostroem},
  {Bouma}, {Brammer}, {Bray}, {Breytenbach}, {Buddelmeijer}, {Burke},
  {Calderone}, {Cano Rodr{\'\i}guez}, {Cara}, {Cardoso}, {Cheedella}, {Copin},
  {Corrales}, {Crichton}, {D'Avella}, {Deil}, {Depagne}, {Dietrich}, {Donath},
  {Droettboom}, {Earl}, {Erben}, {Fabbro}, {Ferreira}, {Finethy}, {Fox},
  {Garrison}, {Gibbons}, {Goldstein}, {Gommers}, {Greco}, {Greenfield},
  {Groener}, {Grollier}, {Hagen}, {Hirst}, {Homeier}, {Horton}, {Hosseinzadeh},
  {Hu}, {Hunkeler}, {Ivezi{\'c}}, {Jain}, {Jenness}, {Kanarek}, {Kendrew},
  {Kern}, {Kerzendorf}, {Khvalko}, {King}, {Kirkby}, {Kulkarni}, {Kumar},
  {Lee}, {Lenz}, {Littlefair}, {Ma}, {Macleod}, {Mastropietro}, {McCully},
  {Montagnac}, {Morris}, {Mueller}, {Mumford}, {Muna}, {Murphy}, {Nelson},
  {Nguyen}, {Ninan}, {N{\"o}the}, {Ogaz}, {Oh}, {Parejko}, {Parley}, {Pascual},
  {Patil}, {Patil}, {Plunkett}, {Prochaska}, {Rastogi}, {Reddy Janga},
  {Sabater}, {Sakurikar}, {Seifert}, {Sherbert}, {Sherwood-Taylor}, {Shih},
  {Sick}, {Silbiger}, {Singanamalla}, {Singer}, {Sladen}, {Sooley},
  {Sornarajah}, {Streicher}, {Teuben}, {Thomas}, {Tremblay}, {Turner},
  {Terr{\'o}n}, {van Kerkwijk}, {de la Vega}, {Watkins}, {Weaver}, {Whitmore},
  {Woillez}, {Zabalza}, \& {Astropy Contributors}}]{Astropy2018}
{Astropy Collaboration}, {Price-Whelan}, A.~M., {Sip{\H{o}}cz}, B.~M., {et~al.}
  2018, \aj, 156, 123, \dodoi{10.3847/1538-3881/aabc4f}

\bibitem[{{Boehle} {et~al.}(2016){Boehle}, {Ghez}, {Sch{\"o}del}, {Meyer},
  {Yelda}, {Albers}, {Martinez}, {Becklin}, {Do}, {Lu}, {Matthews}, {Morris},
  {Sitarski}, \& {Witzel}}]{Boehle2016}
{Boehle}, A., {Ghez}, A.~M., {Sch{\"o}del}, R., {et~al.} 2016, \apj, 830, 17,
  \dodoi{10.3847/0004-637X/830/1/17}

\bibitem[{{Borkar} {et~al.}(2020){Borkar}, {Eckart}, {Straubmeier}, {Sabha},
  {Sjouwerman}, {Karas}, {Kunneriath}, {Moser}, {Britzen},
  {Valencia-Schneider}, {Donea}, \& {Zensus}}]{Borkar2019}
{Borkar}, A., {Eckart}, A., {Straubmeier}, C., {et~al.} 2020, in Multifrequency
  Behaviour of High Energy Cosmic Sources - XIII. 3-8 June 2019. Palermo, 33.
\newblock \doarXiv{1909.13753}

\bibitem[{{Genzel} {et~al.}(2010){Genzel}, {Eisenhauer}, \&
  {Gillessen}}]{Genzel2010}
{Genzel}, R., {Eisenhauer}, F., \& {Gillessen}, S. 2010, Reviews of Modern
  Physics, 82, 3121, \dodoi{10.1103/RevModPhys.82.3121}

\bibitem[{{Ghez} {et~al.}(2008){Ghez}, {Salim}, {Weinberg}, {Lu}, {Do}, {Dunn},
  {Matthews}, {Morris}, {Yelda}, {Becklin}, {Kremenek}, {Milosavljevic}, \&
  {Naiman}}]{Ghez2008}
{Ghez}, A.~M., {Salim}, S., {Weinberg}, N.~N., {et~al.} 2008, \apj, 689, 1044,
  \dodoi{10.1086/592738}

\bibitem[{{Gonidakis} {et~al.}(2010){Gonidakis}, {Diamond}, \&
  {Kemball}}]{Gonidakis2010}
{Gonidakis}, I., {Diamond}, P.~J., \& {Kemball}, A.~J. 2010, \mnras, 406, 395,
  \dodoi{10.1111/j.1365-2966.2010.16716.x}

\bibitem[{{Gravity Collaboration} {et~al.}(2018){Gravity Collaboration},
  {Abuter}, {Amorim}, {Anugu}, {Baub{\"o}ck}, {Benisty}, {Berger}, {Blind},
  {Bonnet}, {Brandner}, {Buron}, {Collin}, {Chapron}, {Cl{\'e}net}, {Coud{\'e}
  Du Foresto}, {de Zeeuw}, {Deen}, {Delplancke-Str{\"o}bele}, {Dembet},
  {Dexter}, {Duvert}, {Eckart}, {Eisenhauer}, {Finger}, {F{\"o}rster
  Schreiber}, {F{\'e}dou}, {Garcia}, {Garcia Lopez}, {Gao}, {Gendron},
  {Genzel}, {Gillessen}, {Gordo}, {Habibi}, {Haubois}, {Haug}, {Hau{\ss}mann},
  {Henning}, {Hippler}, {Horrobin}, {Hubert}, {Hubin}, {Jimenez Rosales},
  {Jochum}, {Jocou}, {Kaufer}, {Kellner}, {Kendrew}, {Kervella}, {Kok},
  {Kulas}, {Lacour}, {Lapeyr{\`e}re}, {Lazareff}, {Le Bouquin}, {L{\'e}na},
  {Lippa}, {Lenzen}, {M{\'e}rand}, {M{\"u}ler}, {Neumann}, {Ott}, {Palanca},
  {Paumard}, {Pasquini}, {Perraut}, {Perrin}, {Pfuhl}, {Plewa}, {Rabien},
  {Ram{\'\i}rez}, {Ramos}, {Rau}, {Rodr{\'\i}guez-Coira}, {Rohloff}, {Rousset},
  {Sanchez-Bermudez}, {Scheithauer}, {Sch{\"o}ller}, {Schuler}, {Spyromilio},
  {Straub}, {Straubmeier}, {Sturm}, {Tacconi}, {Tristram}, {Vincent}, {von
  Fellenberg}, {Wank}, {Waisberg}, {Widmann}, {Wieprecht}, {Wiest},
  {Wiezorrek}, {Woillez}, {Yazici}, {Ziegler}, \&
  {Zins}}]{GRAVITYCollaboration2018}
{Gravity Collaboration}, {Abuter}, R., {Amorim}, A., {et~al.} 2018, \aap, 615,
  L15, \dodoi{10.1051/0004-6361/201833718}

\bibitem[{{Gravity Collaboration} {et~al.}(2019){Gravity Collaboration},
  {Abuter}, {Amorim}, {Baub{\"o}ck}, {Berger}, {Bonnet}, {Brandner},
  {Cl{\'e}net}, {Coud{\'e} Du Foresto}, {de Zeeuw}, {Dexter}, {Duvert},
  {Eckart}, {Eisenhauer}, {F{\"o}rster Schreiber}, {Garcia}, {Gao}, {Gendron},
  {Genzel}, {Gerhard}, {Gillessen}, {Habibi}, {Haubois}, {Henning}, {Hippler},
  {Horrobin}, {Jim{\'e}nez-Rosales}, {Jocou}, {Kervella}, {Lacour},
  {Lapeyr{\`e}re}, {Le Bouquin}, {L{\'e}na}, {Ott}, {Paumard}, {Perraut},
  {Perrin}, {Pfuhl}, {Rabien}, {Rodriguez Coira}, {Rousset}, {Scheithauer},
  {Sternberg}, {Straub}, {Straubmeier}, {Sturm}, {Tacconi}, {Vincent}, {von
  Fellenberg}, {Waisberg}, {Widmann}, {Wieprecht}, {Wiezorrek}, {Woillez}, \&
  {Yazici}}]{GRAVITYCollaboration2019}
---. 2019, \aap, 625, L10, \dodoi{10.1051/0004-6361/201935656}

\bibitem[{{Gravity Collaboration} {et~al.}(2020){Gravity Collaboration},
  {Abuter}, {Amorim}, {Baub{\"o}ck}, {Berger}, {Bonnet}, {Brandner}, {Cardoso},
  {Cl{\'e}net}, {de Zeeuw}, {Dexter}, {Eckart}, {Eisenhauer}, {F{\"o}rster
  Schreiber}, {Garcia}, {Gao}, {Gendron}, {Genzel}, {Gillessen}, {Habibi},
  {Haubois}, {Henning}, {Hippler}, {Horrobin}, {Jim{\'e}nez-Rosales}, {Jochum},
  {Jocou}, {Kaufer}, {Kervella}, {Lacour}, {Lapeyr{\`e}re}, {Le Bouquin},
  {L{\'e}na}, {Nowak}, {Ott}, {Paumard}, {Perraut}, {Perrin}, {Pfuhl},
  {Rodr{\'\i}guez-Coira}, {Shangguan}, {Scheithauer}, {Stadler}, {Straub},
  {Straubmeier}, {Sturm}, {Tacconi}, {Vincent}, {von Fellenberg}, {Waisberg},
  {Widmann}, {Wieprecht}, {Wiezorrek}, {Woillez}, {Yazici}, \&
  {Zins}}]{GRAVITYCollaboration2020}
---. 2020, \aap, 636, L5, \dodoi{10.1051/0004-6361/202037813}

\bibitem[{{Harris} {et~al.}(2020){Harris}, {Millman}, {van der Walt},
  {Gommers}, {Virtanen}, {Cournapeau}, {Wieser}, {Taylor}, {Berg}, {Smith},
  {Kern}, {Picus}, {Hoyer}, {van Kerkwijk}, {Brett}, {Haldane}, {del R{\'\i}o},
  {Wiebe}, {Peterson}, {G{\'e}rard-Marchant}, {Sheppard}, {Reddy}, {Weckesser},
  {Abbasi}, {Gohlke}, \& {Oliphant}}]{Harris2020}
{Harris}, C.~R., {Millman}, K.~J., {van der Walt}, S.~J., {et~al.} 2020, \nat,
  585, 357, \dodoi{10.1038/s41586-020-2649-2}

\bibitem[{{Hunter}(2007)}]{Hunter2007}
{Hunter}, J.~D. 2007, Computing in Science and Engineering, 9, 90,
  \dodoi{10.1109/MCSE.2007.55}

\bibitem[{{Jackson} {et~al.}(2002){Jackson}, {Ivezi{\'c}}, \&
  {Knapp}}]{Jackson2002}
{Jackson}, T., {Ivezi{\'c}}, {\v{Z}}., \& {Knapp}, G.~R. 2002, \mnras, 337,
  749, \dodoi{10.1046/j.1365-8711.2002.05980.x}

\bibitem[{{Kemball}(2007)}]{Kemball2007}
{Kemball}, A.~J. 2007, in Astrophysical Masers and their Environments, ed.
  J.~M. {Chapman} \& W.~A. {Baan}, Vol. 242, 236--245,
  \dodoi{10.1017/S1743921307013063}

\bibitem[{{Lacroix}(2018)}]{Lacroix2018}
{Lacroix}, T. 2018, \aap, 619, A46, \dodoi{10.1051/0004-6361/201832652}

\bibitem[{{Li} {et~al.}(2010){Li}, {An}, {Shen}, \& {Miyazaki}}]{Li2010}
{Li}, J., {An}, T., {Shen}, Z.-Q., \& {Miyazaki}, A. 2010, \apjl, 720, L56,
  \dodoi{10.1088/2041-8205/720/1/L56}

\bibitem[{{Mart{\'\i}-Vidal} {et~al.}(2014){Mart{\'\i}-Vidal}, {Vlemmings},
  {Muller}, \& {Casey}}]{Marti-Vidal2014}
{Mart{\'\i}-Vidal}, I., {Vlemmings}, W.~H.~T., {Muller}, S., \& {Casey}, S.
  2014, \aap, 563, A136, \dodoi{10.1051/0004-6361/201322633}

\bibitem[{{McMillan}(2017)}]{McMillan2016}
{McMillan}, P.~J. 2017, \mnras, 465, 76, \dodoi{10.1093/mnras/stw2759}

\bibitem[{{McMullin} {et~al.}(2007){McMullin}, {Waters}, {Schiebel}, {Young},
  \& {Golap}}]{McMullin2007}
{McMullin}, J.~P., {Waters}, B., {Schiebel}, D., {Young}, W., \& {Golap}, K.
  2007, in Astronomical Society of the Pacific Conference Series, Vol. 376,
  Astronomical Data Analysis Software and Systems XVI, ed. R.~A. {Shaw},
  F.~{Hill}, \& D.~J. {Bell}, 127

\bibitem[{{Menten} {et~al.}(1997){Menten}, {Reid}, {Eckart}, \&
  {Genzel}}]{Menten1997}
{Menten}, K.~M., {Reid}, M.~J., {Eckart}, A., \& {Genzel}, R. 1997, \apjl, 475,
  L111, \dodoi{10.1086/310472}

\bibitem[{{Pardo} {et~al.}(2004){Pardo}, {Alcolea}, {Bujarrabal}, {Colomer},
  {del Romero}, \& {de Vicente}}]{Pardo2004}
{Pardo}, J.~R., {Alcolea}, J., {Bujarrabal}, V., {et~al.} 2004, \aap, 424, 145,
  \dodoi{10.1051/0004-6361:20040309}

\bibitem[{{Pesce} {et~al.}(2020){Pesce}, {Braatz}, {Reid}, {Condon}, {Gao},
  {Henkel}, {Kuo}, {Lo}, \& {Zhao}}]{Pesce2020}
{Pesce}, D.~W., {Braatz}, J.~A., {Reid}, M.~J., {et~al.} 2020, \apj, 890, 118,
  \dodoi{10.3847/1538-4357/ab6bcd}

\bibitem[{{Plewa} {et~al.}(2015){Plewa}, {Gillessen}, {Eisenhauer}, {Ott},
  {Pfuhl}, {George}, {Dexter}, {Habibi}, {Genzel}, {Reid}, \&
  {Menten}}]{Plewa2015}
{Plewa}, P.~M., {Gillessen}, S., {Eisenhauer}, F., {et~al.} 2015, \mnras, 453,
  3234, \dodoi{10.1093/mnras/stv1910}

\bibitem[{{Reid} {et~al.}(2013){Reid}, {Braatz}, {Condon}, {Lo}, {Kuo},
  {Impellizzeri}, \& {Henkel}}]{Reid2013}
{Reid}, M.~J., {Braatz}, J.~A., {Condon}, J.~J., {et~al.} 2013, \apj, 767, 154,
  \dodoi{10.1088/0004-637X/767/2/154}

\bibitem[{{Reid} {et~al.}(2003){Reid}, {Menten}, {Genzel}, {Ott},
  {Sch{\"o}del}, \& {Eckart}}]{Reid2003}
{Reid}, M.~J., {Menten}, K.~M., {Genzel}, R., {et~al.} 2003, \apj, 587, 208,
  \dodoi{10.1086/368074}

\bibitem[{{Reid} {et~al.}(2007){Reid}, {Menten}, {Trippe}, {Ott}, \&
  {Genzel}}]{Reid2007}
{Reid}, M.~J., {Menten}, K.~M., {Trippe}, S., {Ott}, T., \& {Genzel}, R. 2007,
  \apj, 659, 378, \dodoi{10.1086/511744}

\bibitem[{{Sakai} {et~al.}(2019){Sakai}, {Lu}, {Ghez}, {Jia}, {Do}, {Witzel},
  {Gautam}, {Hees}, {Becklin}, {Matthews}, \& {Hosek}}]{Sakai2019}
{Sakai}, S., {Lu}, J.~R., {Ghez}, A., {et~al.} 2019, \apj, 873, 65,
  \dodoi{10.3847/1538-4357/ab0361}

\bibitem[{{Sch{\"o}del} {et~al.}(2018){Sch{\"o}del}, {Gallego-Cano}, {Dong},
  {Nogueras-Lara}, {Gallego-Calvente}, {Amaro-Seoane}, \&
  {Baumgardt}}]{Schodel2018}
{Sch{\"o}del}, R., {Gallego-Cano}, E., {Dong}, H., {et~al.} 2018, \aap, 609,
  A27, \dodoi{10.1051/0004-6361/201730452}

\bibitem[{{Trippe} {et~al.}(2008){Trippe}, {Gillessen}, {Gerhard}, {Bartko},
  {Fritz}, {Maness}, {Eisenhauer}, {Martins}, {Ott}, {Dodds-Eden}, \&
  {Genzel}}]{Trippe2008}
{Trippe}, S., {Gillessen}, S., {Gerhard}, O.~E., {et~al.} 2008, \aap, 492, 419,
  \dodoi{10.1051/0004-6361:200810191}

\bibitem[{{Virtanen} {et~al.}(2020){Virtanen}, {Gommers}, {Oliphant},
  {Haberland}, {Reddy}, {Cournapeau}, {Burovski}, {Peterson}, {Weckesser},
  {Bright}, {van der Walt}, {Brett}, {Wilson}, {Millman}, {Mayorov}, {Nelson},
  {Jones}, {Kern}, {Larson}, {Carey}, {Polat}, {Feng}, {Moore}, {VanderPlas},
  {Laxalde}, {Perktold}, {Cimrman}, {Henriksen}, {Quintero}, {Harris},
  {Archibald}, {Ribeiro}, {Pedregosa}, {van Mulbregt}, \& {SciPy 1. 0
  Contributors}}]{Virtanen2020}
{Virtanen}, P., {Gommers}, R., {Oliphant}, T.~E., {et~al.} 2020, Nature
  Methods, 17, 261, \dodoi{10.1038/s41592-019-0686-2}

\bibitem[{{Yelda} {et~al.}(2010){Yelda}, {Lu}, {Ghez}, {Clarkson}, {Anderson},
  {Do}, \& {Matthews}}]{Yelda2010}
{Yelda}, S., {Lu}, J.~R., {Ghez}, A.~M., {et~al.} 2010, \apj, 725, 331,
  \dodoi{10.1088/0004-637X/725/1/331}

\end{thebibliography}
\bibliographystyle{aasjournal}

\appendix

\section{Spectra and Doppler fitting}\label{app:spectra}

Figure \ref{fig:spectra and fits} shows maser spectra and radial velocity and radial acceleration fits described in Section \ref{sec:doppler}. 

\begin{figure}[ht!]
\gridline{ \fig{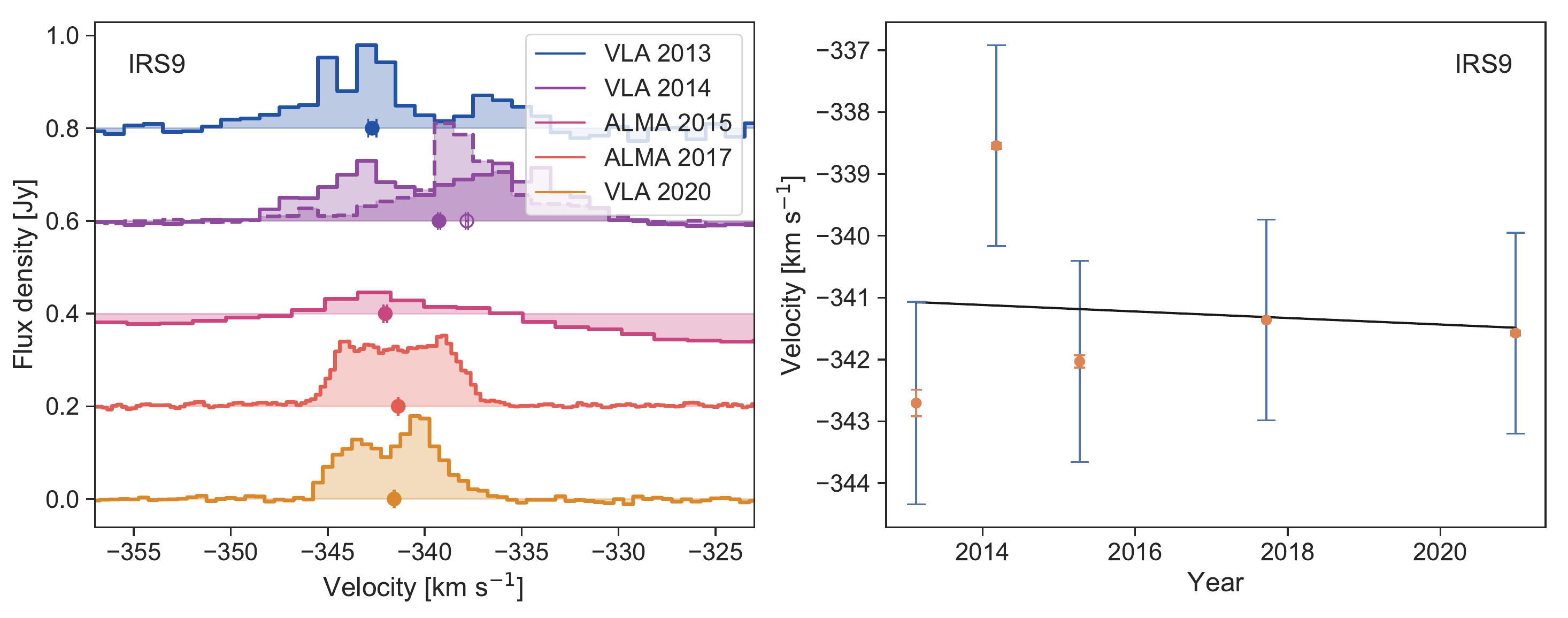}{\textwidth}{}}
\vspace*{-10mm}
\gridline{ \fig{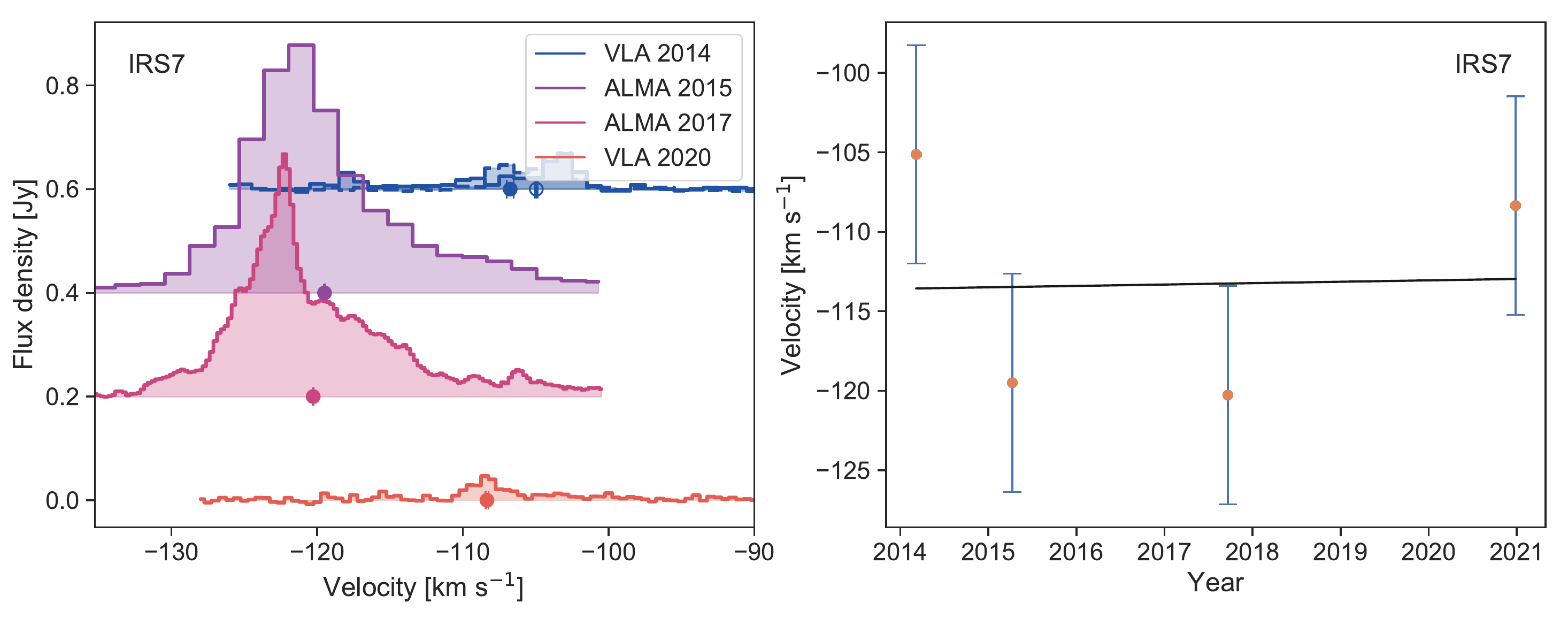}{\textwidth}{}}
\vspace*{-10mm}
\gridline{ \fig{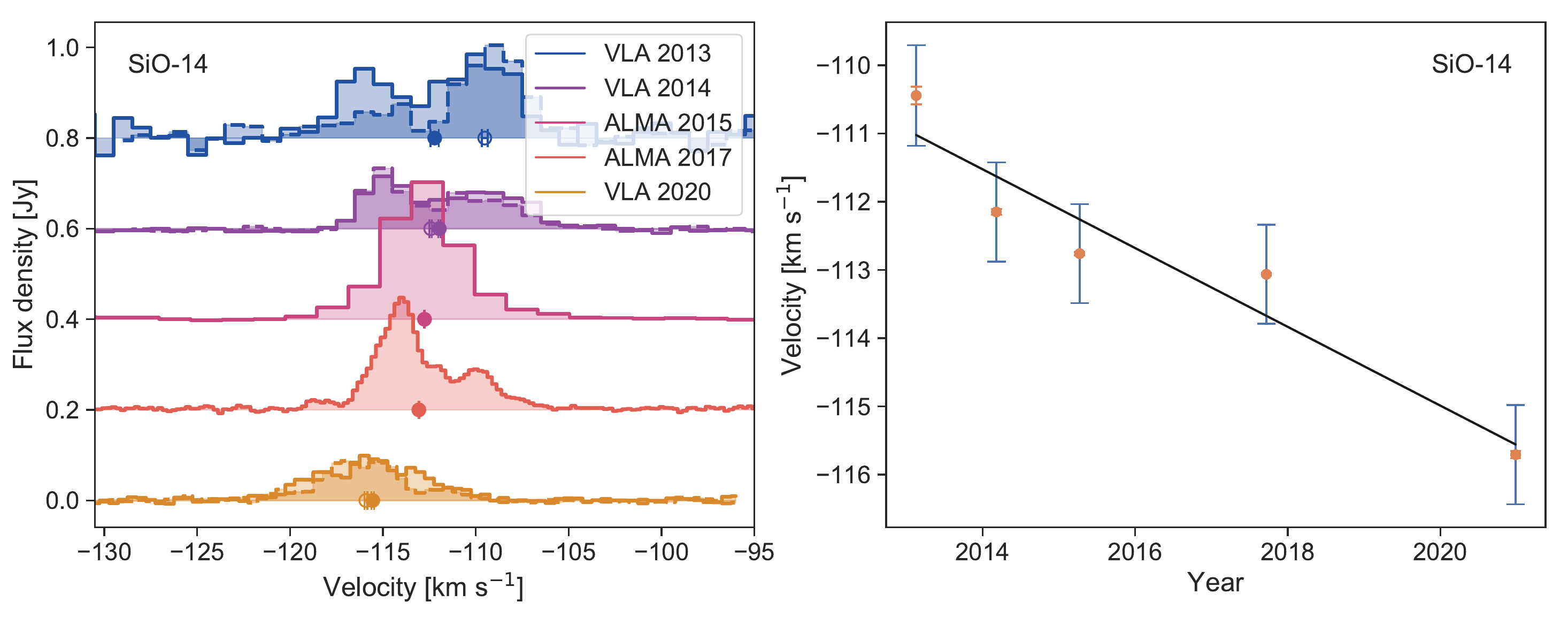}{\textwidth}{}}
\vspace*{-10mm}
\caption{Left: Maser spectra for each epoch when detected. For VLA epochs, solid lines indicate $J=1-0$, $v=1$ spectra and dashed lines are $J=1-0$, $v=2$ spectra. Error bars indicate the centroid velocity for each spectrum, with filled circles for $v=1$ spectra and open circles for $v=2$. Right: Velocity measurements as a function of time for each maser. Velocities have been averaged per epoch. Orange error bars show the statistical uncertainty on the velocity and blue error bars show the additional systematic uncertainty. Solid lines are the acceleration fits described in Section~\ref{sec:doppler}.  Dotted lines are the error-weighted mean velocities for masers with too few data points for an acceleration fit. } \label{fig:spectra and fits}
\end{figure}

\setcounter{figure}{4}
\begin{figure}[ht!]
\gridline{ \fig{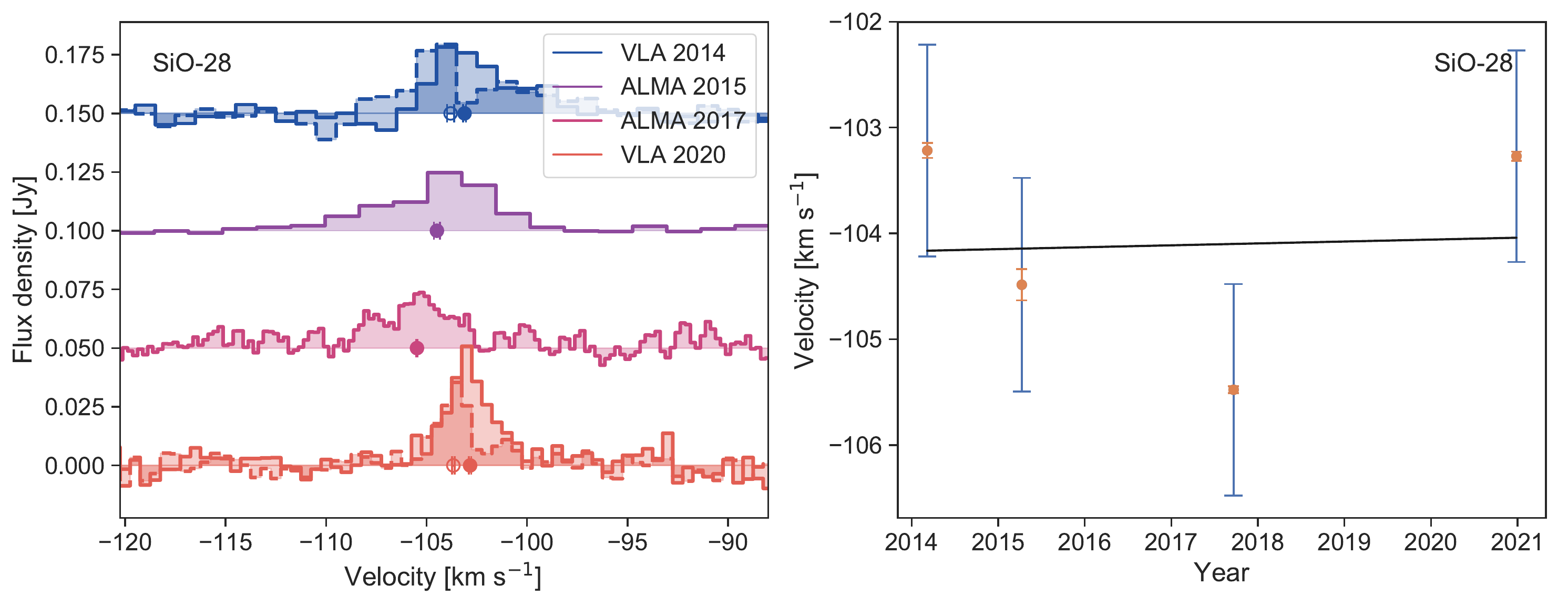}{\textwidth}{} }
\vspace*{-10mm}
\gridline{ \fig{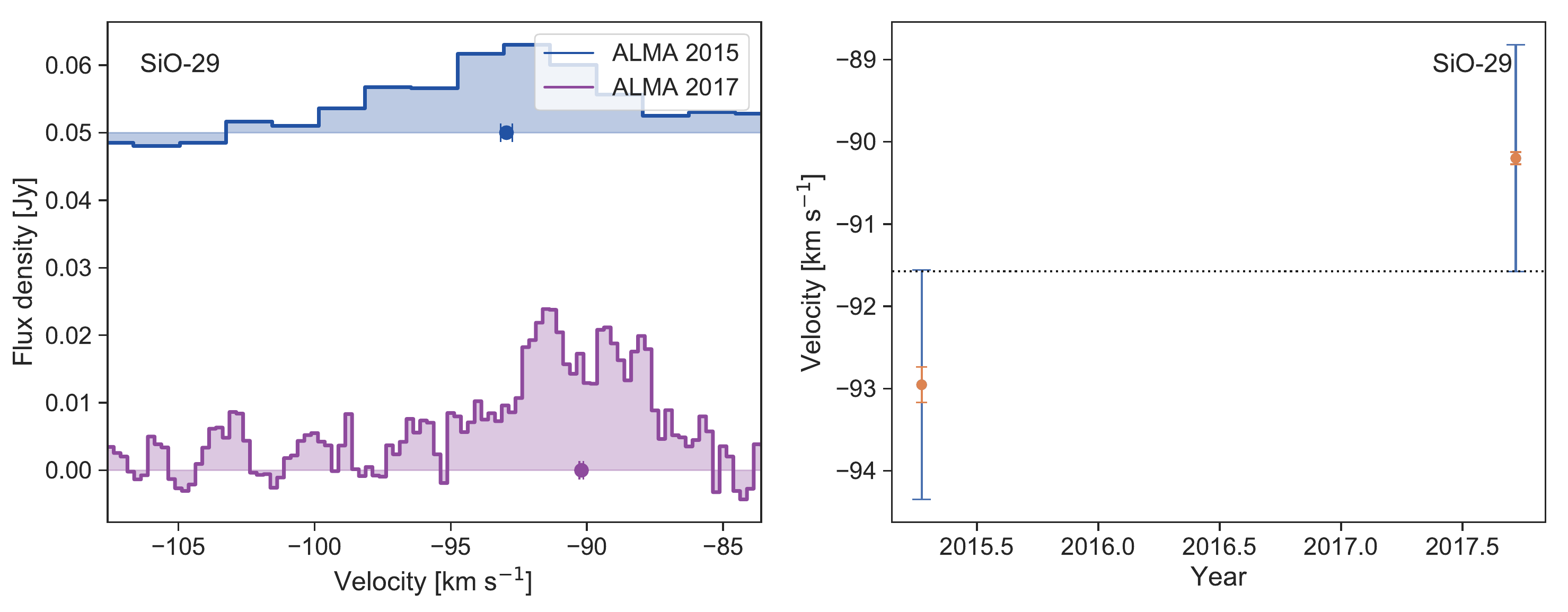}{\textwidth}{} }
\vspace*{-10mm}
\gridline{ \fig{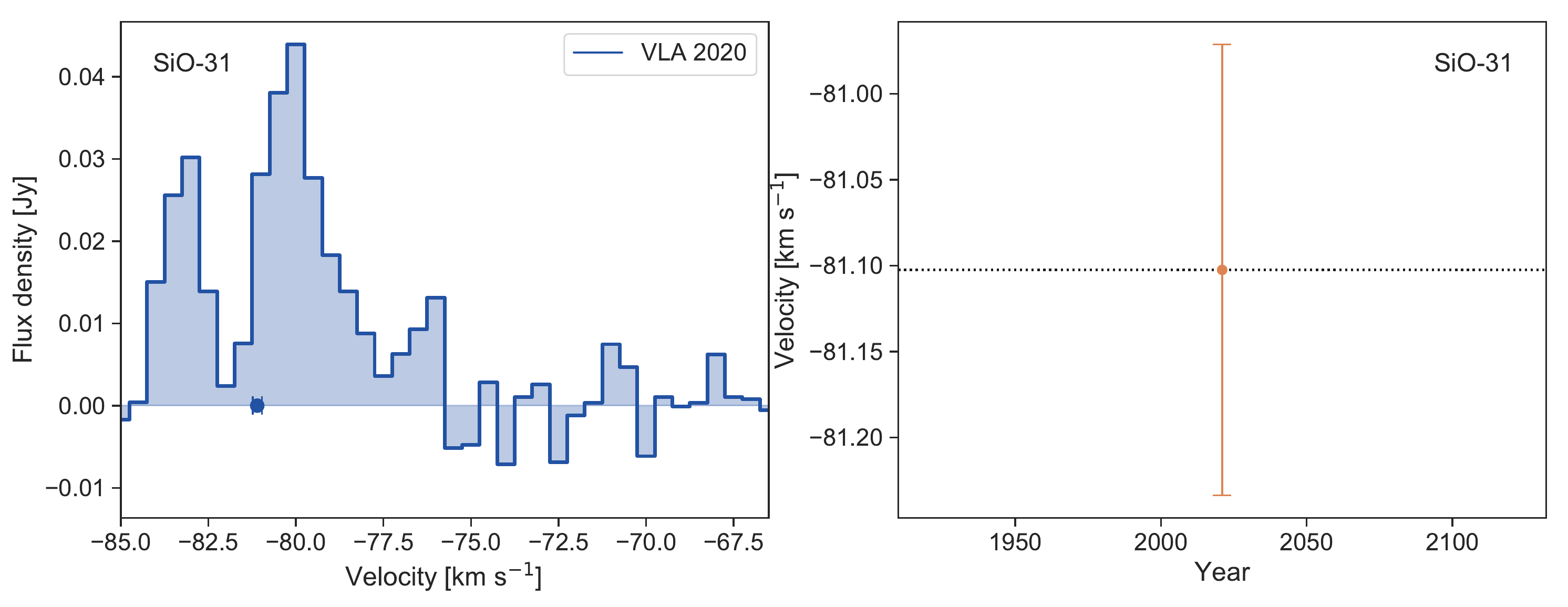}{\textwidth}{} }
\caption{(Continued)}
\end{figure}

\setcounter{figure}{4}
\begin{figure}[ht!]
\gridline{ \fig{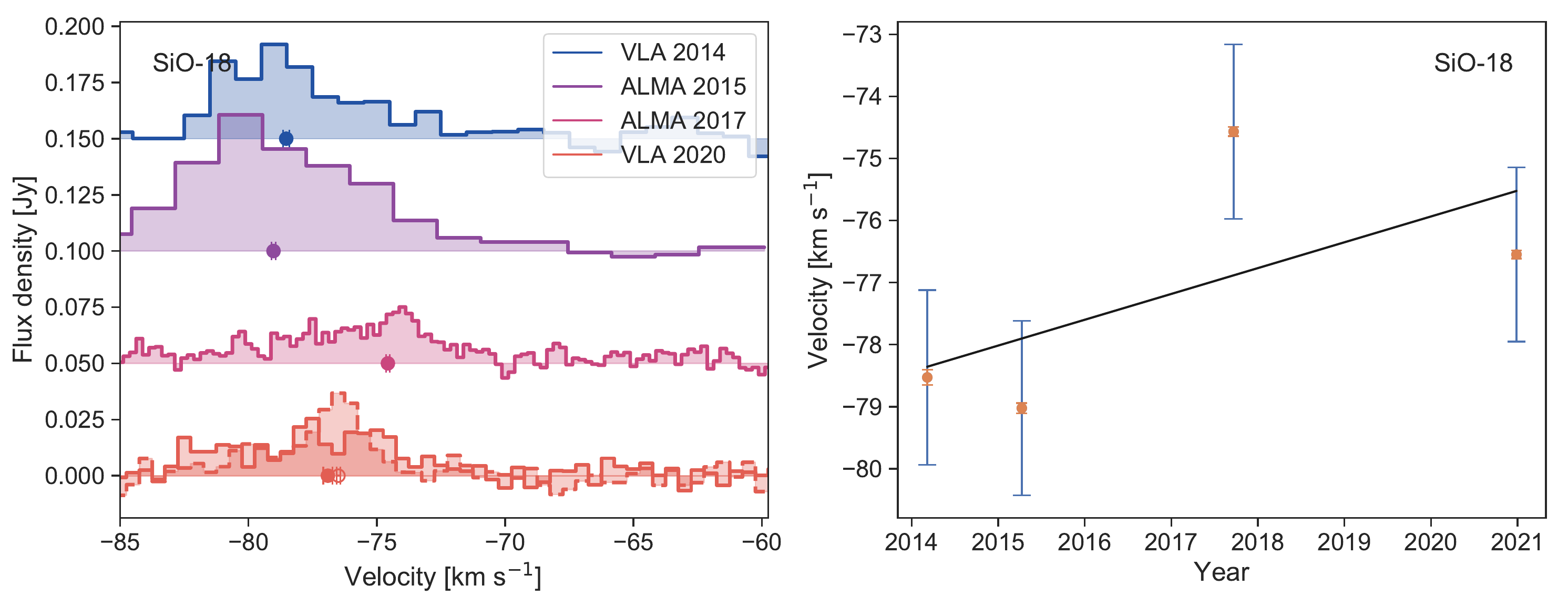}{\textwidth}{} }
\vspace*{-10mm}
\gridline{ \fig{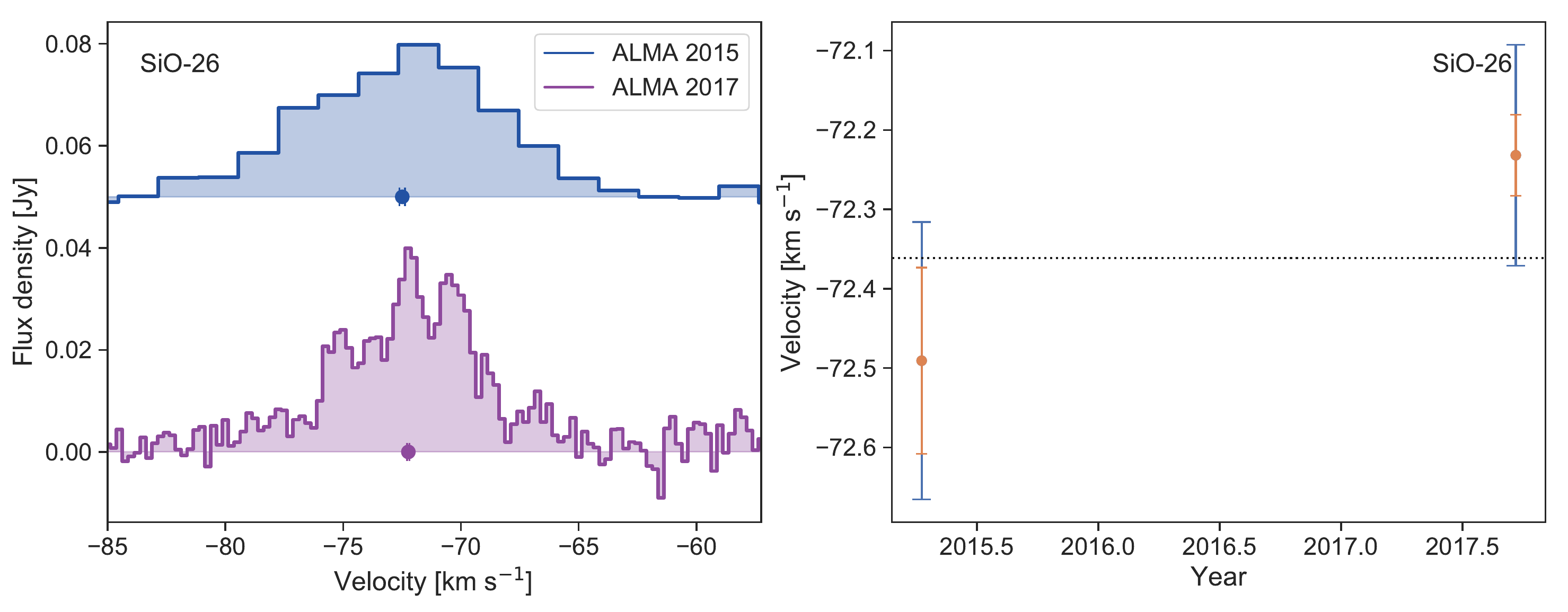}{\textwidth}{} }
\vspace*{-10mm}
\gridline{ \fig{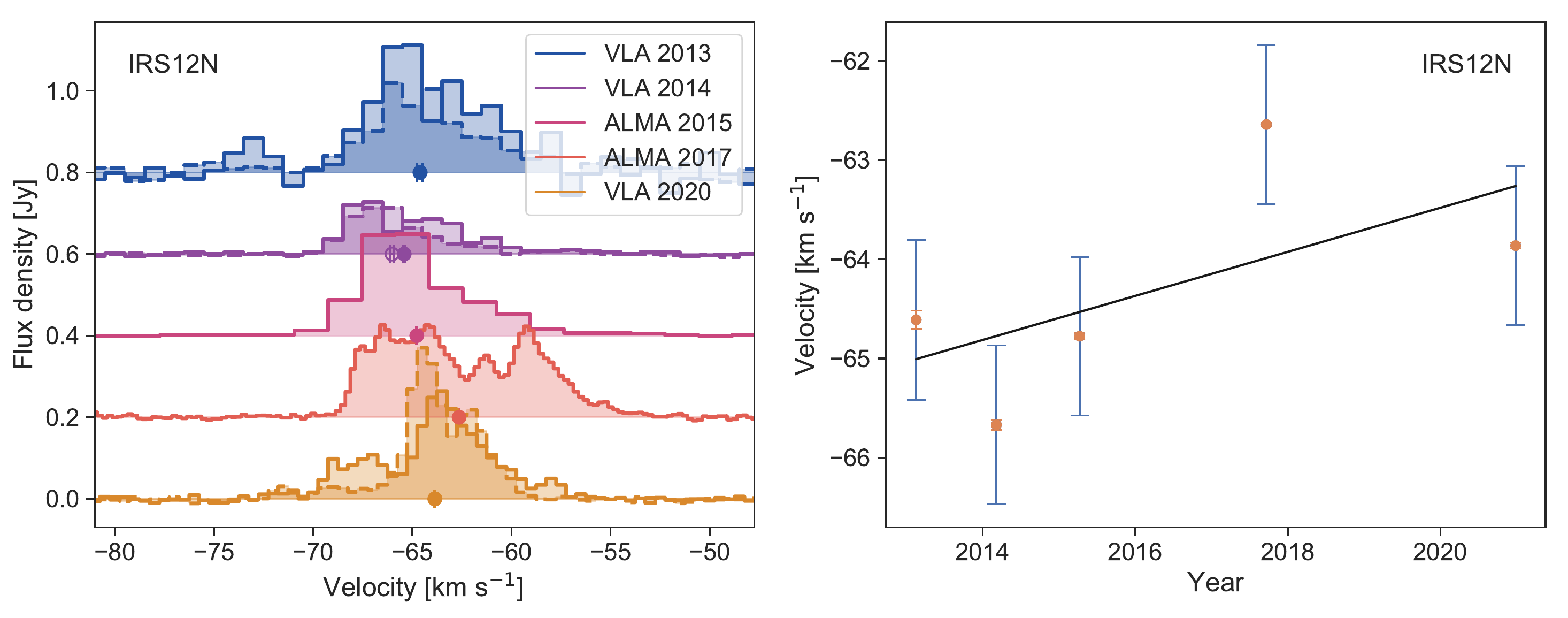}{\textwidth}{} }
\caption{(Continued)}
\end{figure}

\setcounter{figure}{4}
\begin{figure}[ht!]
\gridline{ \fig{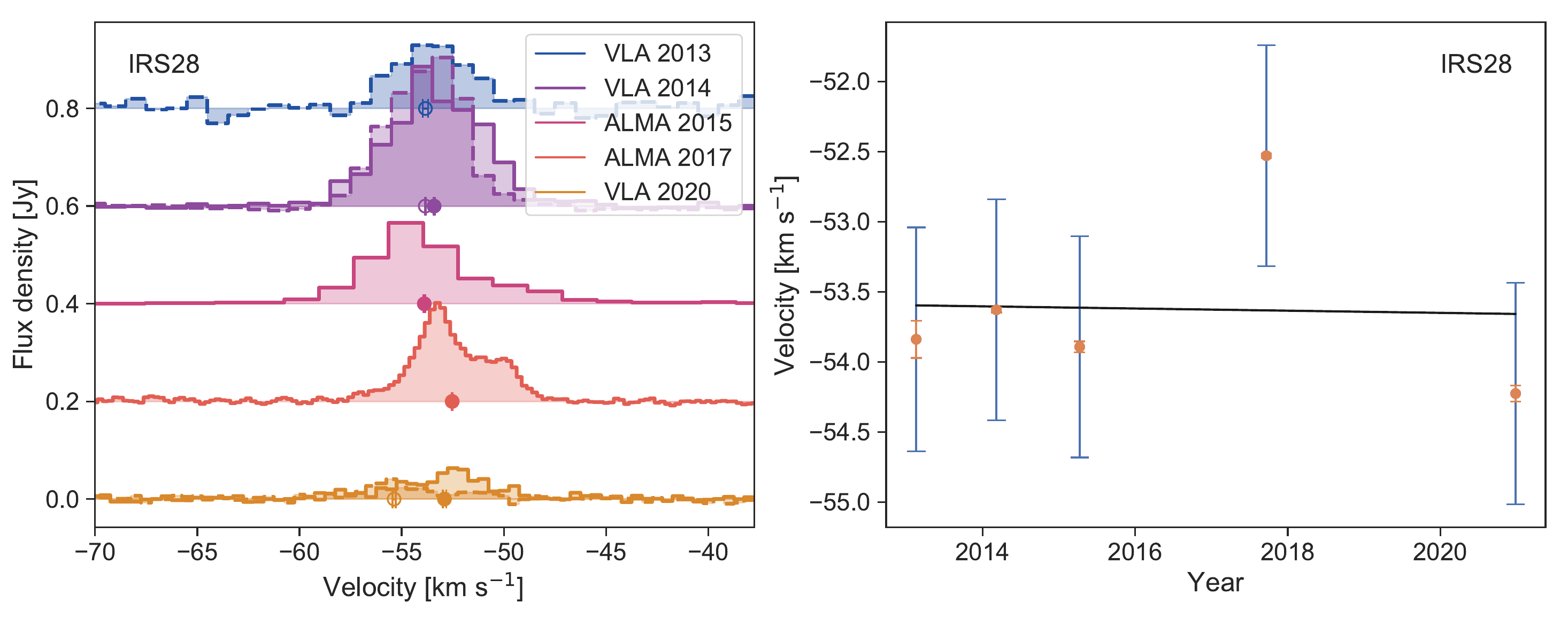}{\textwidth}{} }
\vspace*{-10mm}
\gridline{ \fig{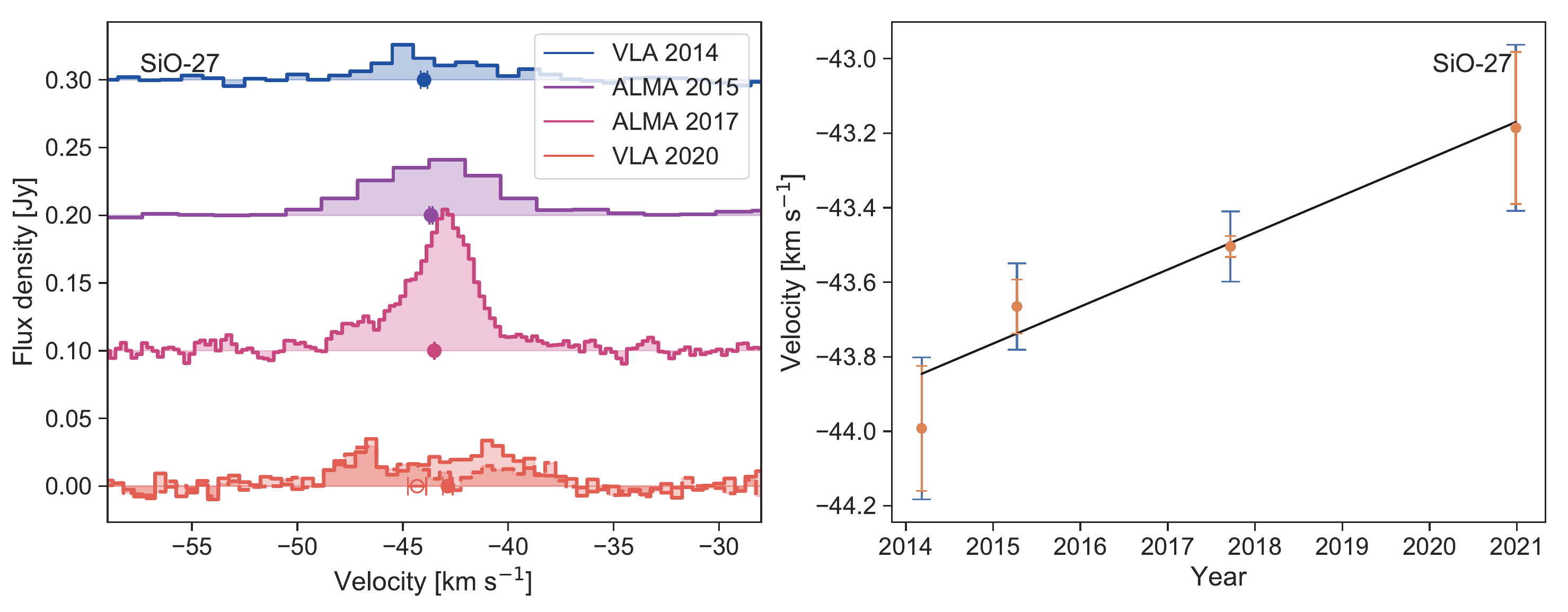}{\textwidth}{} }
\vspace*{-10mm}
\gridline{ \fig{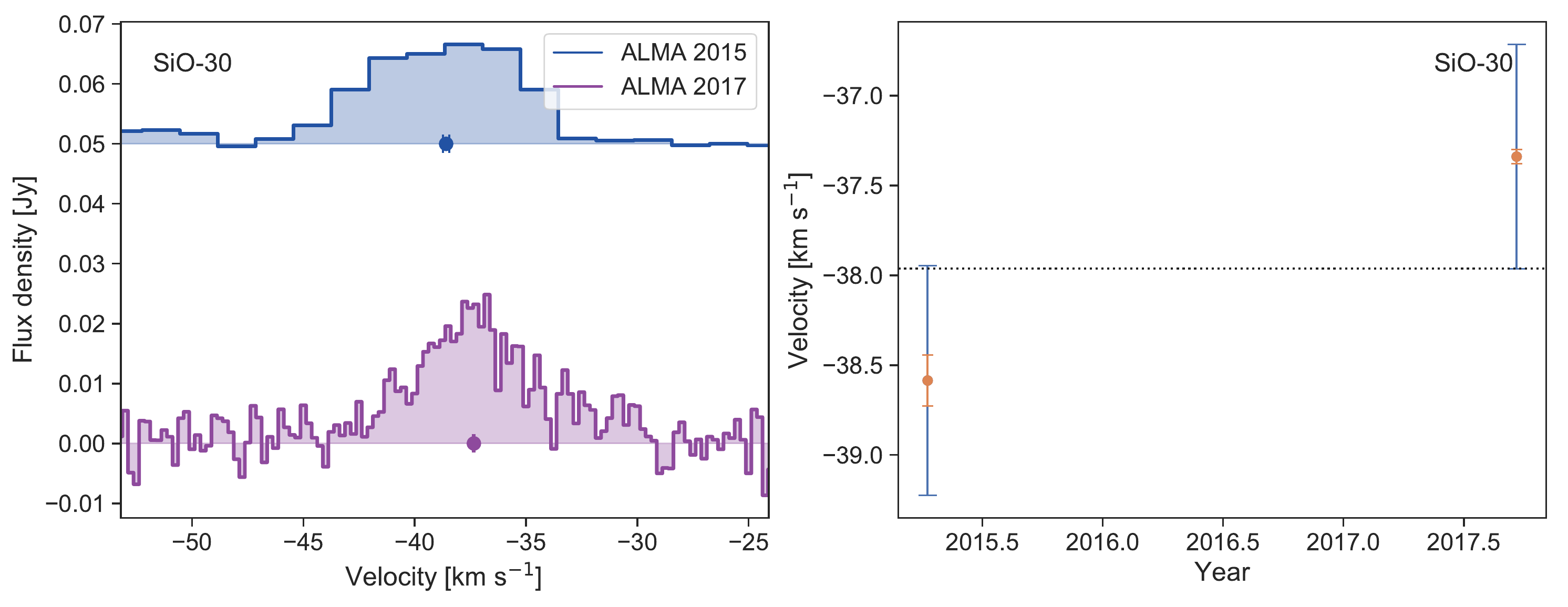}{\textwidth}{} }
\caption{(Continued)}
\end{figure}

\setcounter{figure}{4}
\begin{figure}[ht!]
\gridline{ \fig{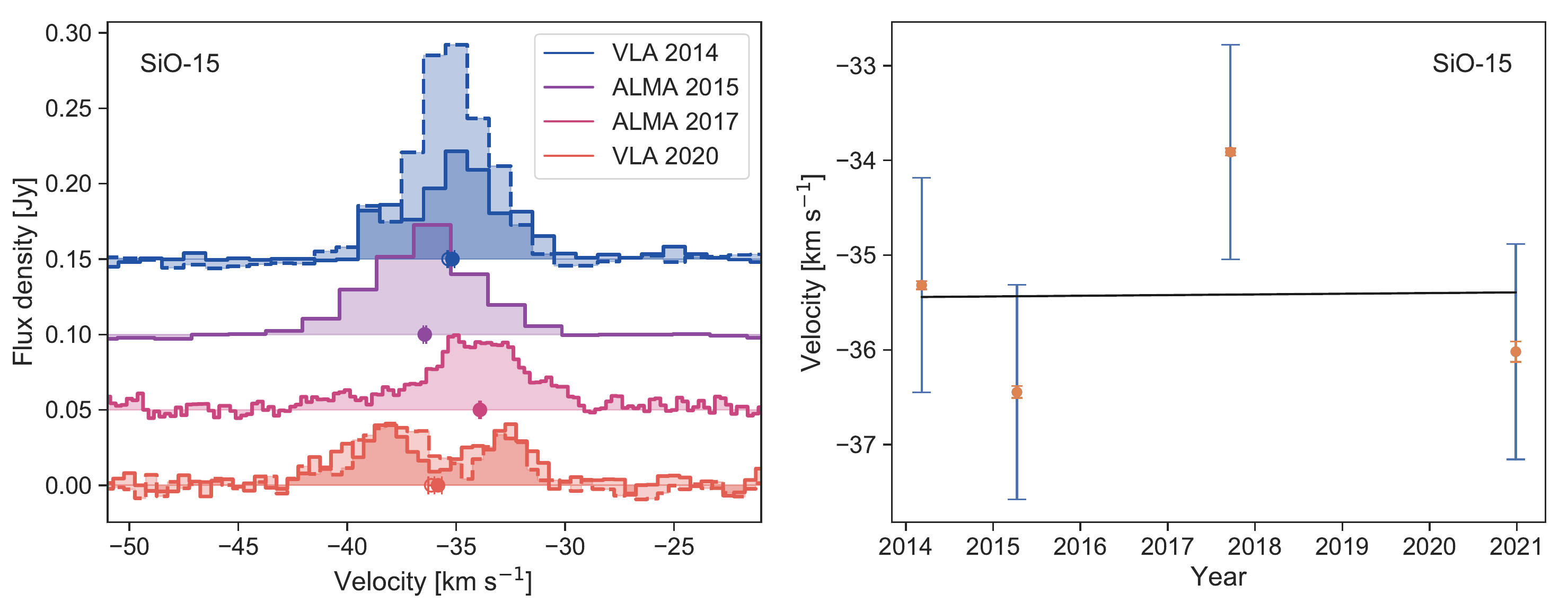}{\textwidth}{} }
\vspace*{-10mm}
\gridline{ \fig{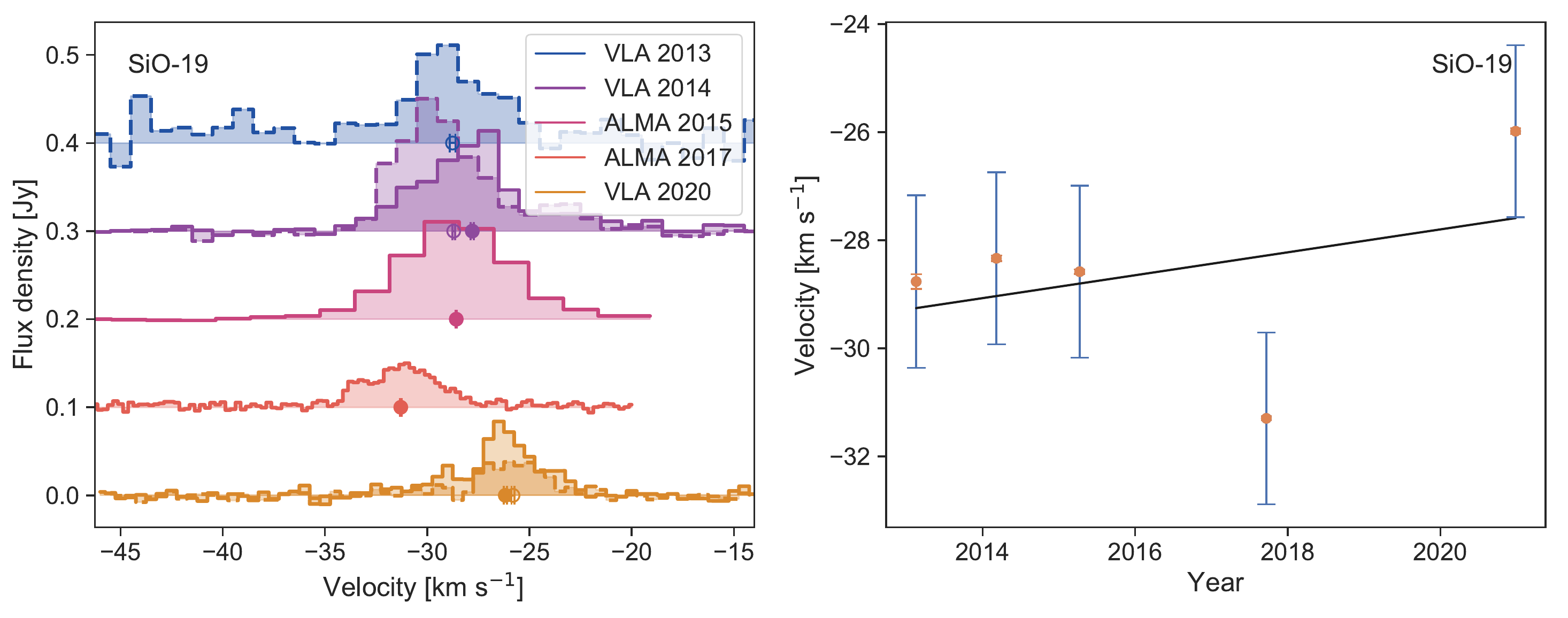}{\textwidth}{} }
\vspace*{-10mm}
\gridline{ \fig{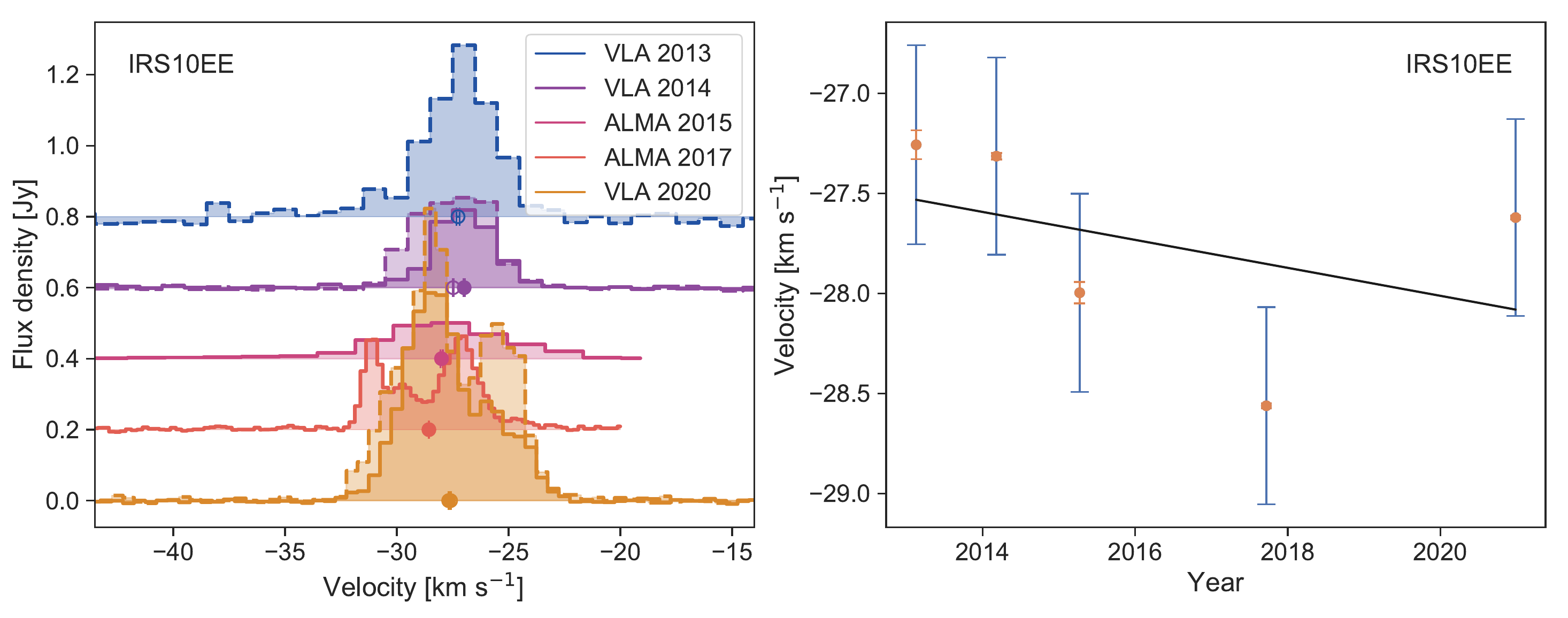}{\textwidth}{} }
\caption{(Continued)}
\end{figure}

\setcounter{figure}{4}
\begin{figure}[ht!]
\gridline{ \fig{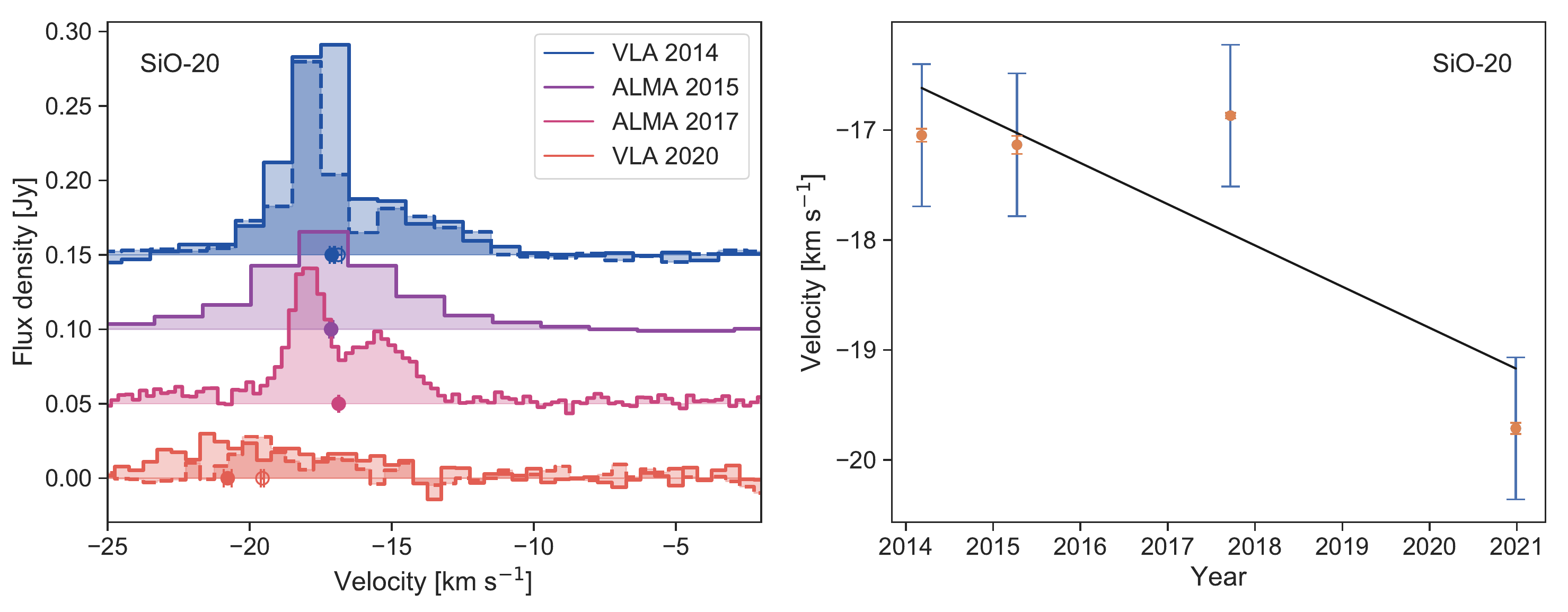}{\textwidth}{} }
\vspace*{-10mm}
\gridline{ \fig{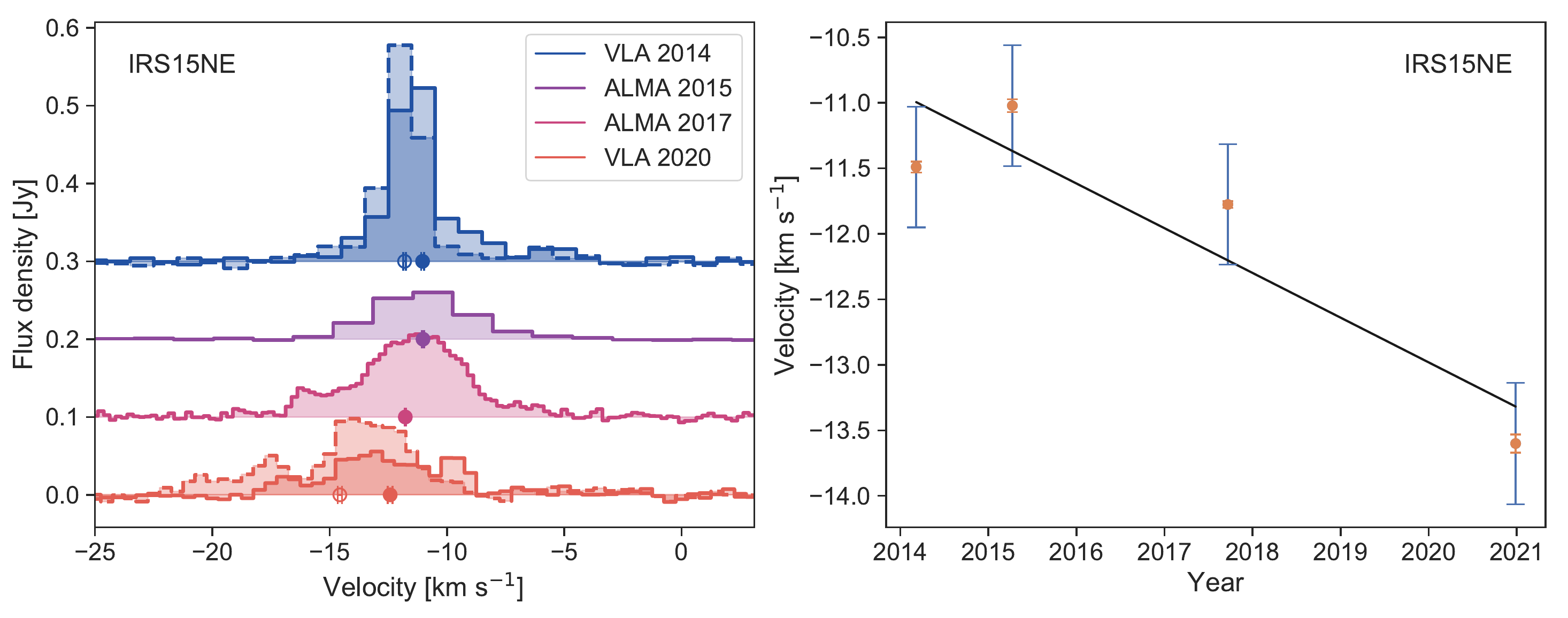}{\textwidth}{} }
\vspace*{-10mm}
\gridline{ \fig{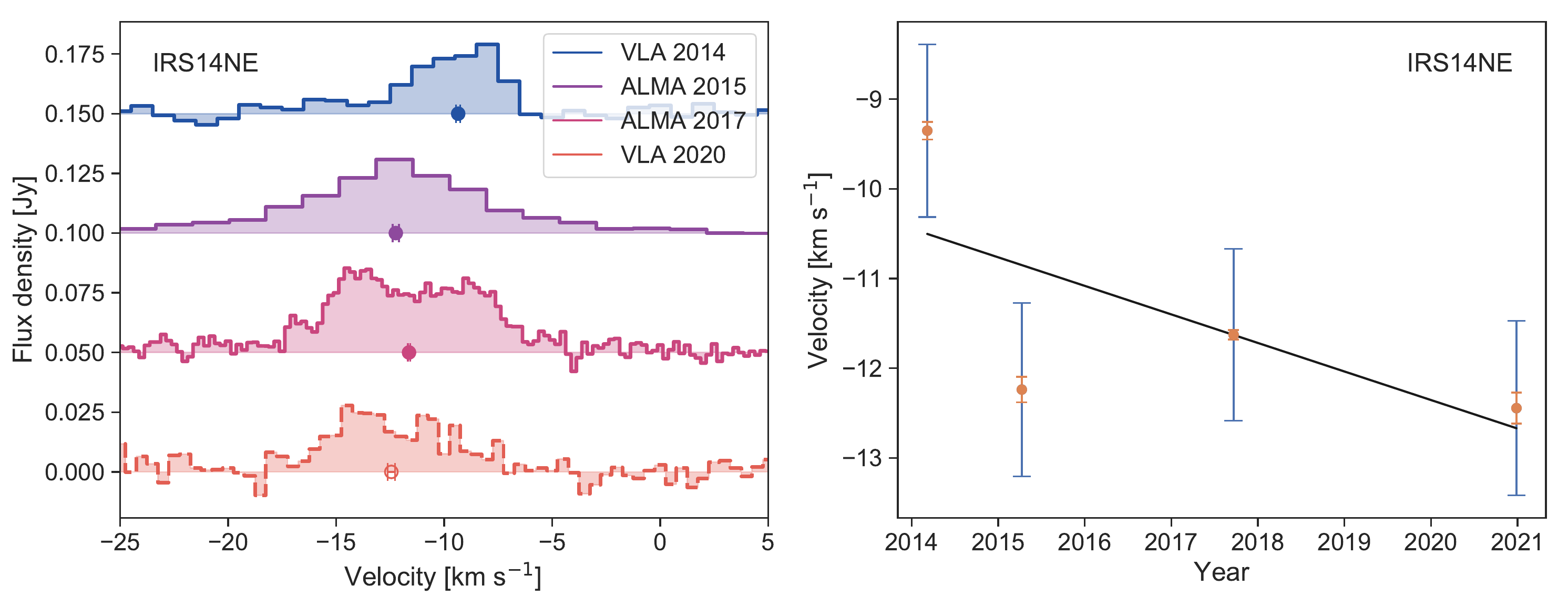}{\textwidth}{} }
\caption{(Continued)}
\end{figure}

\setcounter{figure}{4}
\begin{figure}[ht!]
\gridline{ \fig{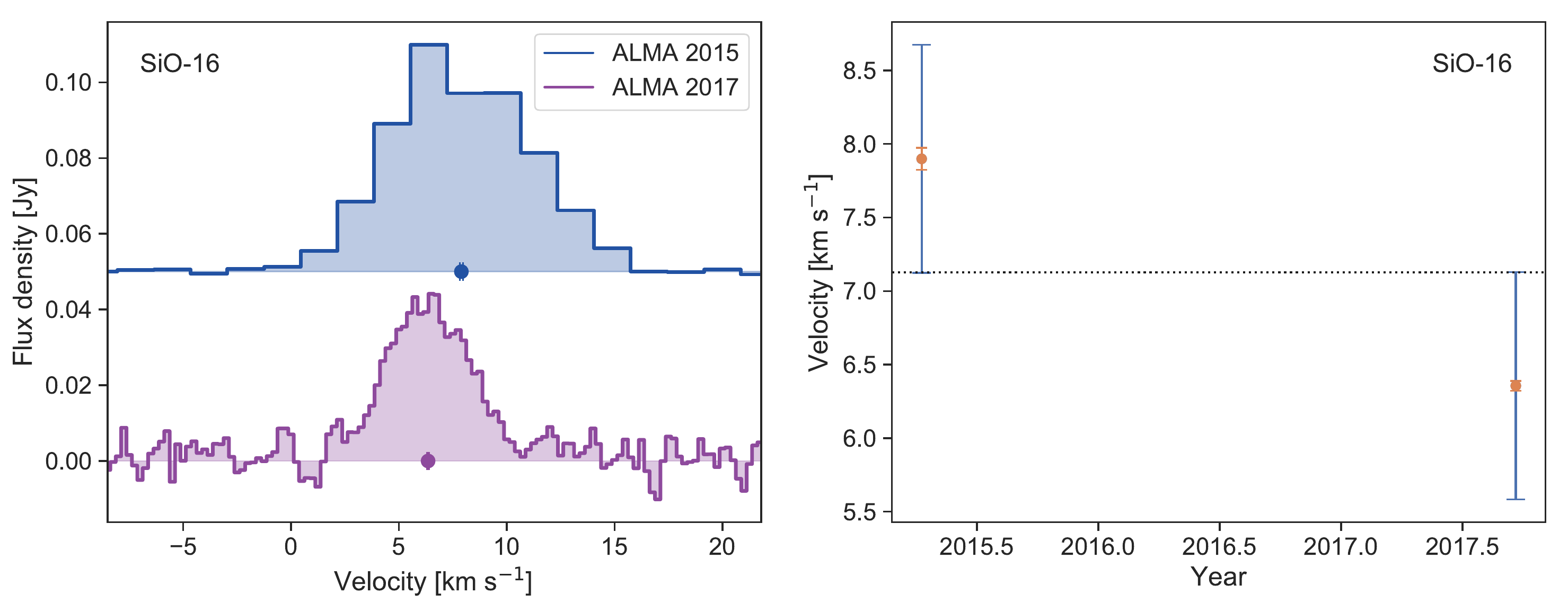}{\textwidth}{} }
\vspace*{-10mm}
\gridline{ \fig{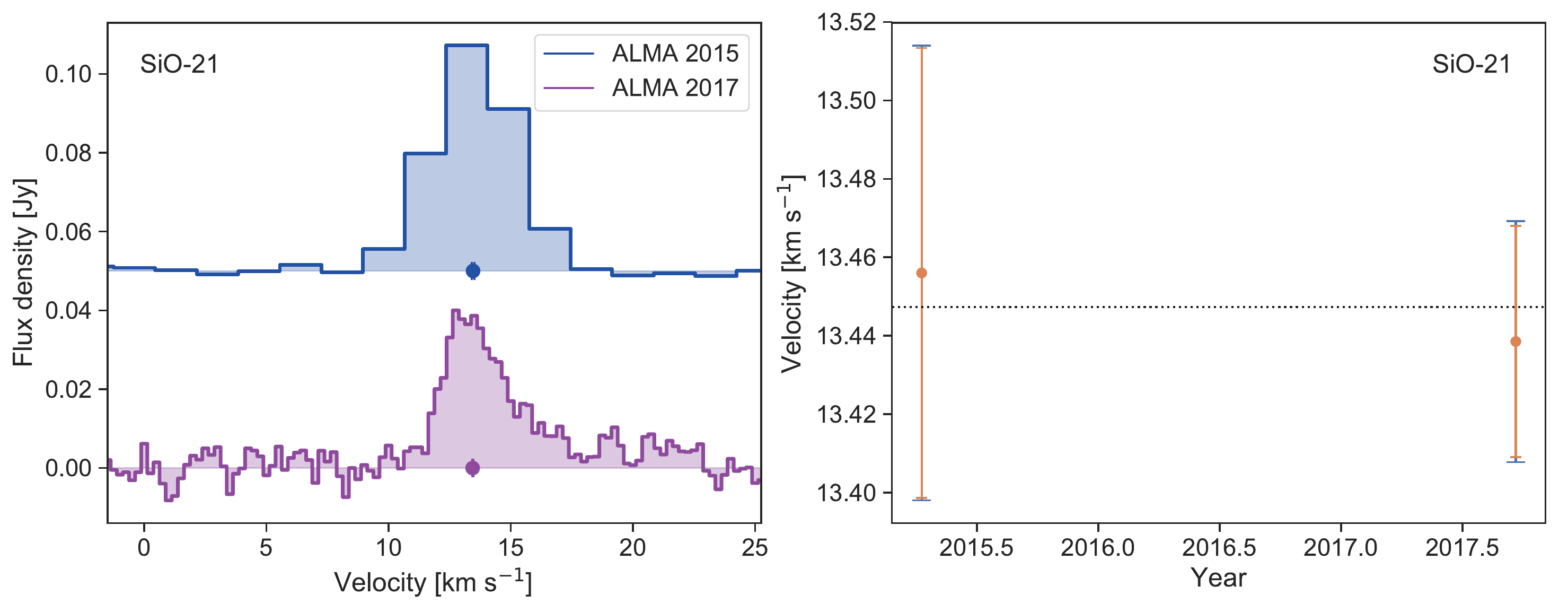}{\textwidth}{} }
\vspace*{-10mm}
\gridline{ \fig{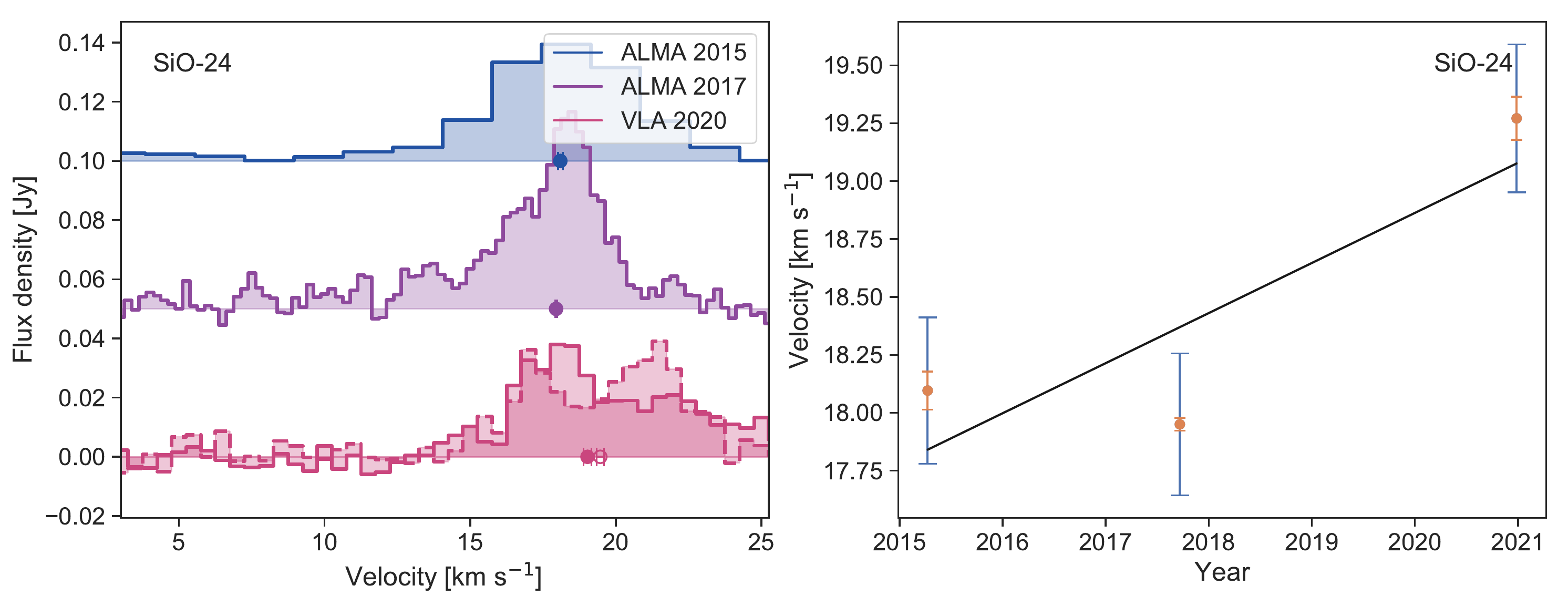}{\textwidth}{} }
\caption{(Continued)}
\end{figure}

\setcounter{figure}{4}
\begin{figure}[ht!]
\gridline{ \fig{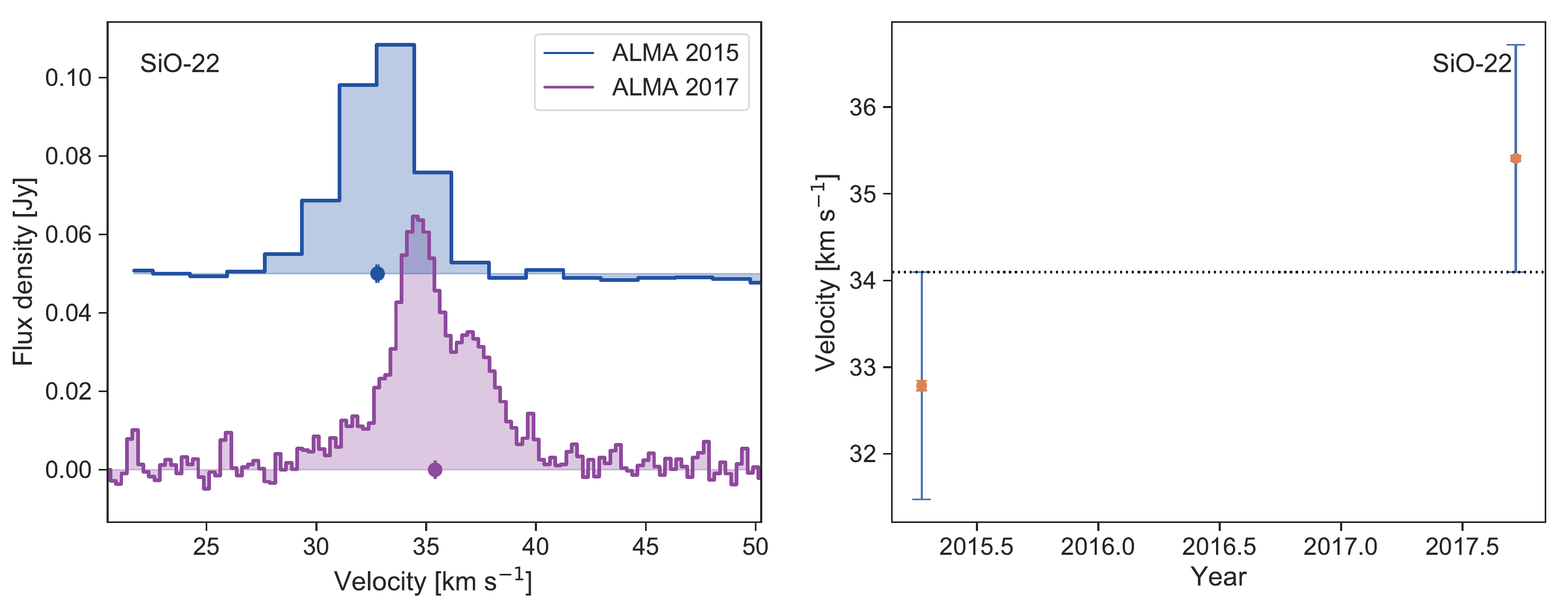}{\textwidth}{} }
\vspace*{-10mm}
\gridline{ \fig{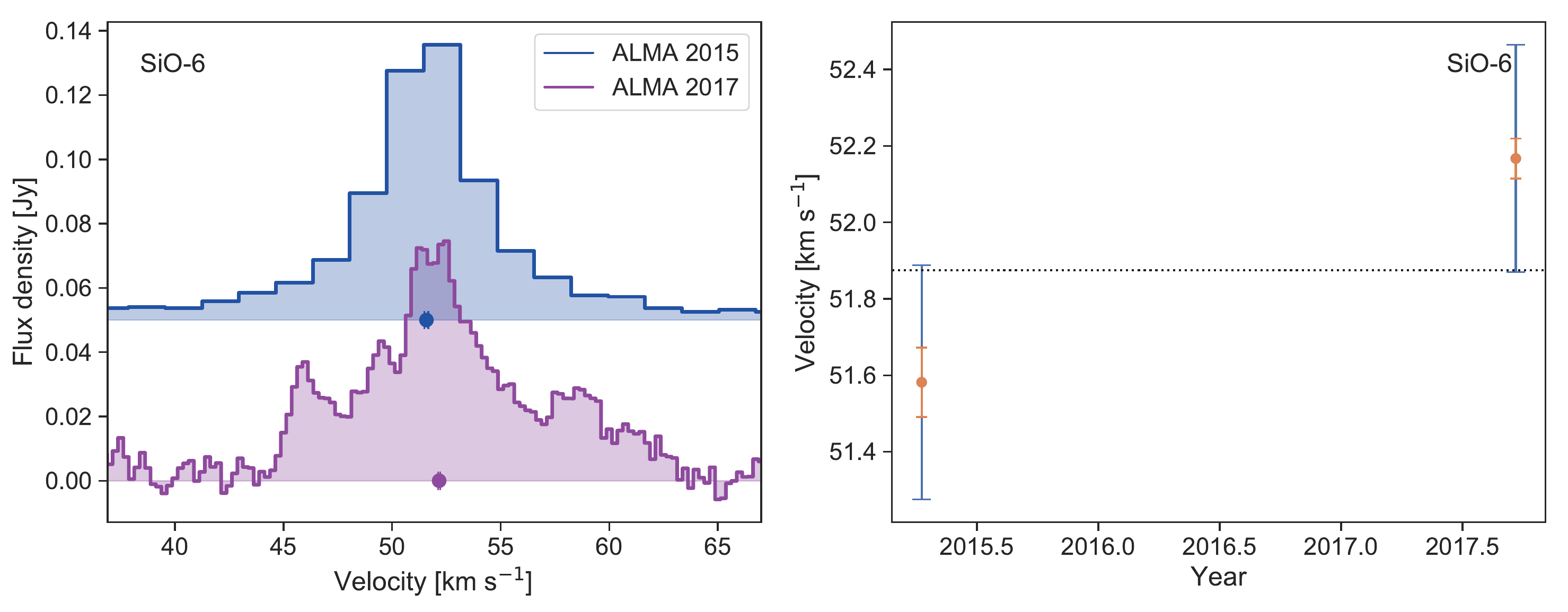}{\textwidth}{} }
\vspace*{-10mm}
\gridline{ \fig{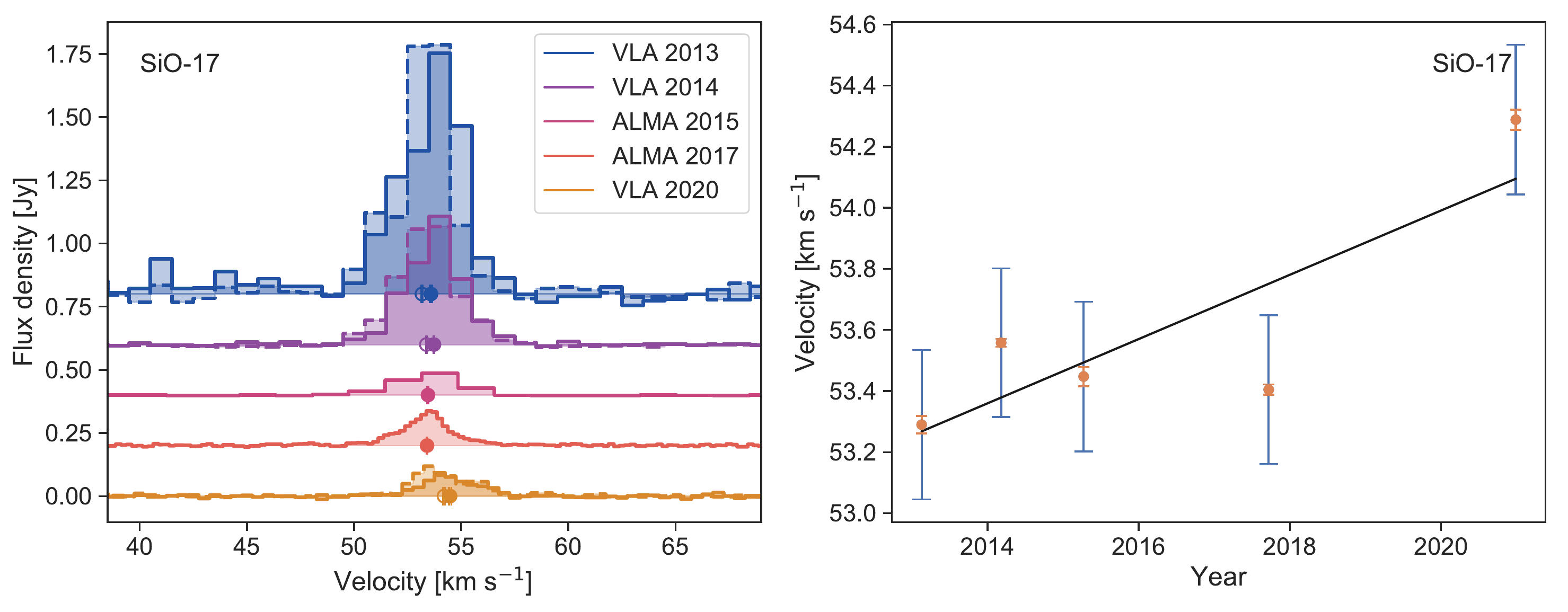}{\textwidth}{} }
\caption{(Continued)}
\end{figure}

\setcounter{figure}{4}
\begin{figure}[ht!]
\gridline{ \fig{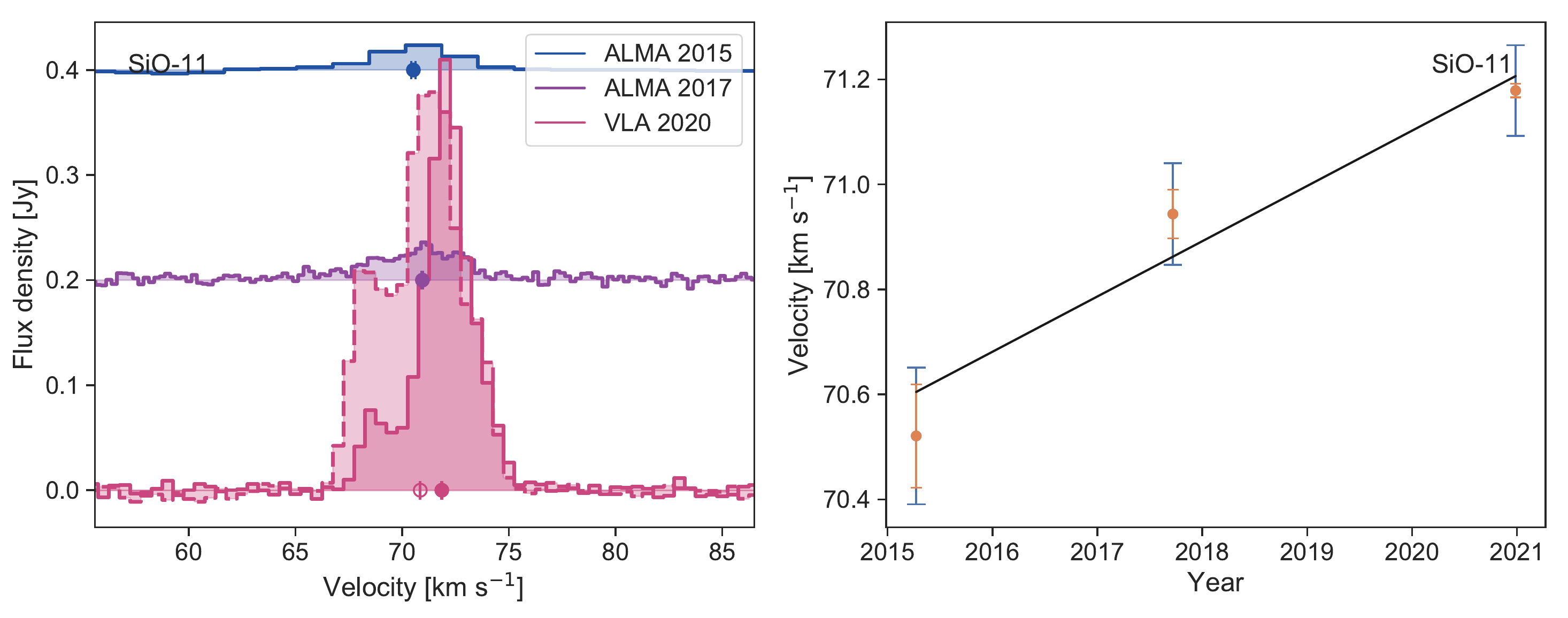}{\textwidth}{} }
\vspace*{-10mm}
\gridline{ \fig{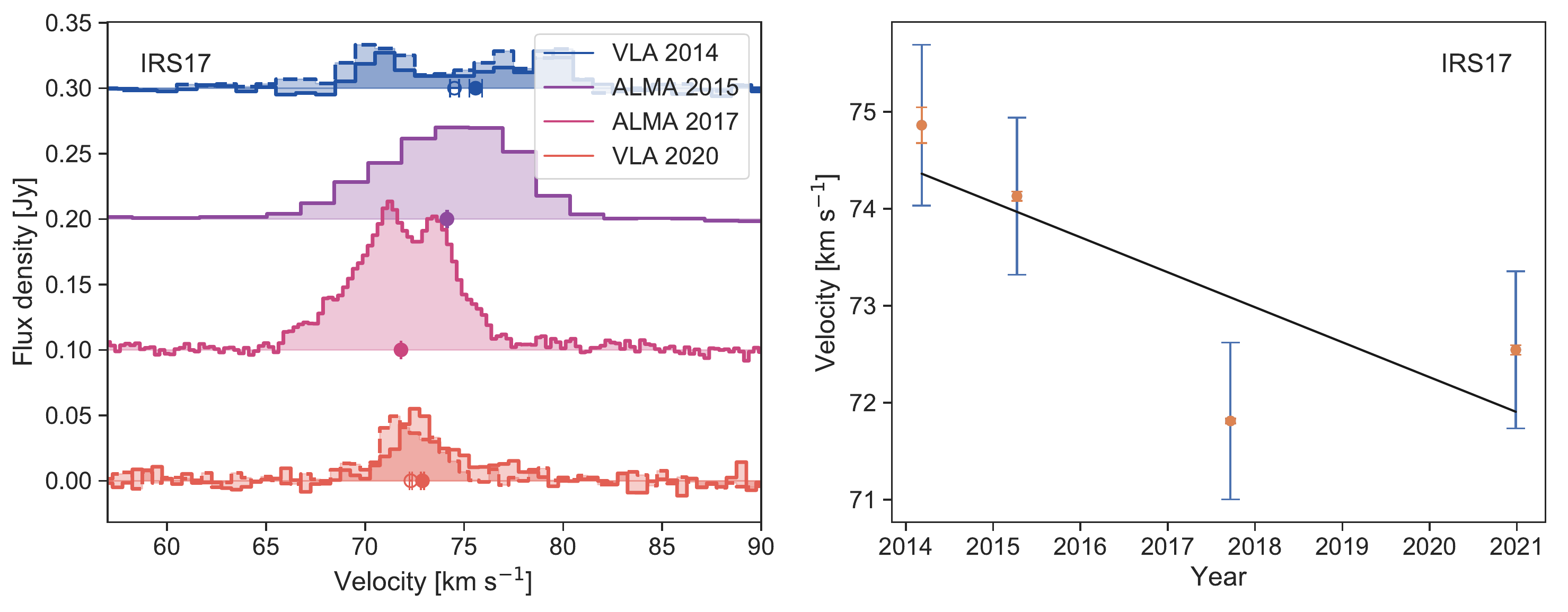}{\textwidth}{} }
\vspace*{-10mm}
\gridline{ \fig{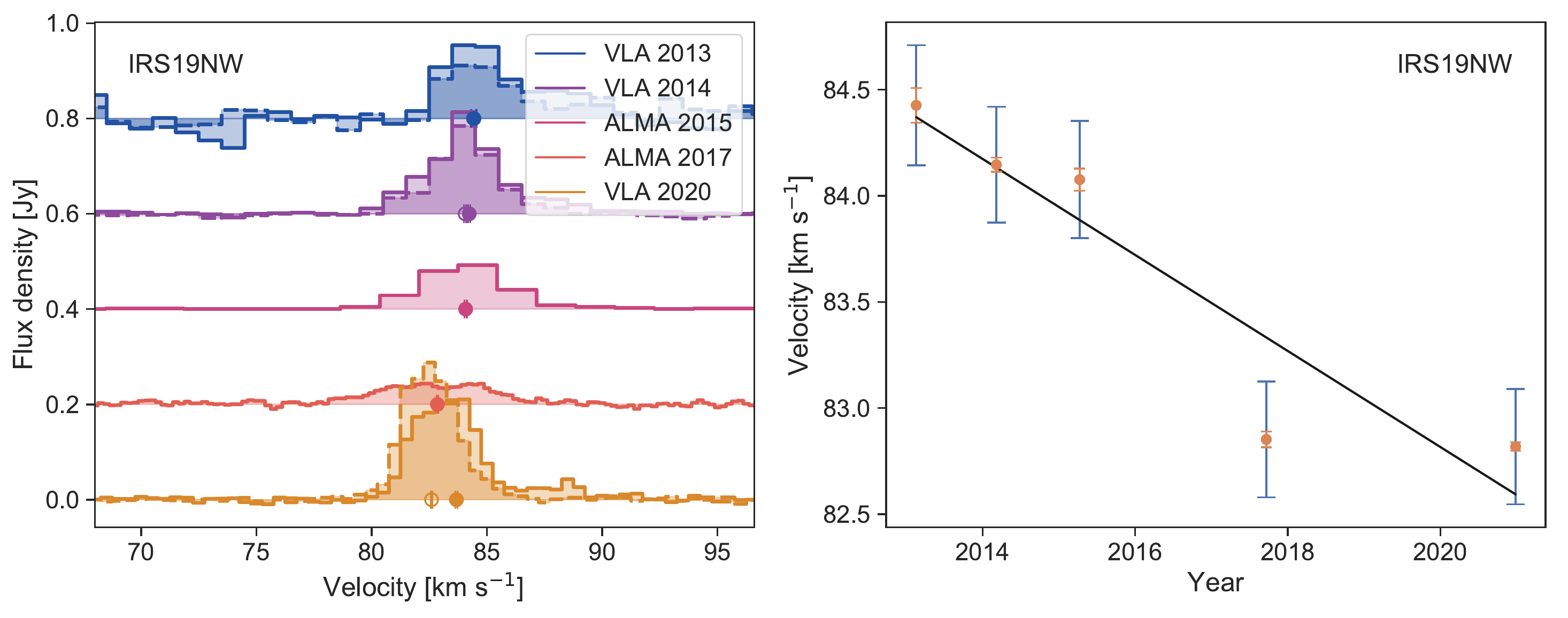}{\textwidth}{} }
\caption{(Continued)}
\end{figure}

\setcounter{figure}{4}
\begin{figure}[ht!]
\gridline{ \fig{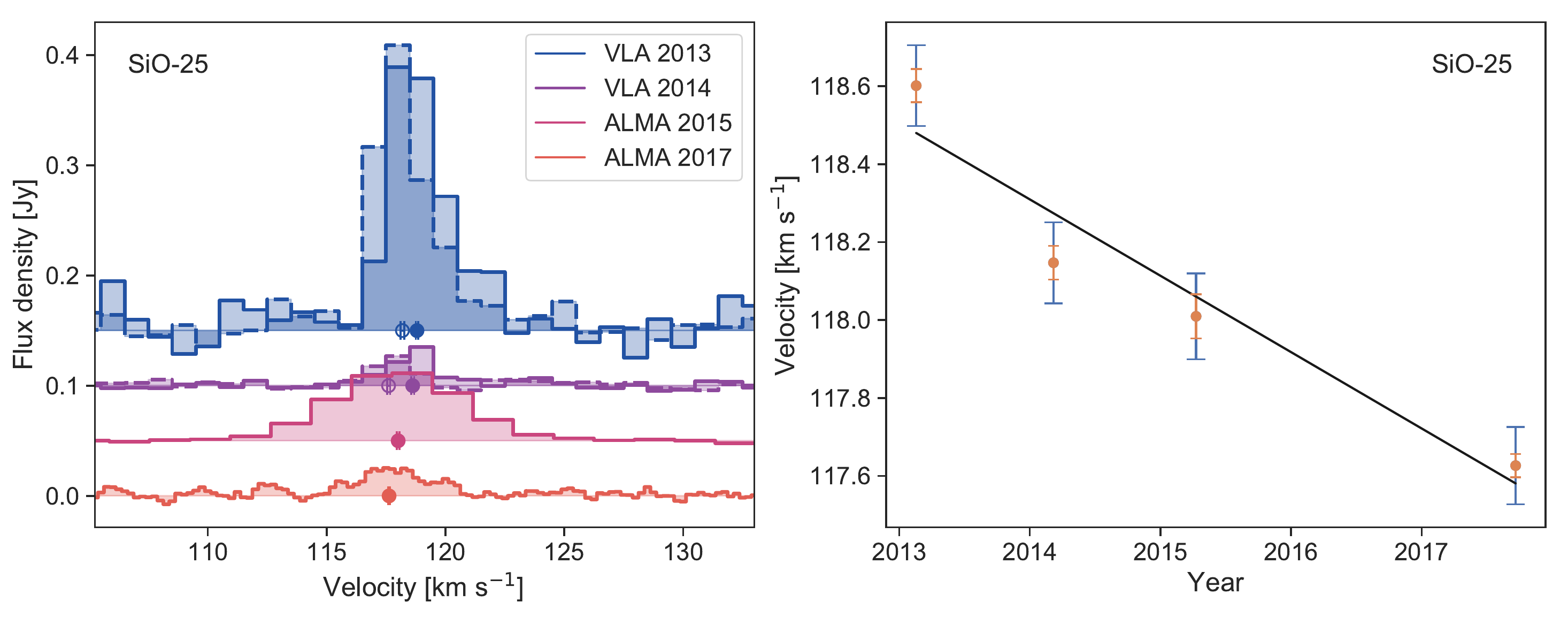}{\textwidth}{} }
\caption{(Continued)}
\end{figure}

\section{Proper motion and proper acceleration fits}\label{app:pm fits}

Figure \ref{fig:pm fits} shows proper motion and proper acceleration fits described in Section \ref{sec:pm}.

\begin{figure}[ht!]
\gridline{ \fig{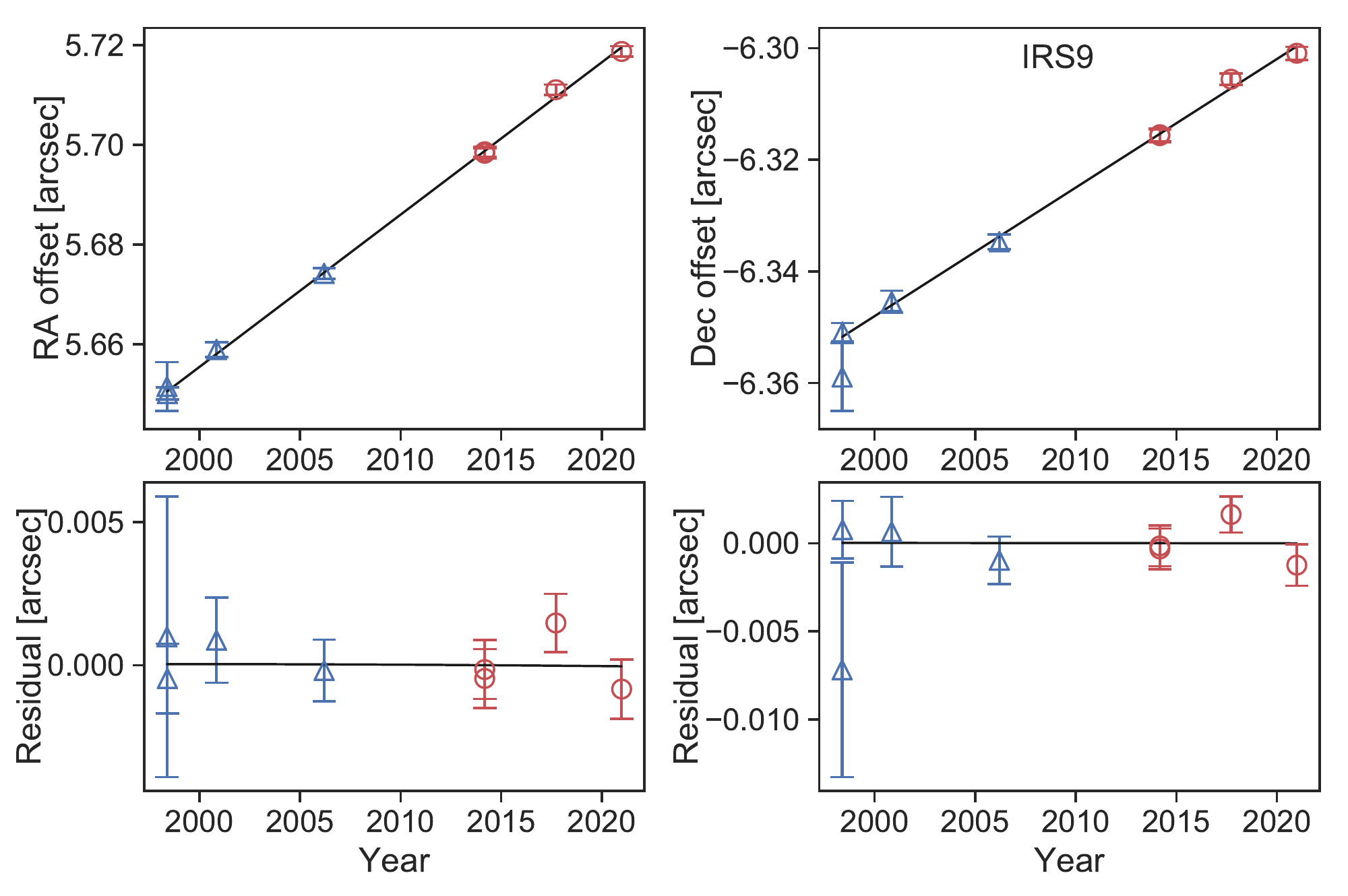}{0.48\textwidth}{}
           \fig{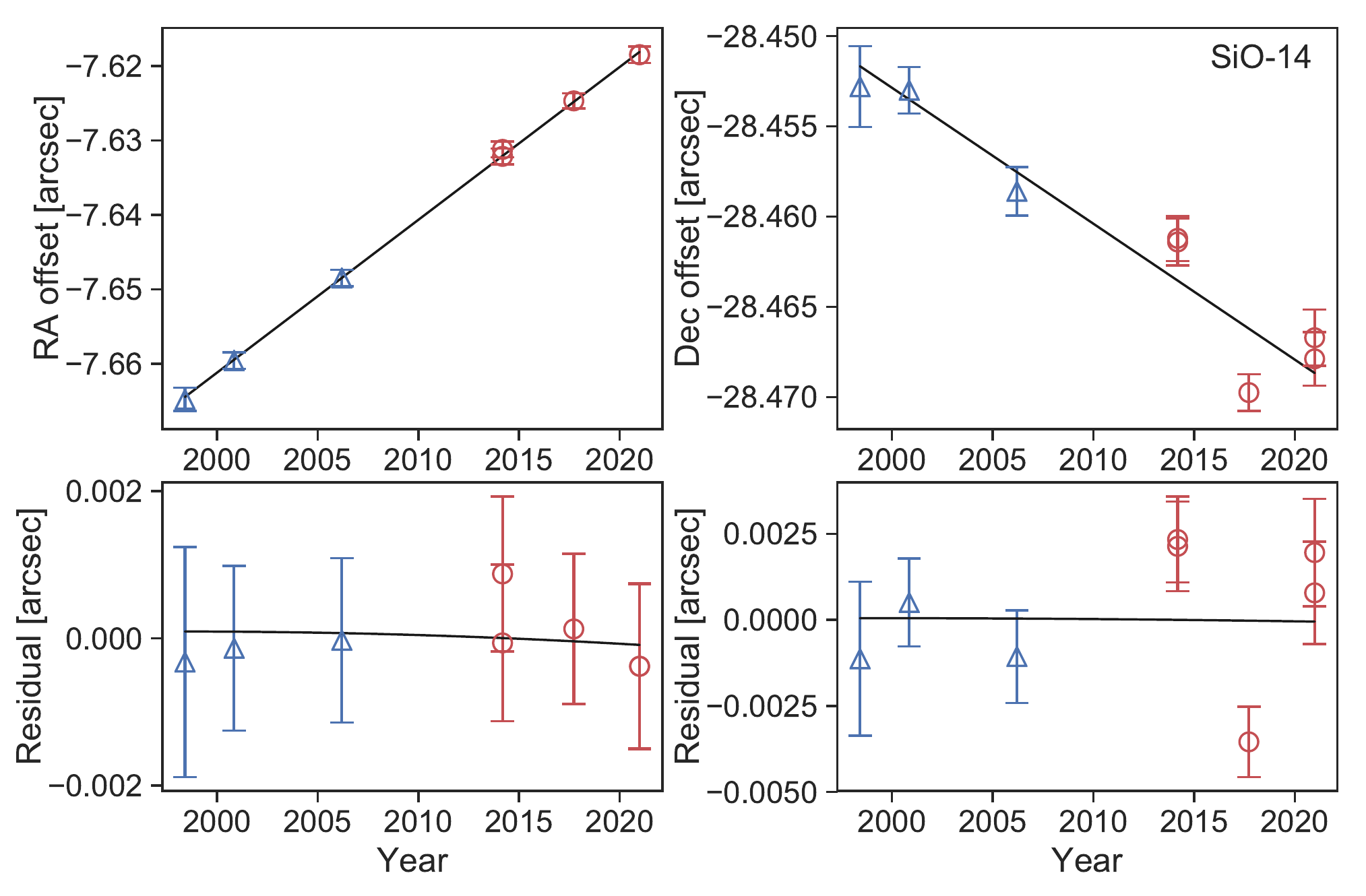}{0.48\textwidth}{} }
\vspace*{-8mm}
\gridline{ \fig{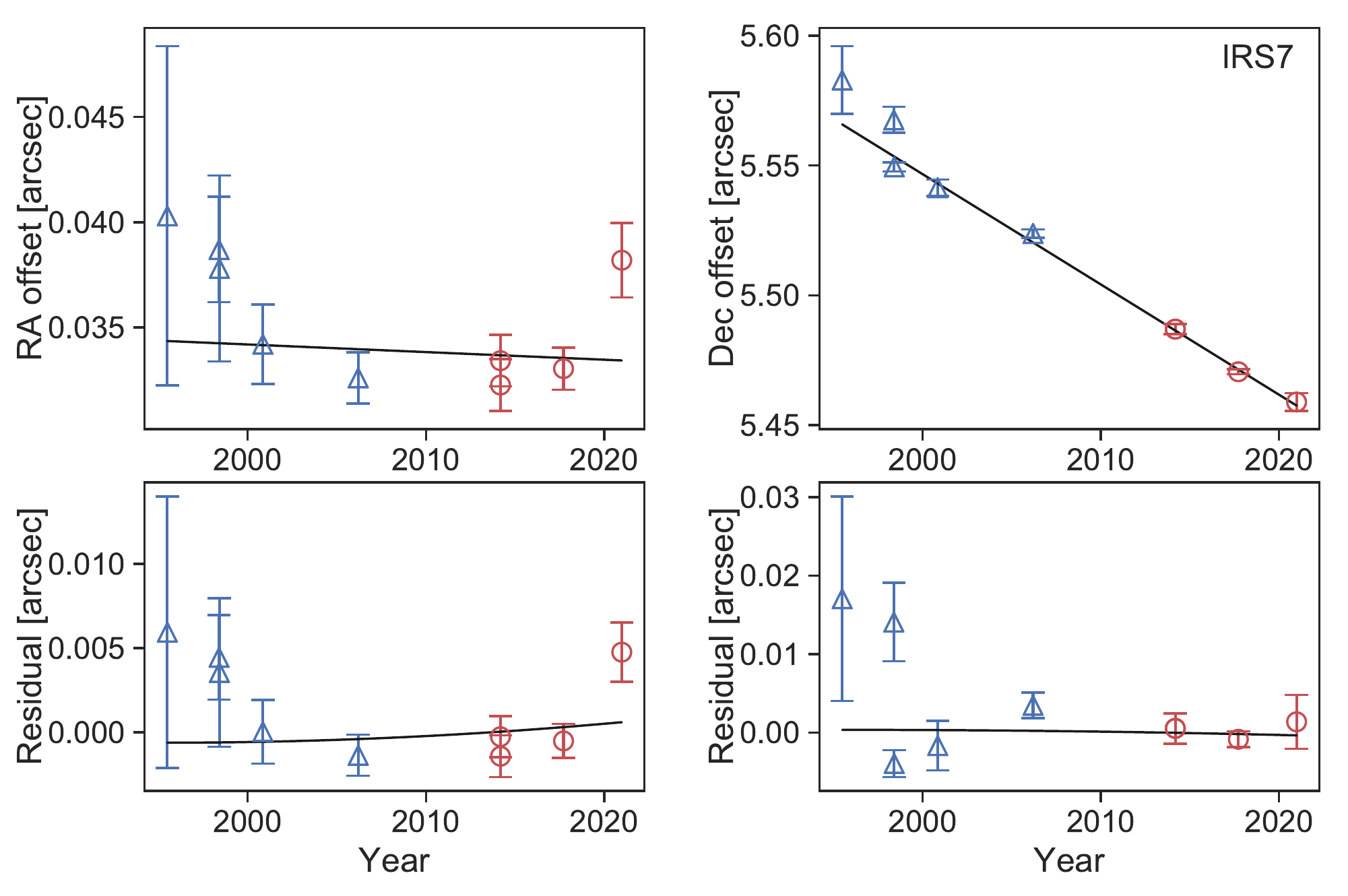}{0.48\textwidth}{}
           \fig{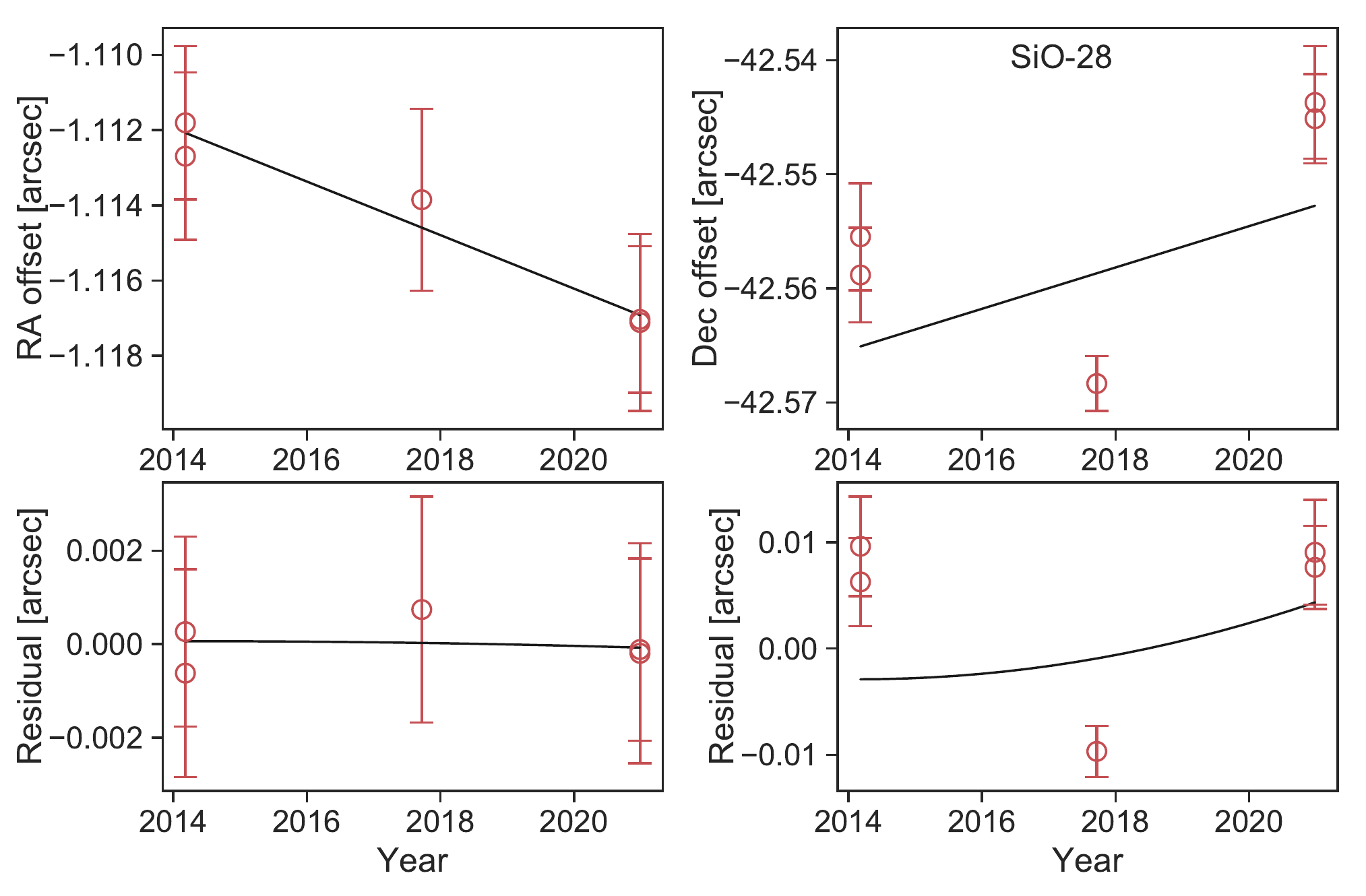}{0.48\textwidth}{} }
\vspace*{-8mm}
\gridline{ \fig{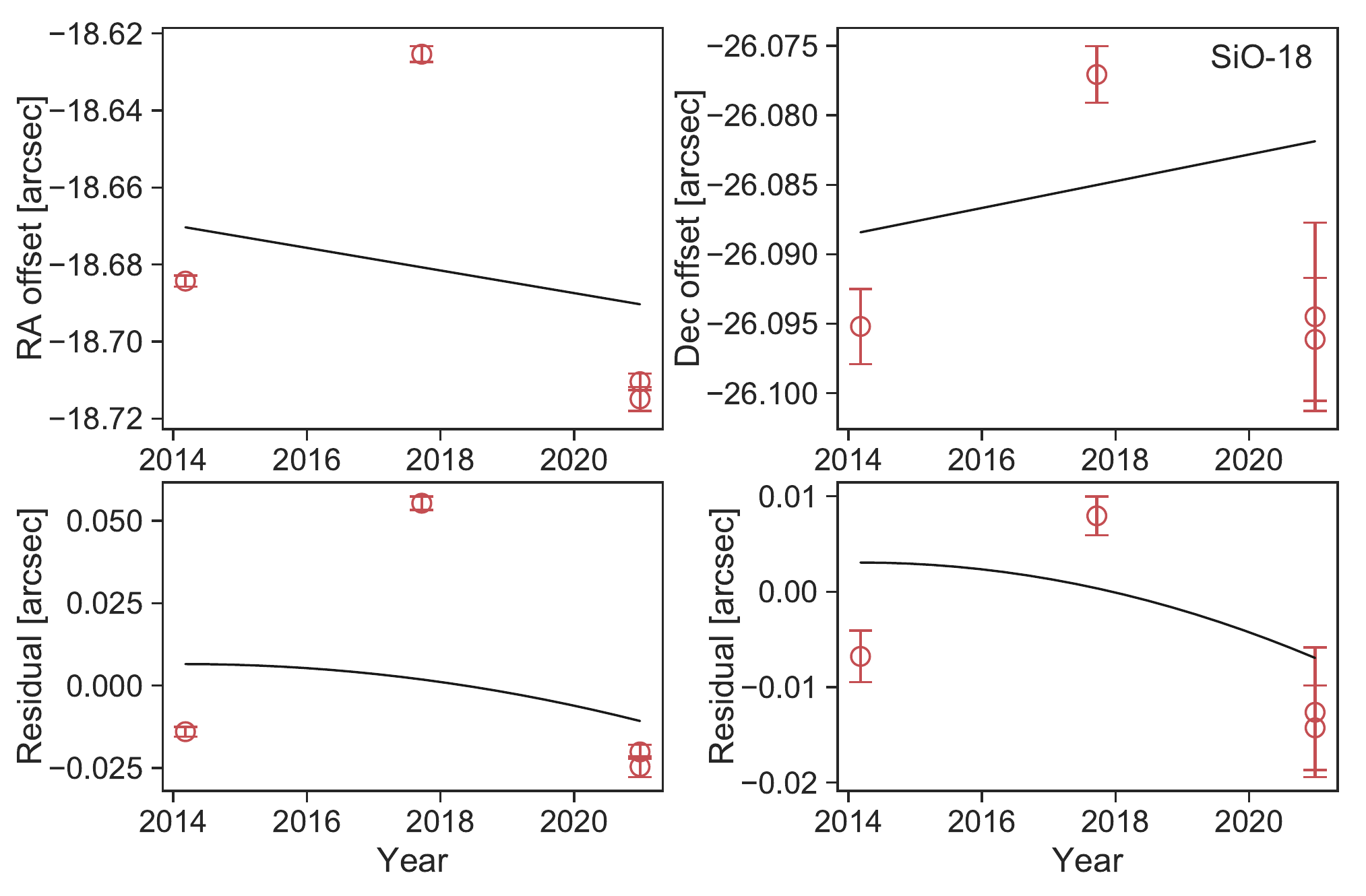}{0.48\textwidth}{}
           \fig{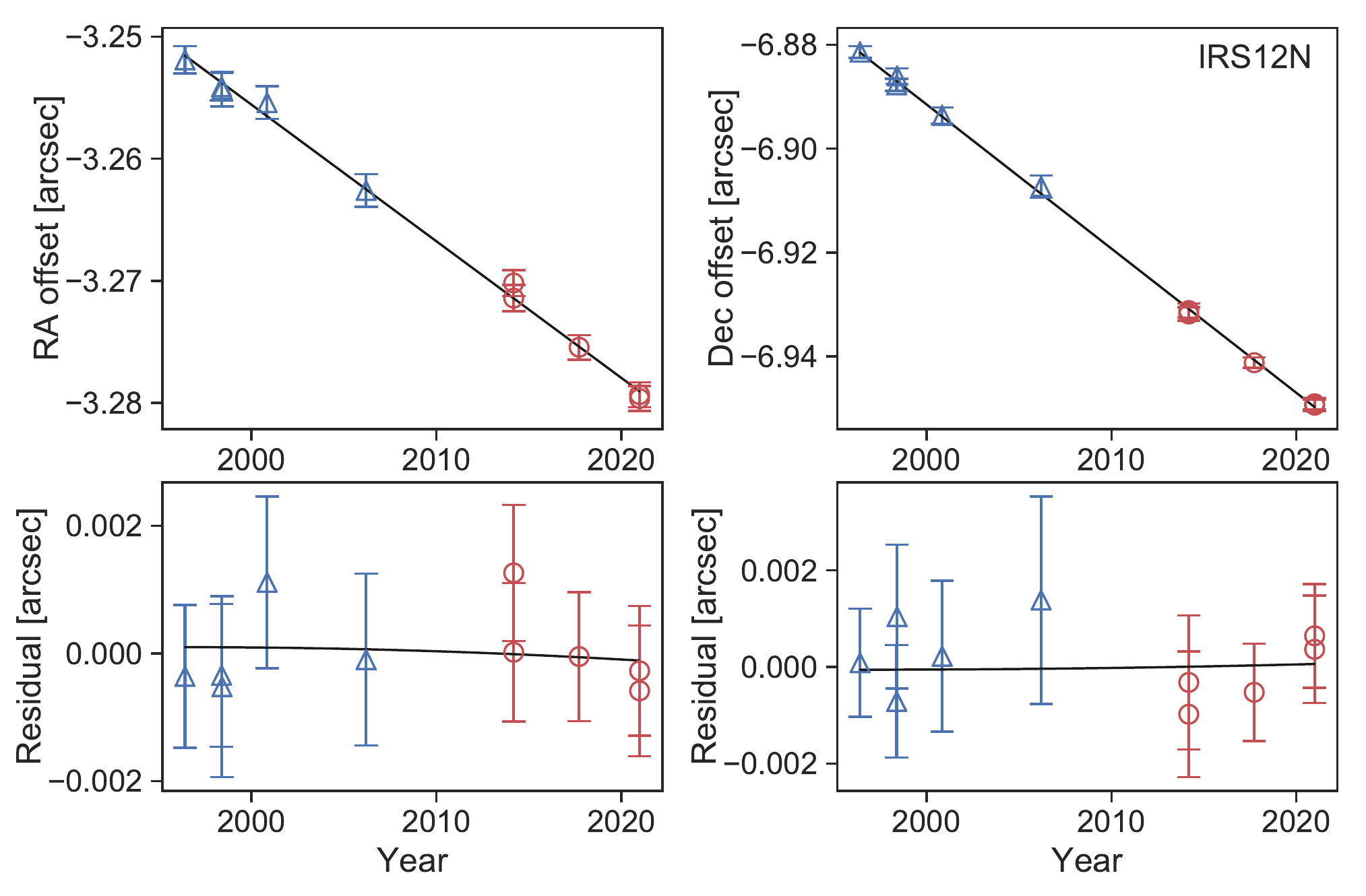}{0.48\textwidth}{} }
\caption{Proper motion and proper acceleration fits for each maser described in Section \ref{sec:pm}. Maser positions are from \citet{Reid2007} (blue triangles), \citet{Li2010} (orange squares), and this work (red circles). The upper panels show linear fits to the position time series to obtain proper motions, and the bottom panels show quadratic fits to the residuals of the linear fits to obtain proper accelerations.  \label{fig:pm fits}}
\end{figure}

\setcounter{figure}{5}
\begin{figure}[ht!]
\gridline{ \fig{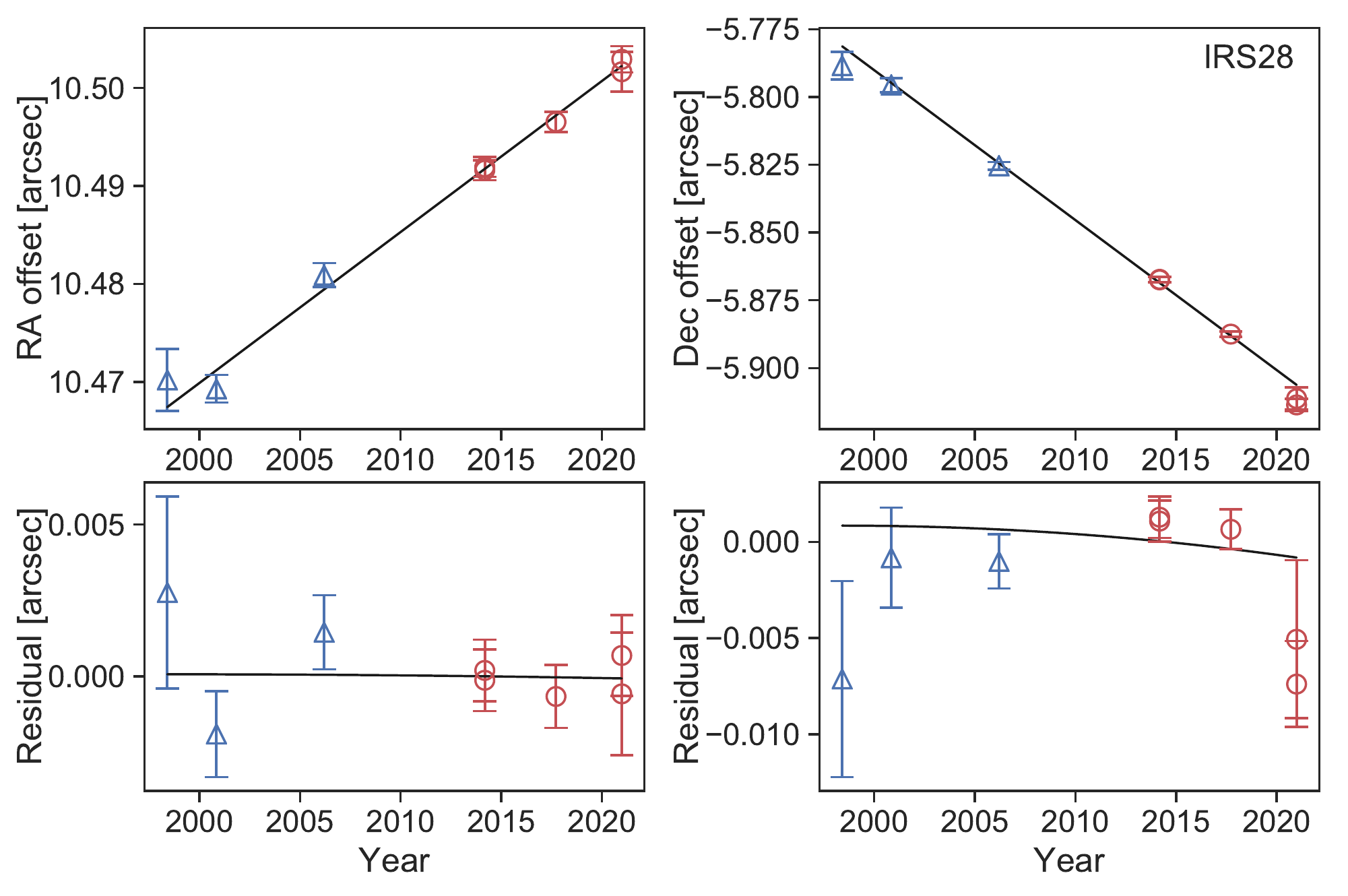}{0.48\textwidth}{}
           \fig{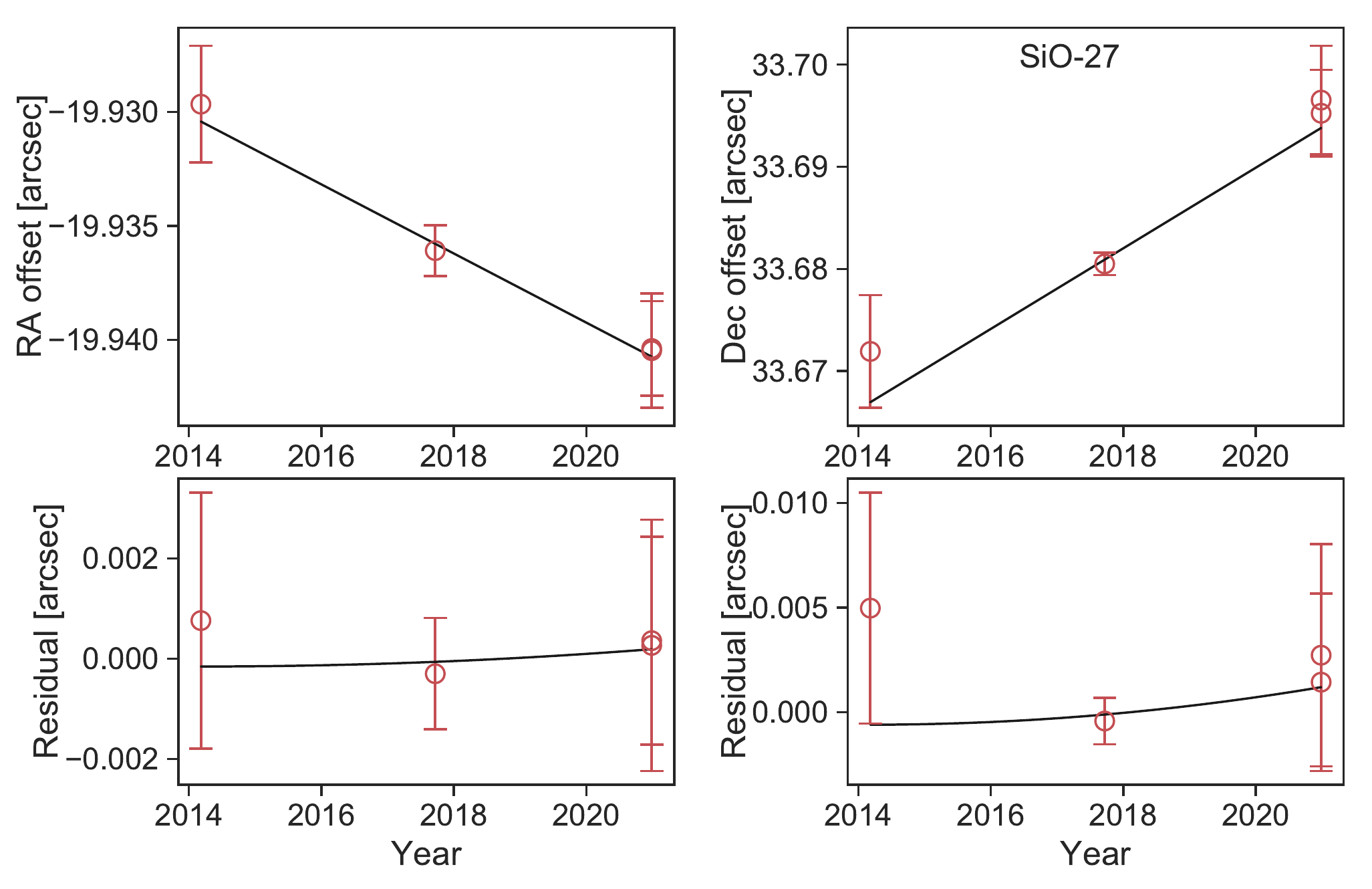}{0.48\textwidth}{} }
\vspace*{-8mm}
\gridline{ \fig{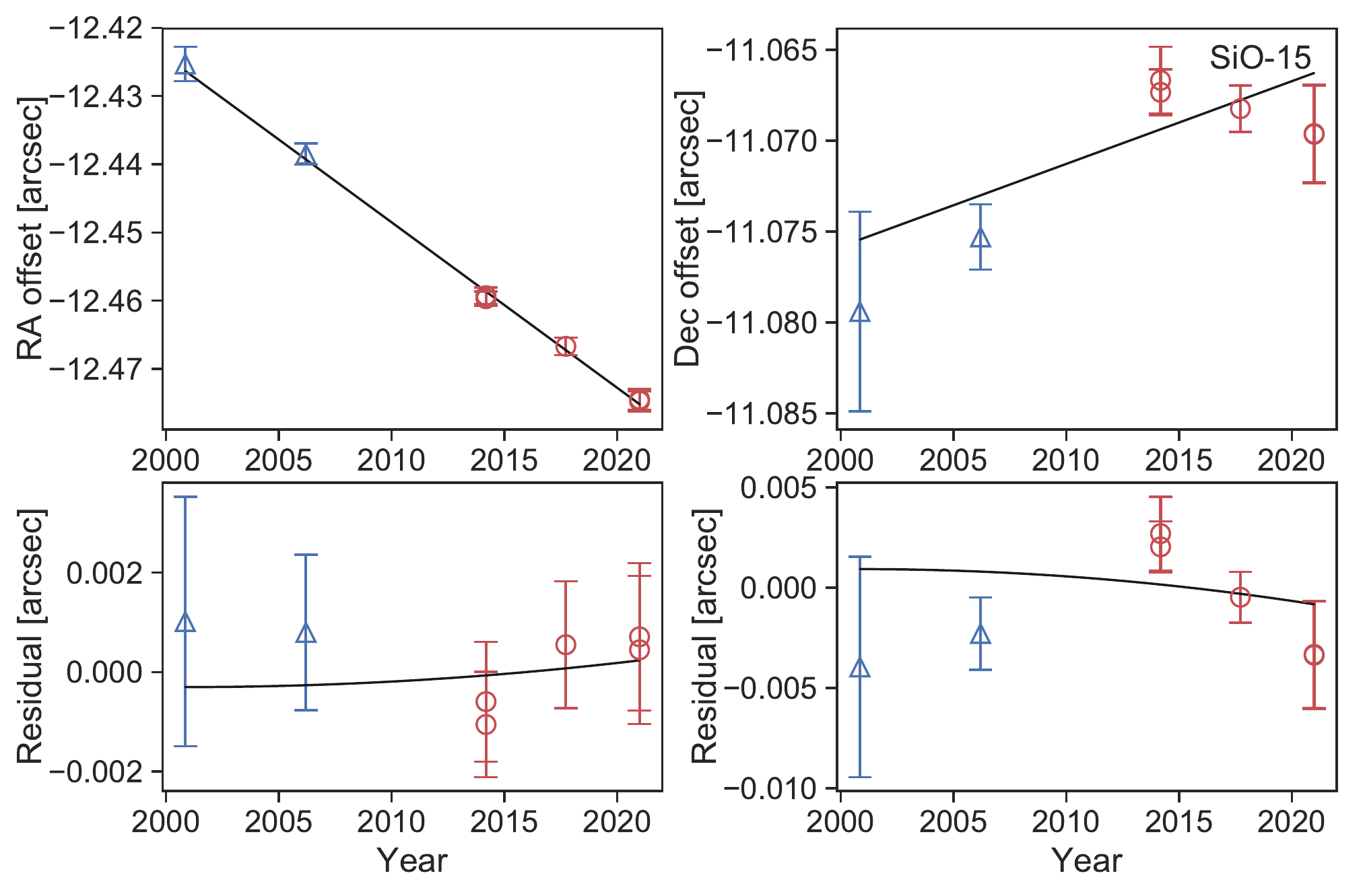}{0.48\textwidth}{}
           \fig{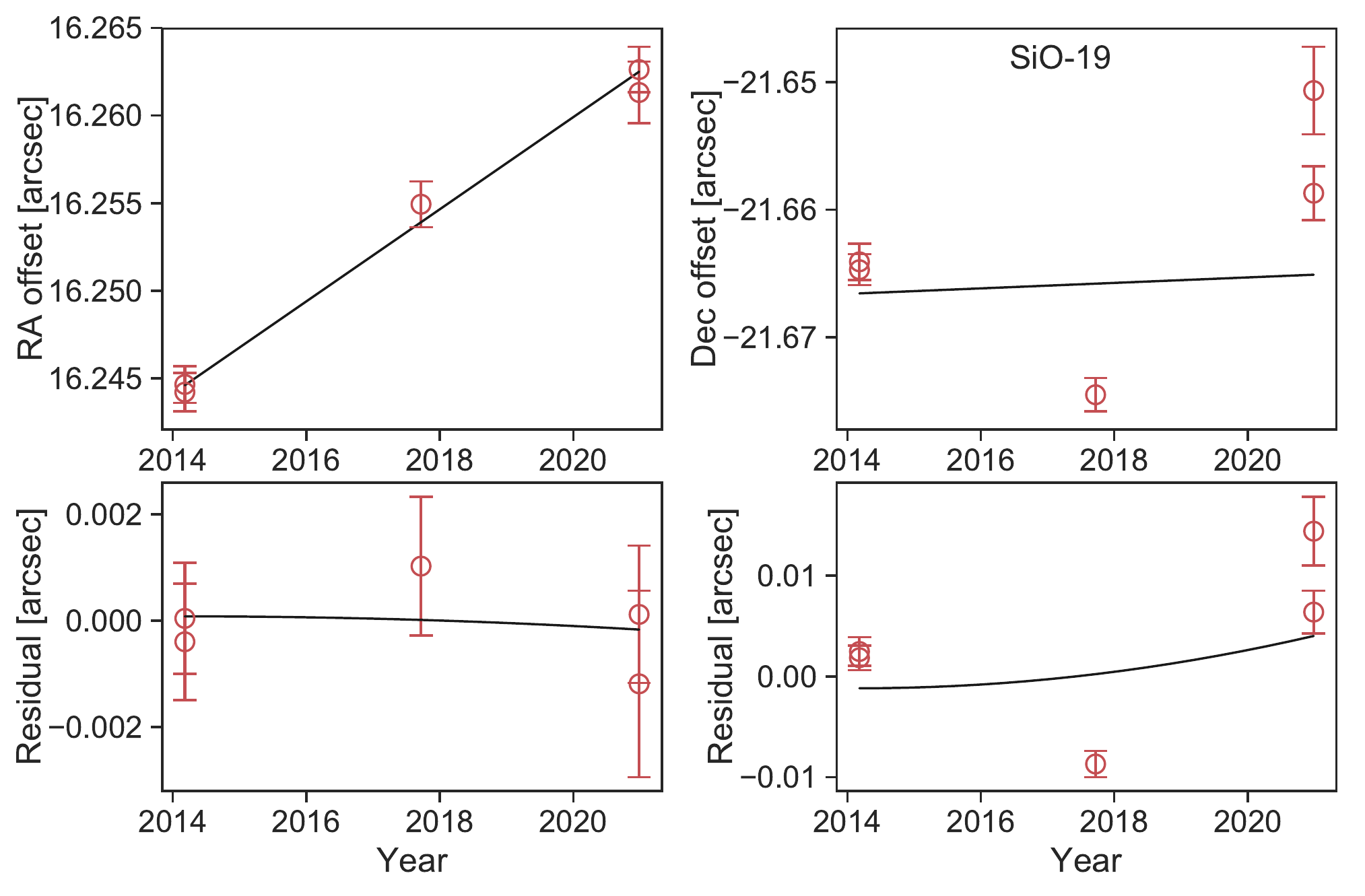}{0.48\textwidth}{} }
\vspace*{-8mm}
\gridline{ \fig{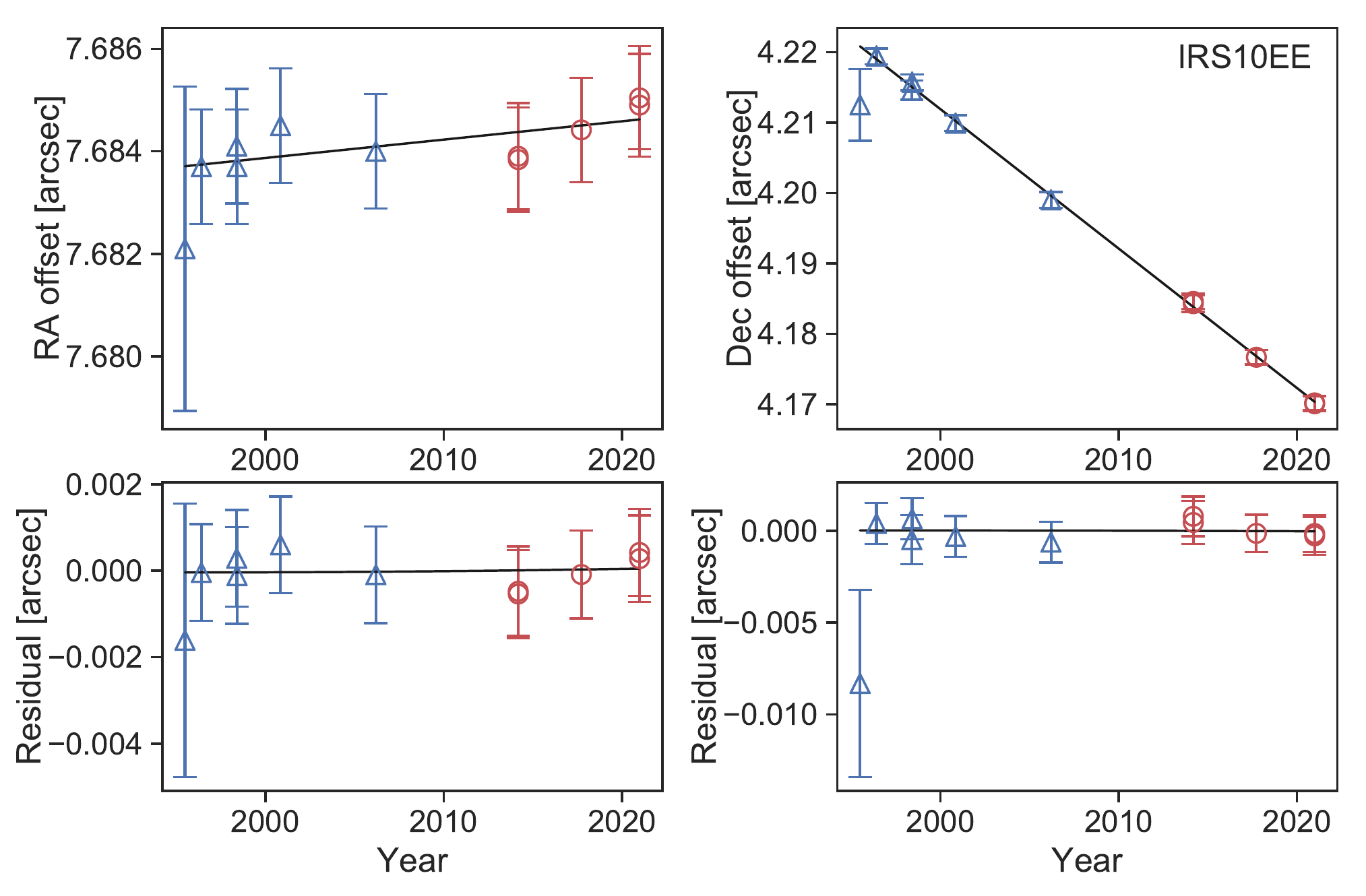}{0.48\textwidth}{}
           \fig{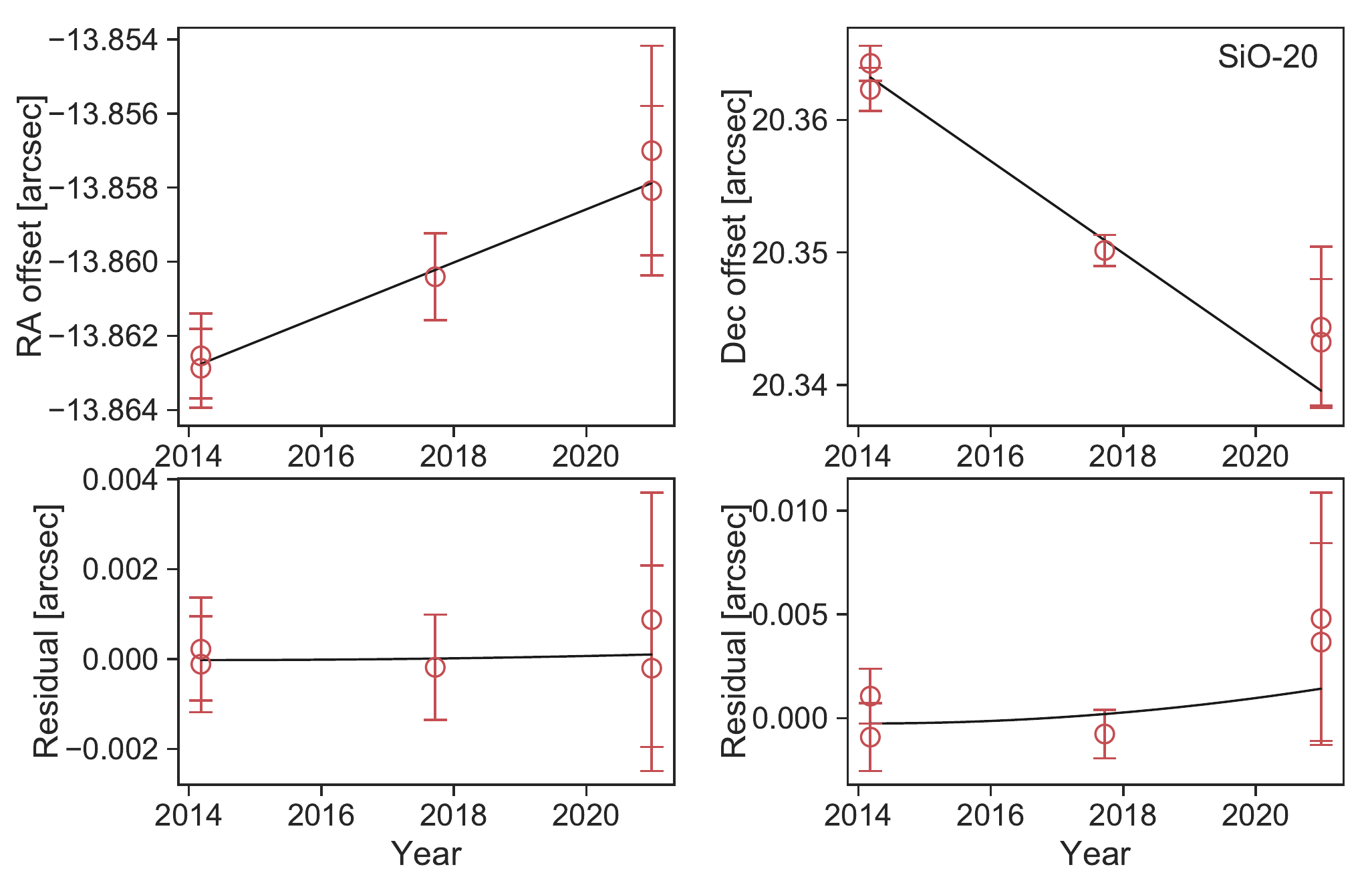}{0.48\textwidth}{} }
\caption{(Continued)}
\end{figure}

\setcounter{figure}{5}
\begin{figure}[ht!]
\gridline{ \fig{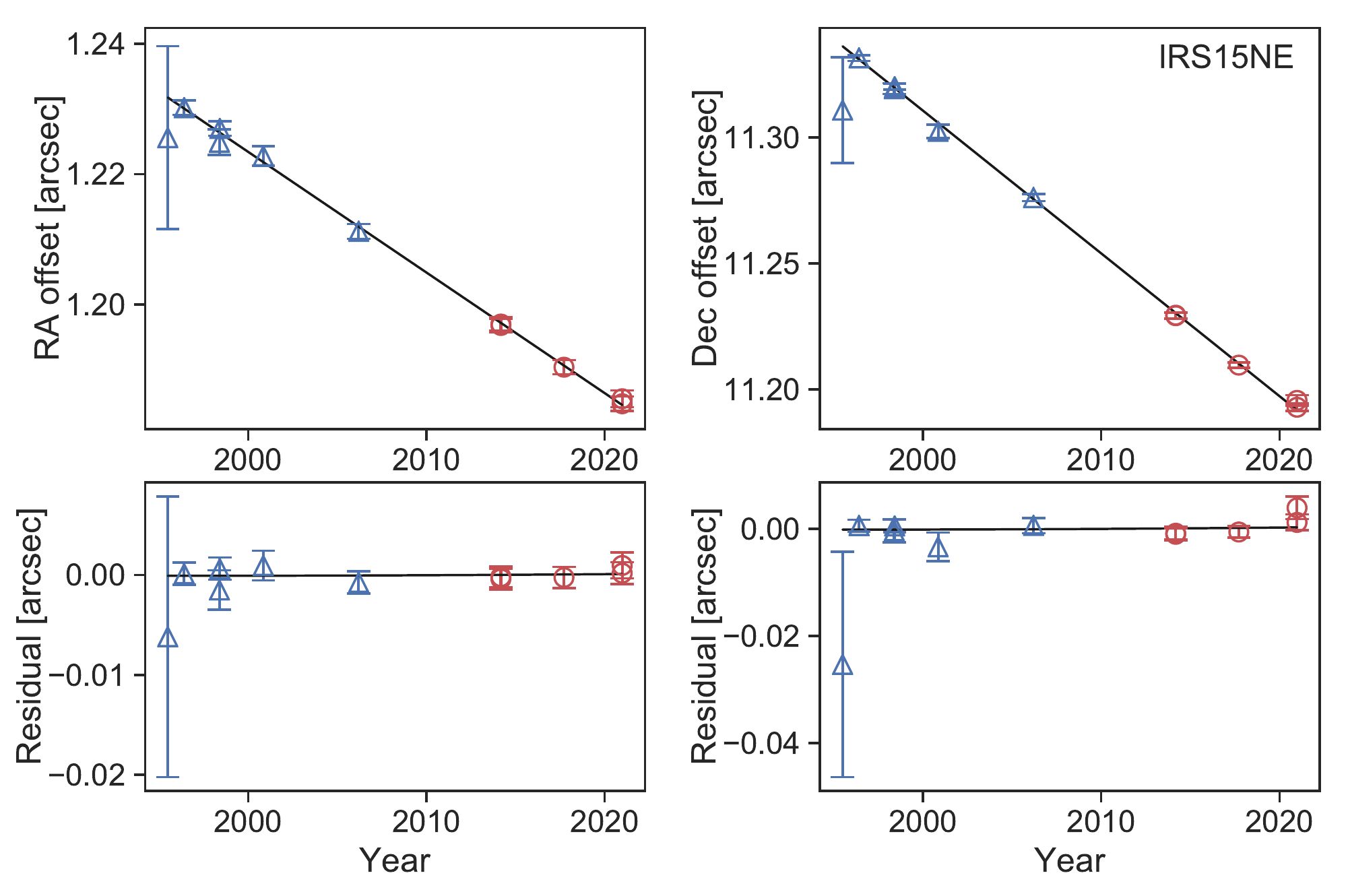}{0.48\textwidth}{}
           \fig{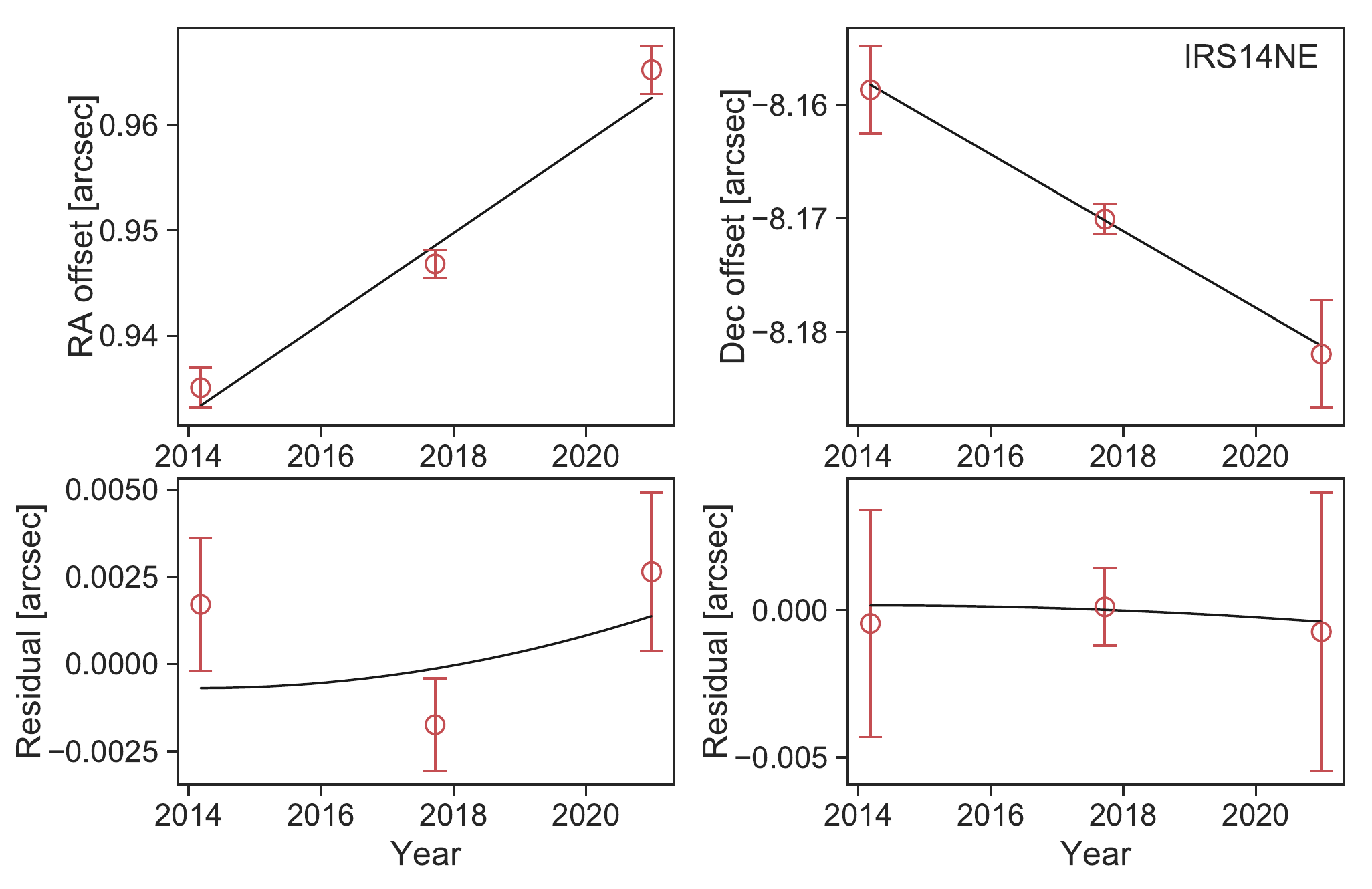}{0.48\textwidth}{} }
\vspace*{-8mm}
\gridline{ \fig{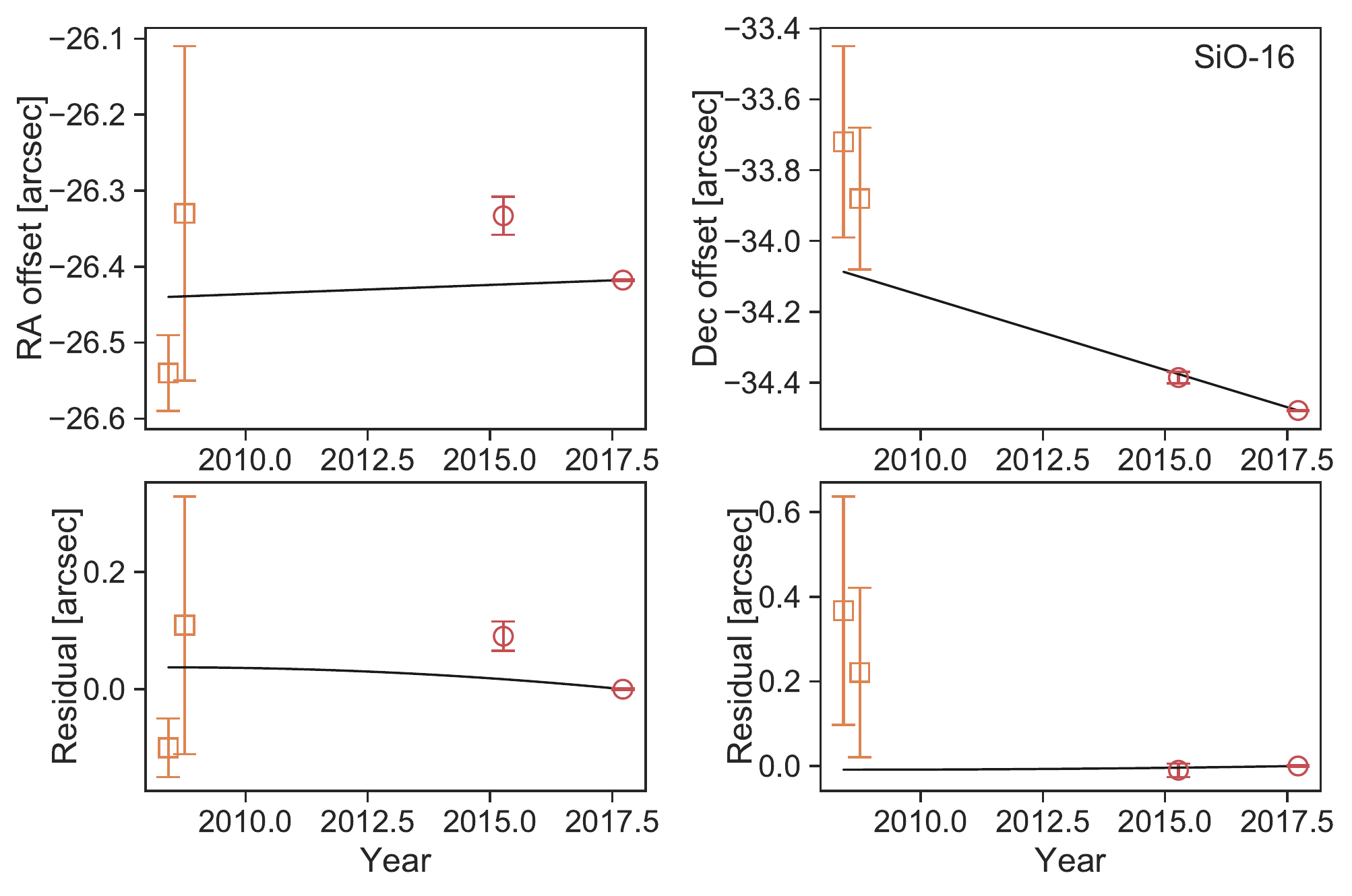}{0.48\textwidth}{}
           \fig{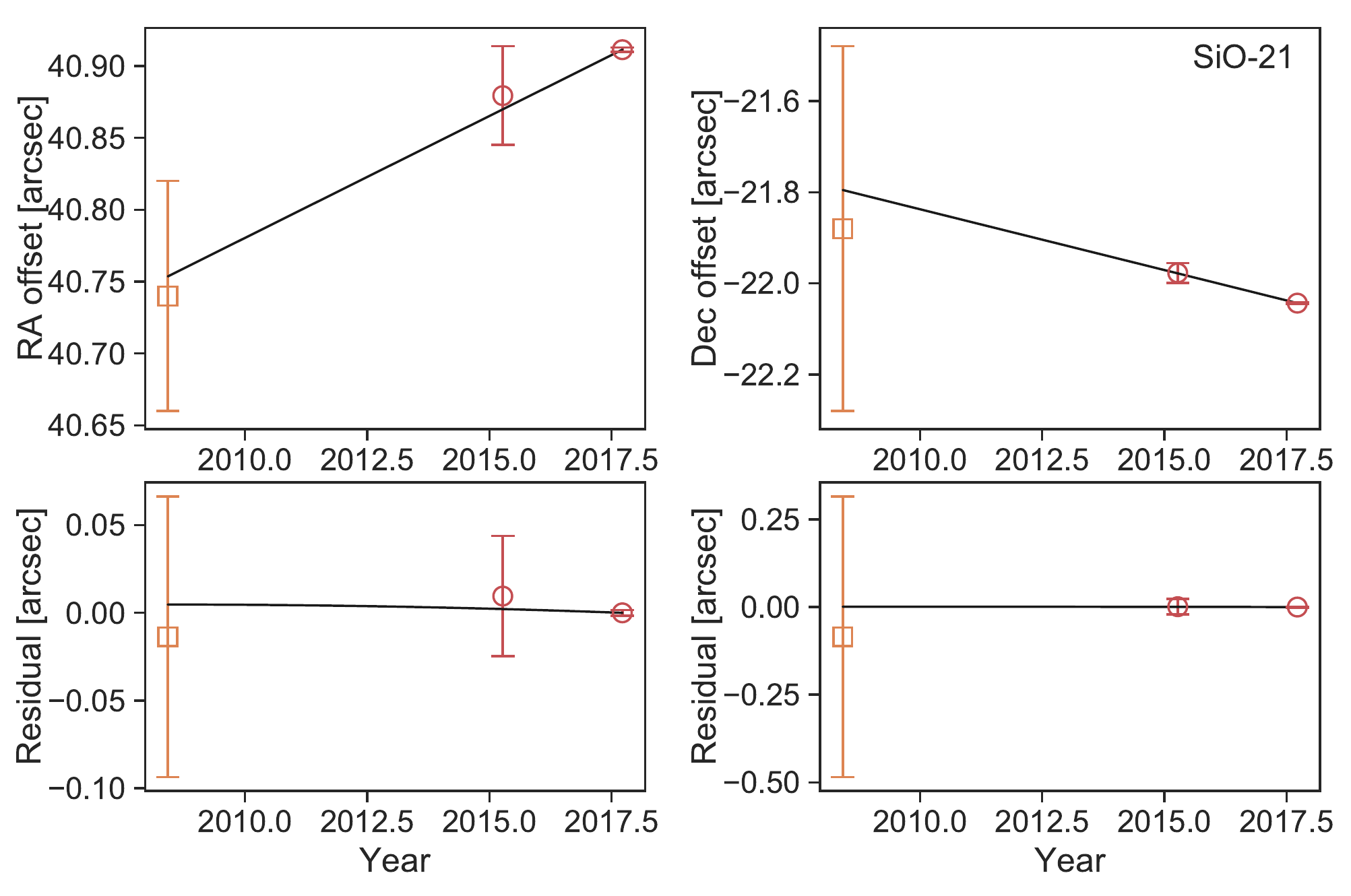}{0.48\textwidth}{} }
\vspace*{-8mm}
\gridline{ \fig{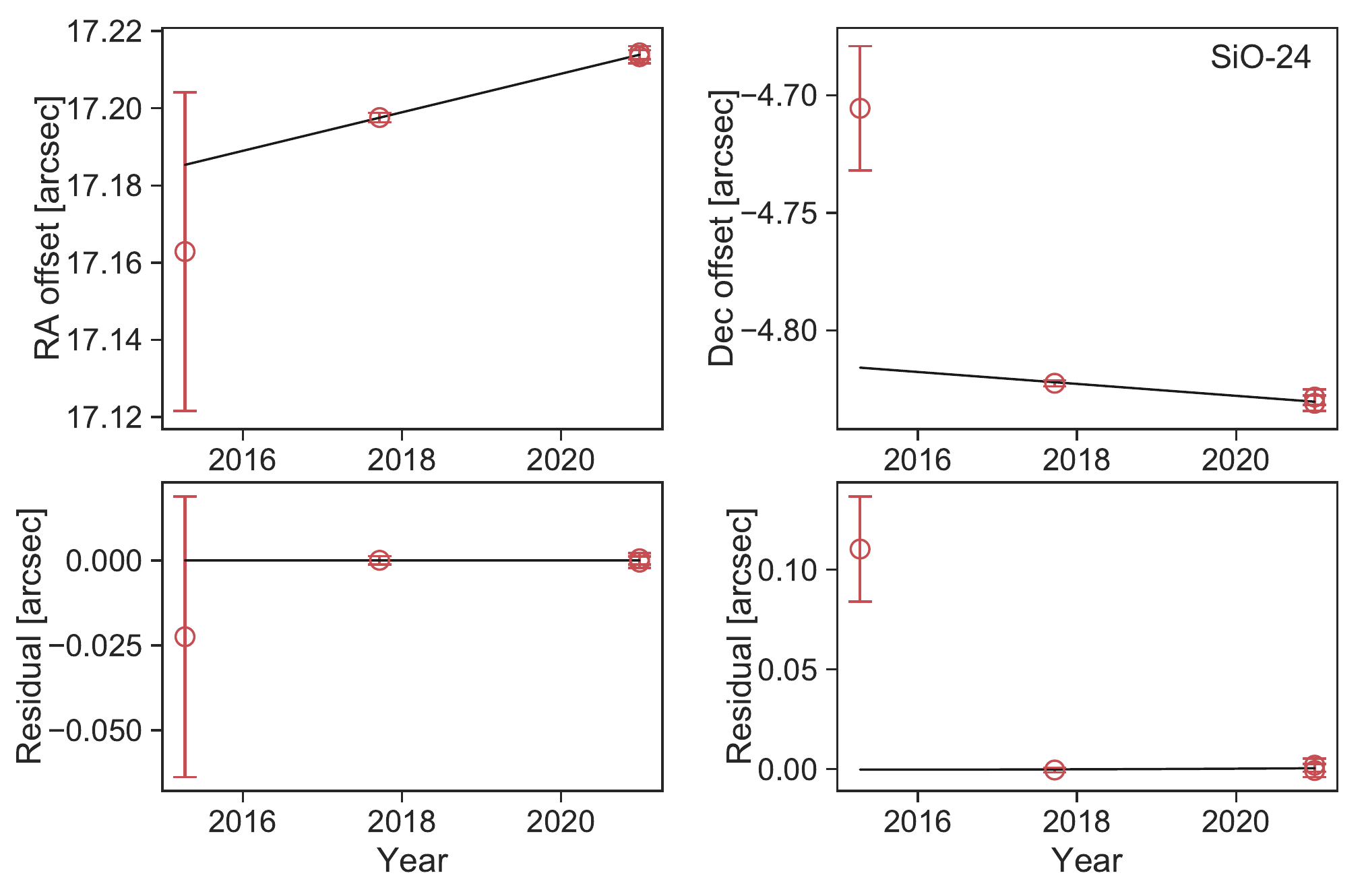}{0.48\textwidth}{}
           \fig{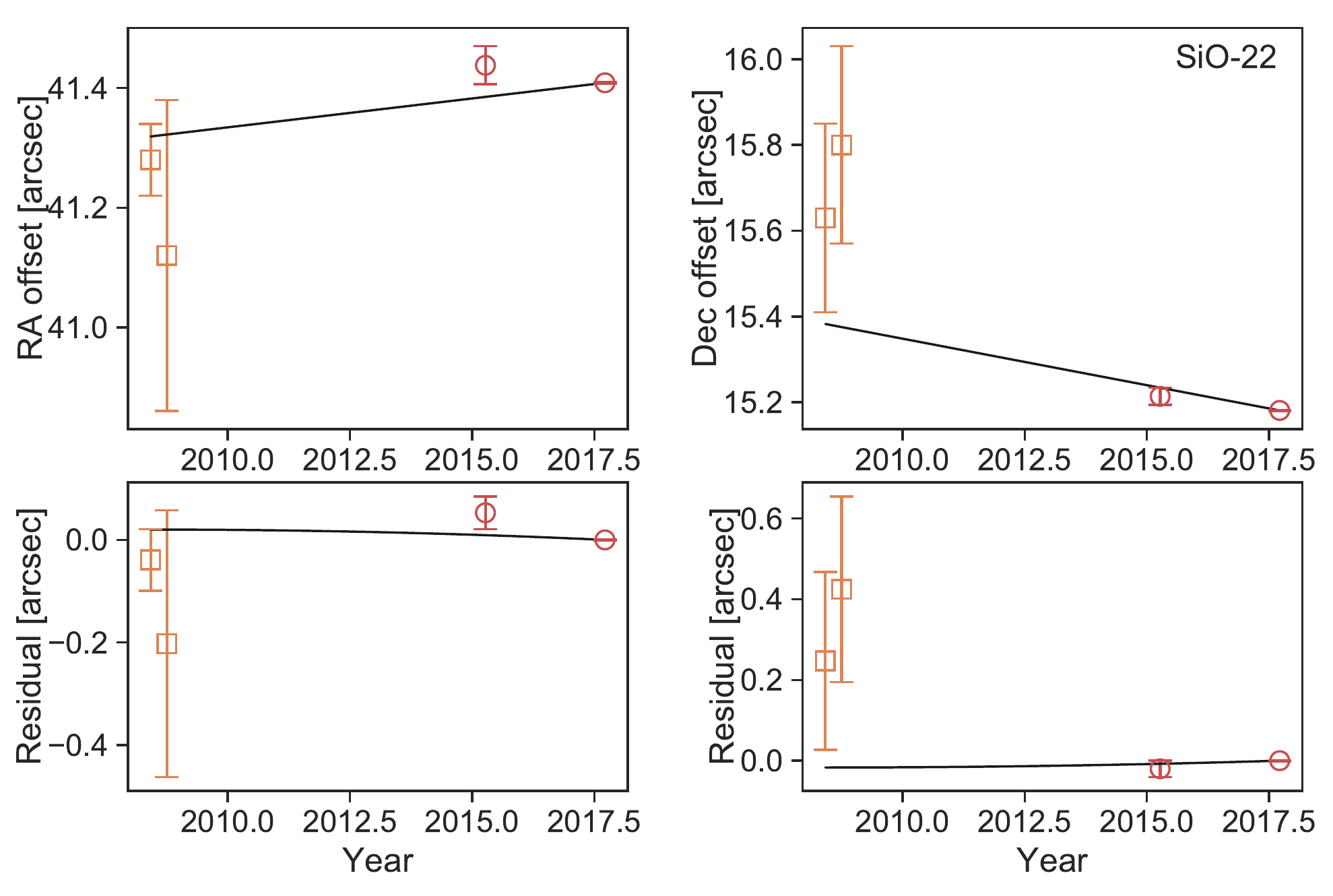}{0.48\textwidth}{} }
\caption{(Continued)}
\end{figure}

\setcounter{figure}{5}
\begin{figure}[ht!]
\gridline{ \fig{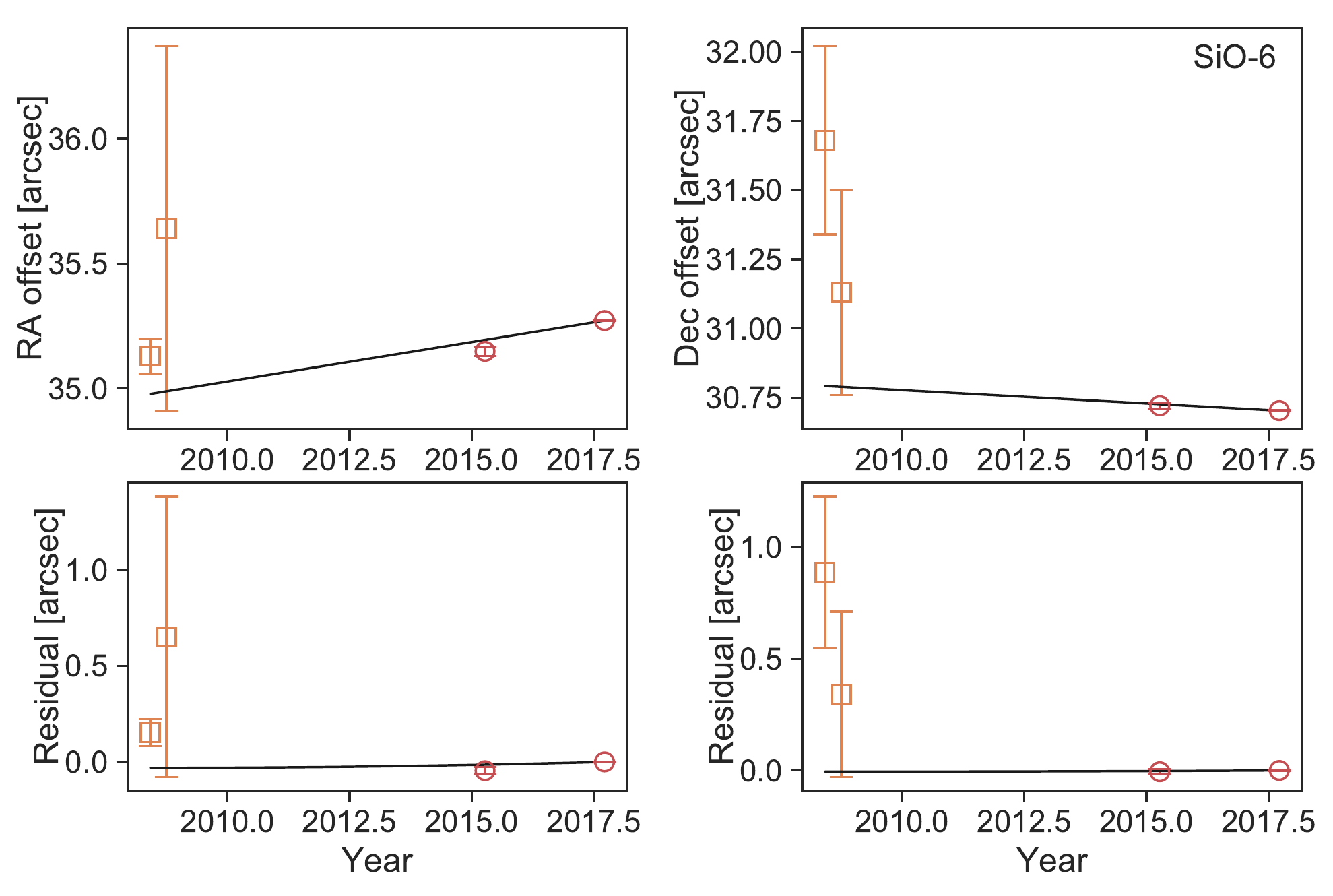}{0.48\textwidth}{}
           \fig{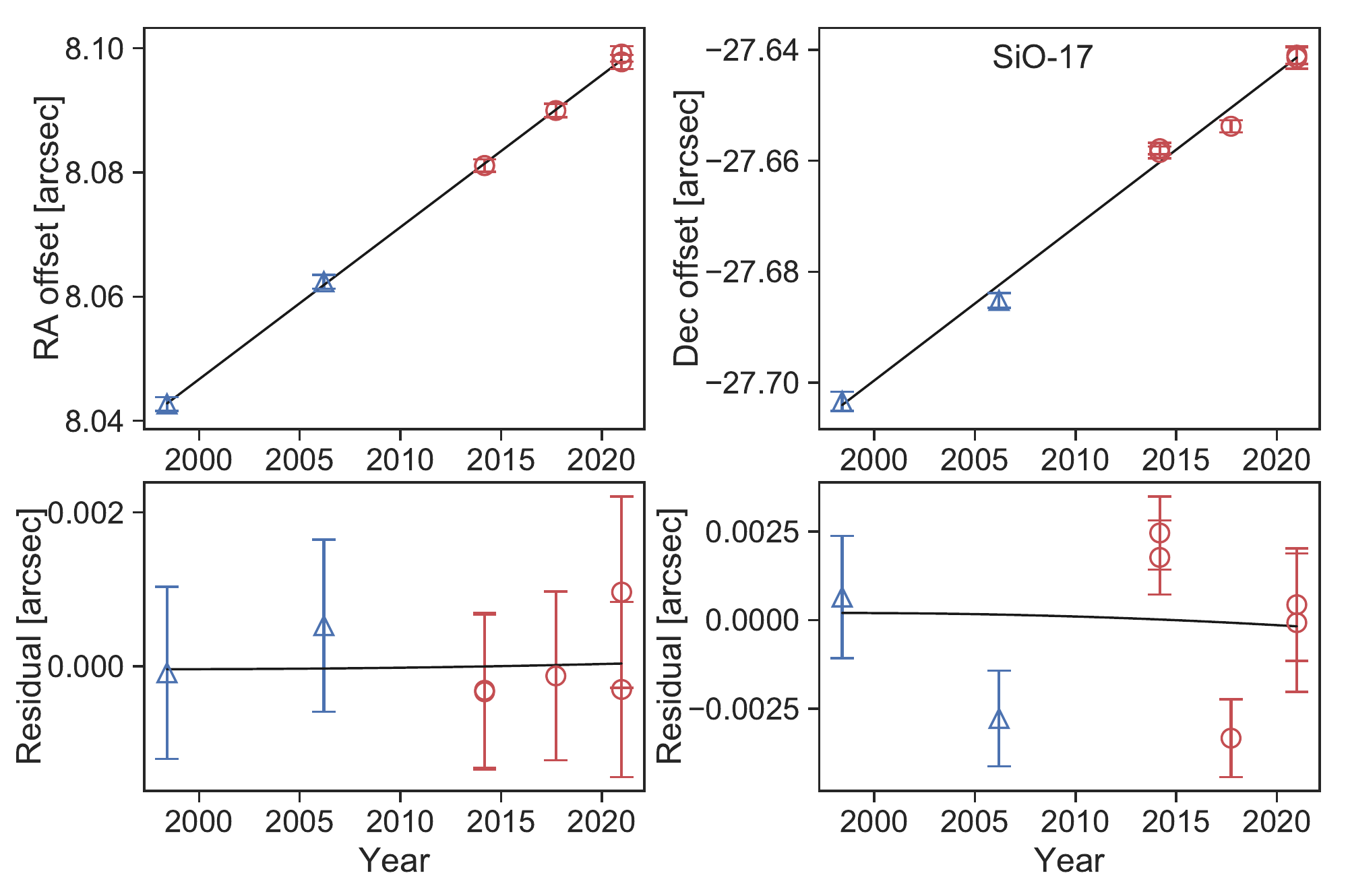}{0.48\textwidth}{} }
\vspace*{-8mm}
\gridline{ \fig{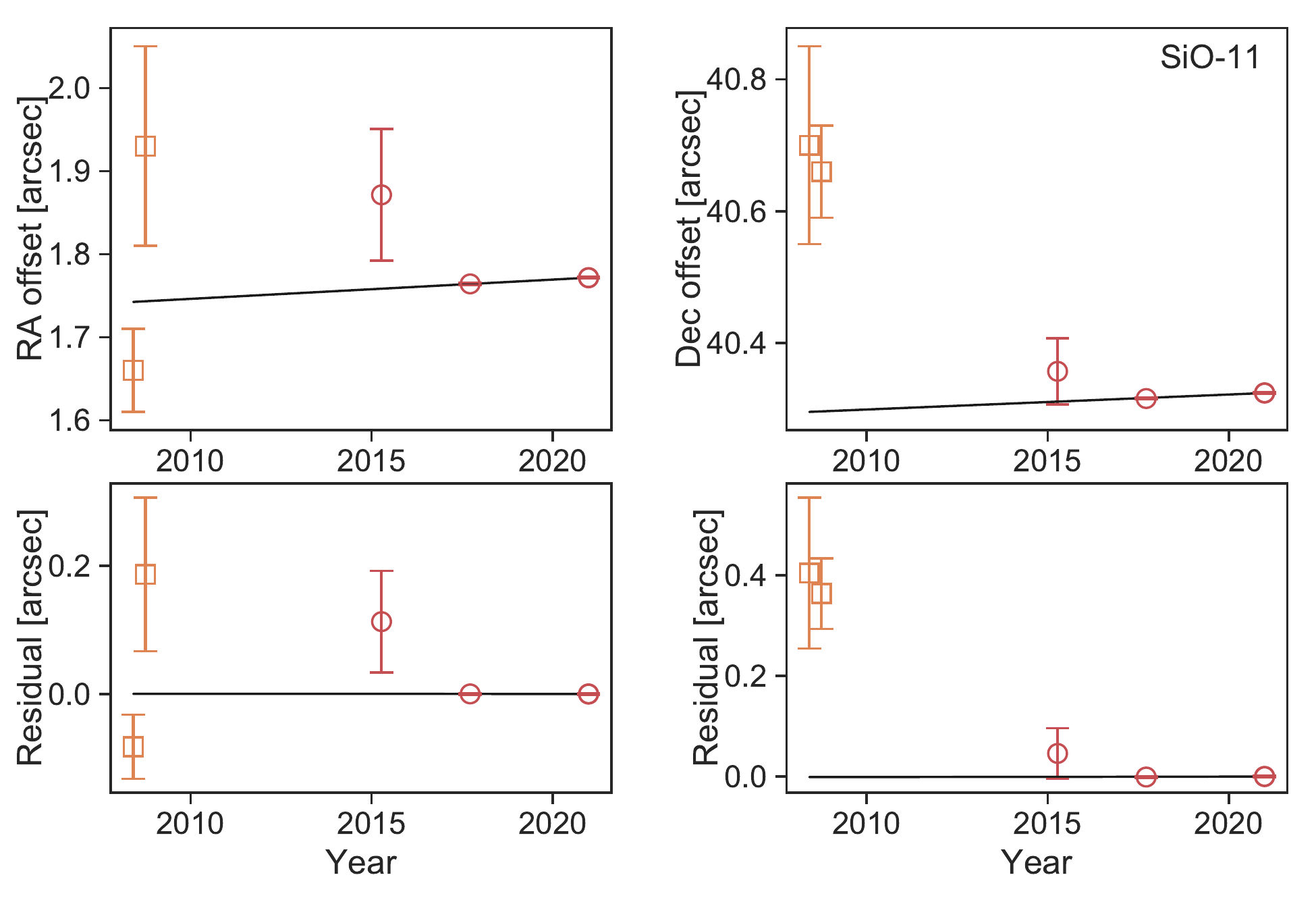}{0.48\textwidth}{}
           \fig{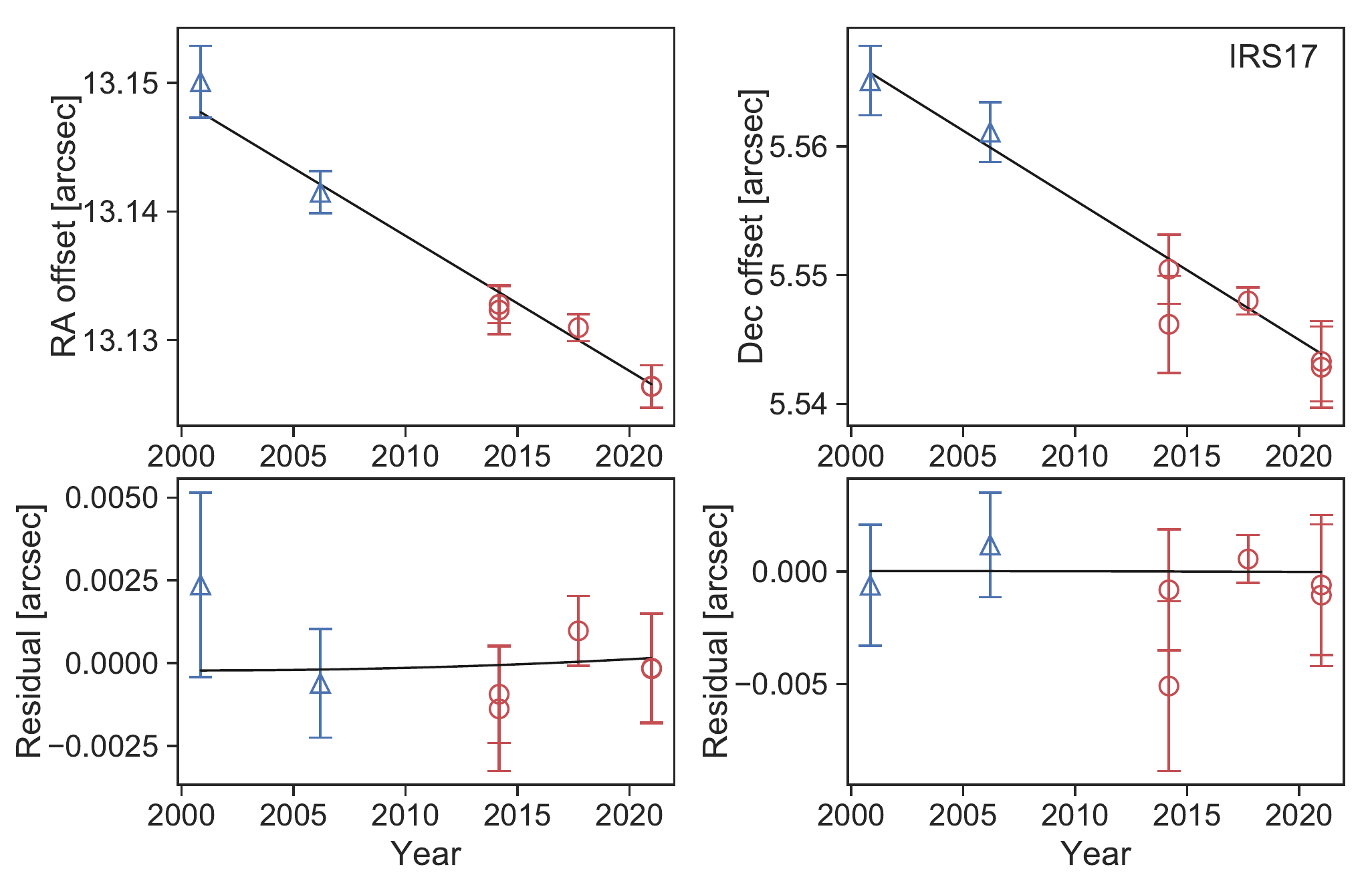}{0.48\textwidth}{} }
\vspace*{-8mm}
\gridline{ \fig{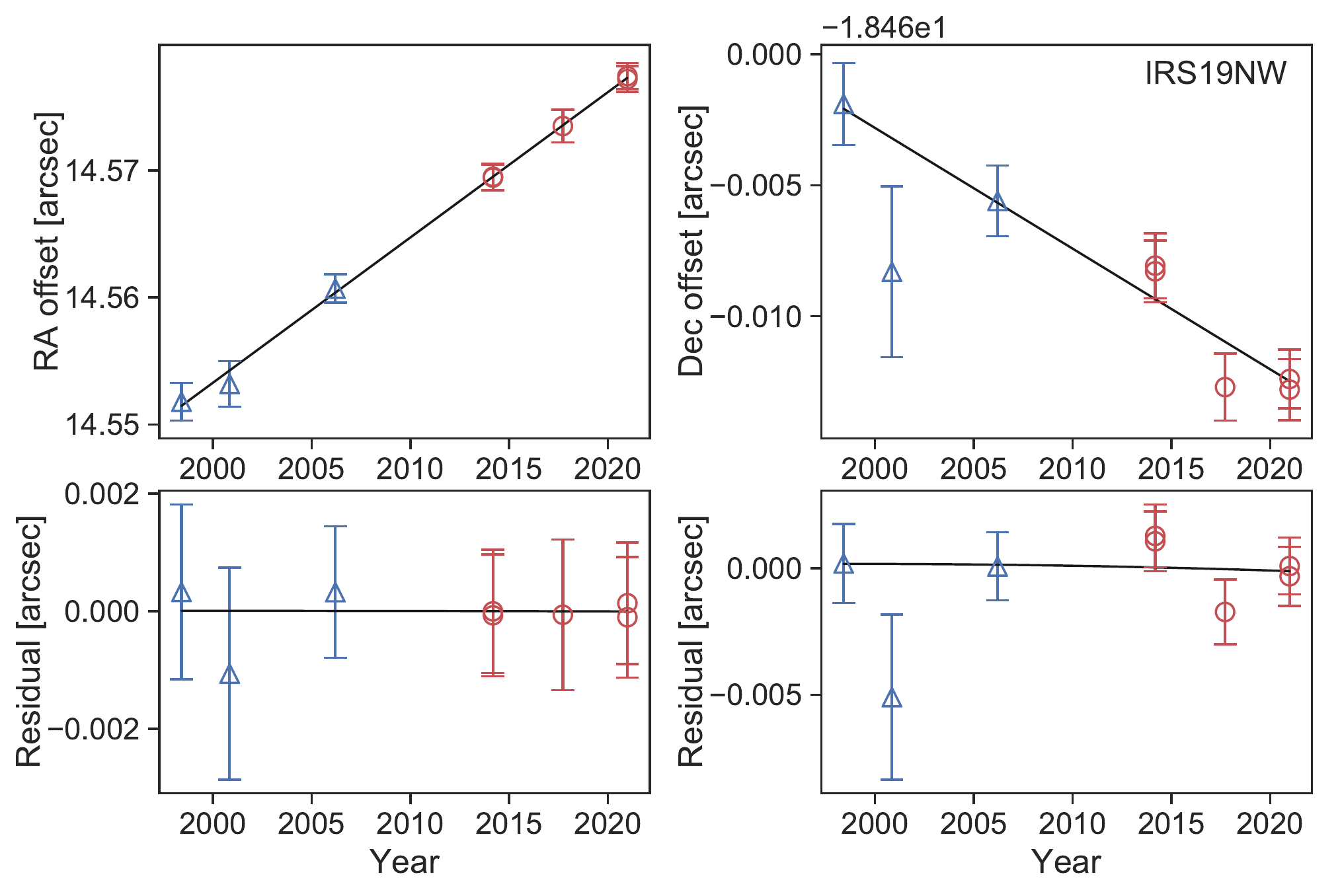}{0.48\textwidth}{}
           \fig{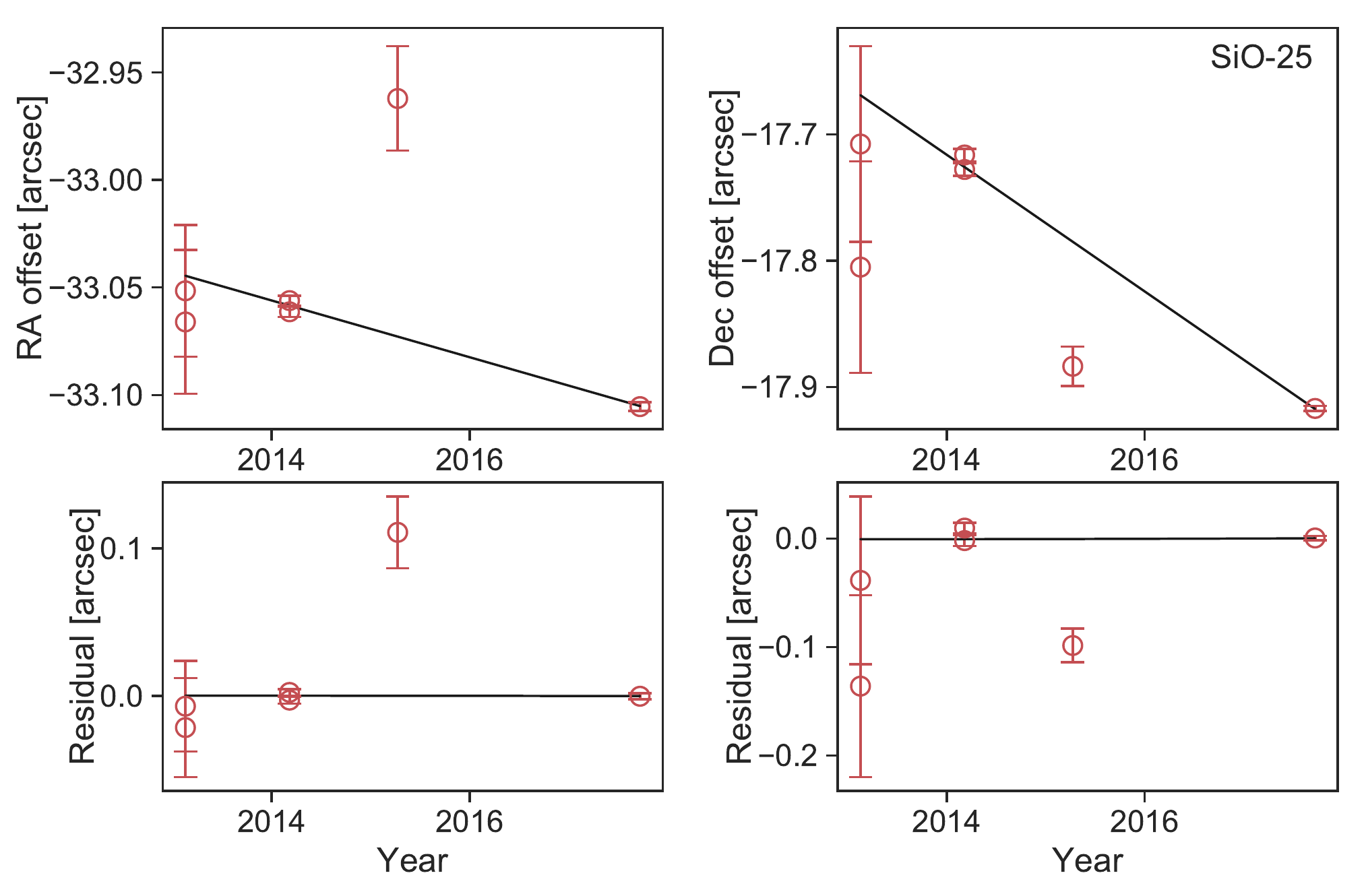}{0.48\textwidth}{} }
\caption{(Continued)}
\end{figure}

\end{document}